\documentclass[aps,prx,reprint,superscriptaddress, nobibnotes]{revtex4-2}

\usepackage[utf8]{inputenc}
\usepackage{amsthm}
\usepackage[normalem]{ulem}
\usepackage{amsmath}
\usepackage{amsfonts}
\usepackage{color}
\usepackage{dsfont}
\usepackage{dsfont}
\usepackage{enumerate}
\usepackage{soul}
\usepackage{hyperref}
\usepackage{graphicx}
\usepackage[dvipsnames]{xcolor}
\usepackage{braket}
\usepackage{comment}
\usepackage{placeins}
\usepackage[caption=false]{subfig}
\theoremstyle{definition}

\makeatletter
\newcommand*\bigcdot{\mathpalette\bigcdot@{.5}}
\newcommand*\bigcdot@[2]{\mathbin{\vcenter{\hbox{\scalebox{#2}{$\m@th#1\bullet$}}}}}
\makeatother

\theoremstyle{definition}
\newcommand{\norm}[1]{\left\lVert#1\right\rVert}
\begin{document}

\newcommand{\liege}{Institut de Physique Nucléaire, Atomique et de Spectroscopie, CESAM, Universit\'e de Li\`ege, 4000 Liège, Belgium}

\newcommand{\munich}{Department of Physics and Arnold Sommerfeld Center for Theoretical Physics,
Ludwig-Maximilians-Universität München, Theresienstr. 37, 80333 Munich, Germany}

\newcommand{\rice}{Department of Physics and Astronomy and Smalley-Curl Institute, Rice University, Houston, Texas 77005, USA}

\newcommand{\caltech}{California Institute of Technology, Pasadena, CA 91125, USA}

\newcommand{\dublinT}{School of Physics, Trinity College Dublin, College Green, Dublin 2, D02K8N4, Ireland}

\newcommand{\dublinR}{Rinn Quantum, School of Physics, Trinity College Dublin}

\title{Generating Multipartite Entanglement via Structured-Bath Engineering}

\author{Baptiste Debecker}
\thanks{These authors contributed equally to this work.}
\affiliation{\liege}
\author{Philipp Westhoff}
\thanks{These authors contributed equally to this work.}
\affiliation{\munich}
\author{Visal So}
\affiliation{\rice}
\affiliation{\caltech}
\author{Sebastian Paeckel}
\affiliation{\munich}
\author{Guido Pagano}
\affiliation{\rice}
\author{John Martin}
\affiliation{\liege}
\author{Mattia Moroder}
\affiliation{\dublinT}
\affiliation{\dublinR}
\author{François Damanet}
\affiliation{\liege}

\begin{abstract}
We present how to dissipatively prepare genuinely multipartite-entangled steady states with mutually independent local reservoirs by harnessing an Abelian strong symmetry and temporal bath correlations. In interacting Ising chains, we demonstrate that, for a broad class of system-environment couplings, flat baths exponentially erase Ising-axis correlations. Structured baths overcome this obstruction by converting temporal bath correlations into classical spatial correlations. Strong symmetry then promotes these correlations to the quantum domain, enabling global multipartite entanglement. Within weak-coupling theory, we obtain exact steady states despite the generator's nonquadratic nature: strong symmetry reduces their construction to a classical problem for arbitrary Ising couplings. This reduction reveals an exact connection between classical distinguishability and entanglement. We then show how to engineer the bath spectrum to increase the correlation length and enable the preparation of GHZ-like states from separable initial conditions. We test the mechanism beyond the analytical regime using exact Markovian embeddings and tensor-network simulations. Our approach is particularly suited to superconducting and trapped-ion platforms, and we propose a concrete implementation for the latter using sympathetically cooled motional modes. 
\end{abstract}

\date{\today}
\maketitle
\begingroup
\renewcommand{\thefootnote}{*}
\footnotetext{These authors contributed equally to this work.}
\endgroup

\section{Introduction} 
Dissipation is conventionally regarded as detrimental to quantum coherence and many-body correlations. Yet, when suitably engineered, it can instead stabilize nontrivial steady states~\cite{Diehl2008Q, Verstraete2009Q, Tindall2019}, accelerate state preparation~\cite{Zhan2026rapid, Pocklington2025A, Bao2025A, Westhoff2025bec}, and even stabilize exotic phases without equilibrium counterparts~\cite{Lee2013Unconventional, Jin2016Cluster}. Experimentally, engineered dissipation has enabled the preparation of correlated quantum gases~\cite{Syassen2008S} and long-range-correlated states in a superconducting array~\cite{Mi2024E}. More recently, trapped-ion platforms have demonstrated impressive control capabilities over engineered dissipation~\cite{So2024T, So2025Q, Sun2025Q}. A particularly ambitious target is the preparation of genuinely multipartite-entangled states, such as Greenberger-Horne-Zeilinger (GHZ) states, which constitute important resources for quantum simulation, metrology, and information processing~\cite{Greenberger1989, PhysRevLett.106.130506, RevModPhys.90.035005,Pont2024-se}.

\begin{figure*}[ht!]
    \centering
    \includegraphics[width=0.9\linewidth]{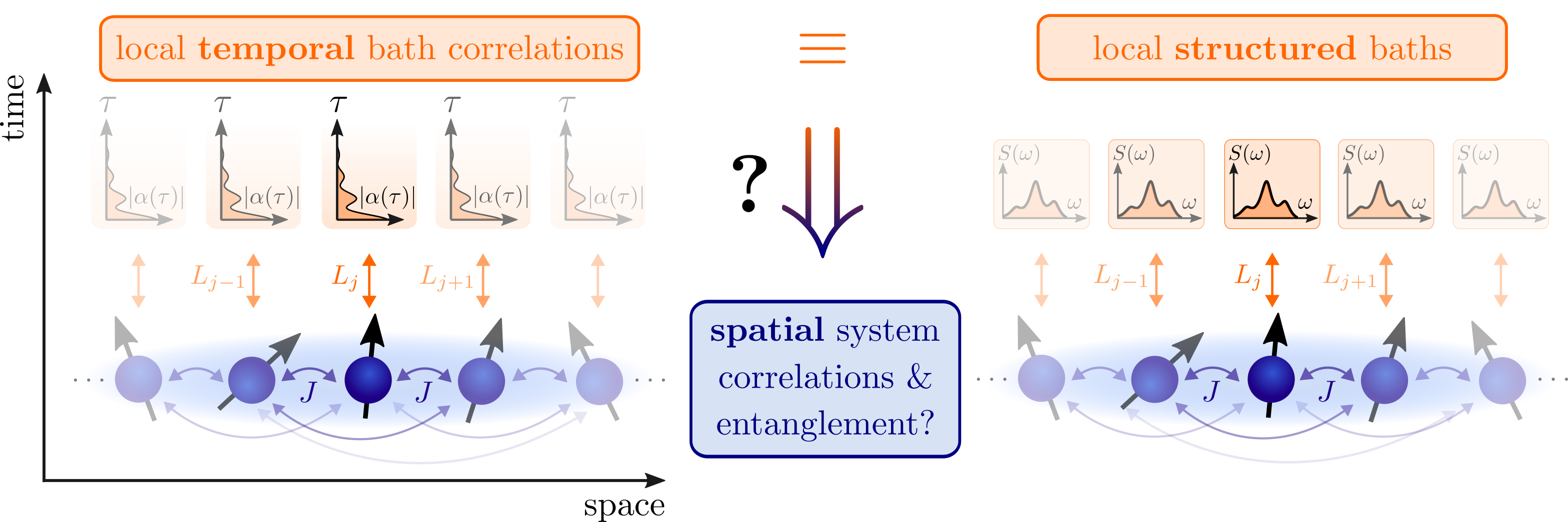}
    \caption{\textbf{Sketch of the central question}: Can independent local reservoirs with temporal correlations $\alpha(\tau)$ (left) --- or, equivalently, with structured spectra $S(\omega)$ (right) --- generate spatial correlations and even stabilize global multipartite entanglement in a one-dimensional system? Here, the system consists of an Ising chain in which each spin interacts with its own local reservoir via the coupling operators $L_j$ ($j = 1, \dotsc, N$).}
    \label{fig::1}
\end{figure*}

Understanding when dissipation suppresses correlations and when it can instead sustain, or even generate, quantum entanglement is therefore a central question in the theory of open quantum systems. An important strategy in dissipative state preparation is to couple different degrees of freedom to the same environment, as in collective or cavity-mediated dissipation~\cite{Chu2026R, Halati2025C, Kirton2019I, Kim2025C, Fallas2026G}. Indeed, a common environment can generate correlations even between otherwise noninteracting systems~\cite{Braun2002C, Benatti2003E}. In particular, dissipative preparation of GHZ states has been pursued using optical pumping through collective modes, engineered multiqubit dissipative maps, and quasi-local jump operators~\cite{PhysRevLett.106.020504, Barreiro2011-jp, Ticozzi2014, Camalet2013S, PhysRevA.96.062315}, as well as with measurement and feedforward protocols \cite{Moses2023A}. In these approaches, the global structure of the target state is encoded either in a shared mediator or directly in the engineered dissipative control. Recent experiments have also stabilized two-qubit entanglement using energy-selective local reservoirs~\cite{Guo2026Entangling}. Here, we ask a more fundamental question: Can mutually independent and strictly local reservoirs, i.e., each coupled to a single site, generate spatial correlations and even stabilize global multipartite entanglement?

At first sight, the answer might appear trivially to be negative since independent reservoirs provide neither shared fluctuations nor any direct channel through which correlations may propagate between different constituents. The reservoirs may, however, possess temporal structure. Equivalently, their response can depend on the frequency (and hence the energy) of a system transition. In an interacting many-body system, the energy cost of a local transition depends on the surrounding configuration, which raises the intriguing possibility of converting temporal bath correlations into spatial correlations and global entanglement---in the long time limit---through the interactions between the constituents, as illustrated in Fig.~\ref{fig::1}.

In this paper, we establish this conversion mechanism in open Ising chains, paradigmatic models that span statistical mechanics, quantum magnetism, critical phenomena, and quantum simulation. Beyond this broad significance, they provide a particularly transparent setting for our purpose because the Ising-axis correlators commute with the Hamiltonian. Thus, for such correlators, the roles of dissipation and coherent dynamics can be disentangled: coherent dynamics leaves them unchanged, so their evolution directly probes the action of the reservoirs.

\begin{figure}[t!]
    \centering
    \includegraphics[width=0.99\linewidth]{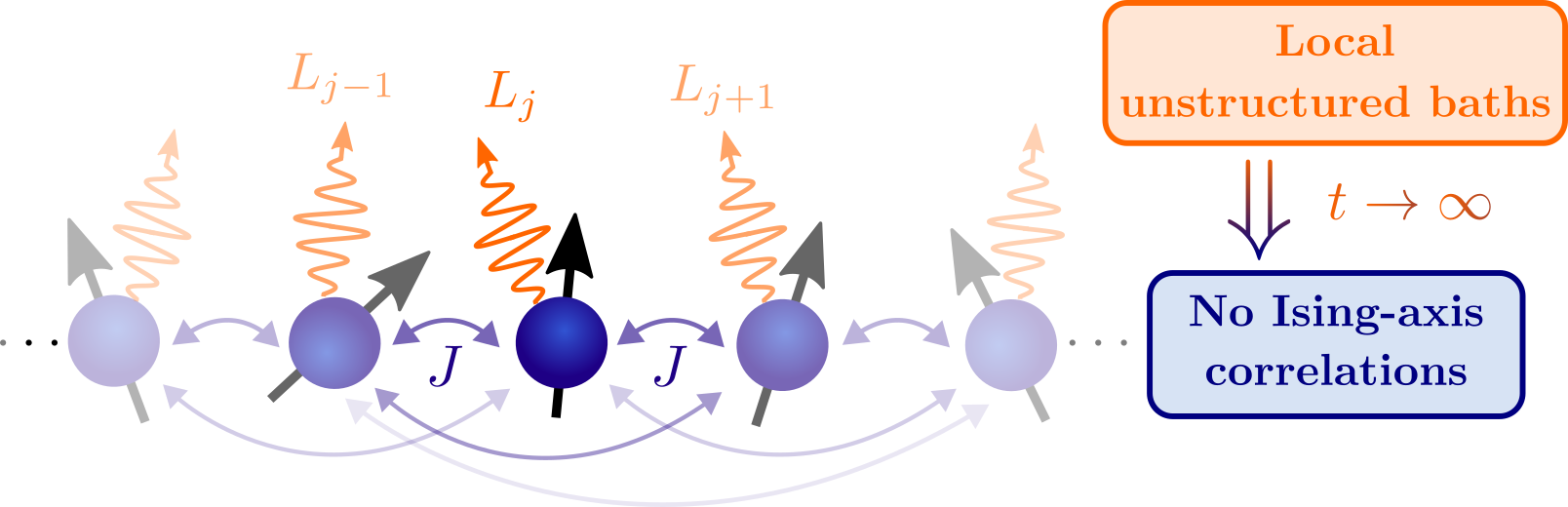}
    \caption{\textbf{No-go theorem}: For flat, unstructured baths, the dissipation is in standard Lindblad form. Our no-go result (see Sec.~\ref{Section::no-go}) applies, forcing the bath to exponentially damp Ising correlations. After an infinite time, all Ising correlations have been washed out.}
    \label{fig::1b}
\end{figure}

We first demonstrate that, for a broad class of local system-bath couplings, local frequency-independent Lindblad dissipation (agnostic of the Hamiltonian spectrum) exponentially erases Ising-axis correlations for arbitrary Ising couplings, as shown in Fig.~\ref{fig::1b}. By contrast, structured reservoirs can lift this obstruction by selectively biasing local spin flips according to their energy cost, thereby converting temporal bath correlations into spatial correlations. For the nearest-neighbor chain, these correlations are controlled by a simple bath-controlled effective temperature in the weak-coupling and secular regime~\cite{BreuerPetruccione2002}. This temperature is universal in the sense that it is set by the bath spectrum and independent of the local coupling operator within the class considered. These classical correlations provide the basis for the generation of multipartite entanglement. 

Crucially, when the system-bath coupling promotes the global $\mathbb{Z}_2$ symmetry from weak to strong, the dynamics of the populations is mirrored by that of the coherences between globally reversed spin configurations. We exploit this mirroring to construct exact steady states of the nonquadratic Redfield generator, without making the secular approximation. For arbitrary Ising couplings, constructing these steady states only requires solving a classical Pauli master equation. Furthermore, for the systems considered here, we uncover an exact connection between classical distinguishability and quantum entanglement: within a fixed symmetry sector, the negativity across any bipartition is determined by how distinguishable the classical stationary distribution is from the distribution obtained by reversing the spins on one side of that bipartition. Importantly, the analytical solution also provides a prescription for shaping the bath spectrum to prepare multipartite entanglement. For finite nearest-neighbor chains, sufficiently low effective temperature enables the preparation of GHZ-like states with genuine multipartite entanglement reached from initial separable states within a fixed symmetry sector. We also show how to bound negativity and certify genuine multipartite entanglement using two-point correlators, provided that the symmetry sector is known. Our mechanism is different from the stabilization of highly entangled stationary states via non-Abelian strong symmetries under unital dynamics from Refs.~\cite{Moharramipour2024S, Li2025H}. Here, non-Abelianity is not required: a simple $\mathbb{Z}_2$ suffices since we harness energy-selective local reservoirs, whose spectral structure breaks unitality.

We also test the robustness of our predictions by investigating the full structured-reservoir dynamics using Lorentzian baths, which admit an exact Markovian embedding in terms of local damped modes.  Combining this representation with state-of-the-art tensor-network simulations based on CLIK-MPS~\cite{Westhoff2025A}, we verify the emergence of spatial correlations and GHZ-like multipartite entanglement without relying on the weak-coupling or secular approximations. 

Finally, we propose a concrete trapped-ion implementation to test our predictions experimentally. In this implementation, individually addressed spins couple selectively to distinct collective vibrational modes, whose damping is provided by sympathetic cooling of ancillary ions. Together, our results establish bath spectral structure and strong symmetry as complementary resources for converting local dissipation into global entanglement. 

The remainder of the paper is organized as follows. In Sec.~\ref{sec::physical setting}, we present the models and motivate the local system-bath couplings. In Sec.~\ref{Section::no-go}, we demonstrate that flat baths exponentially destroy Ising-axis correlations. In Sec.~\ref{sec::frombathmem}, we discuss in detail the mechanism that converts temporal bath correlations into spatial correlations. In particular, we show that --- even if the baths are not thermal --- a simple effective temperature controls a large part of the physics at play.  In Sec.~\ref{sec:sz}, we construct the entangled steady states, establish their connection to classical distinguishability, and discuss simple criteria for detecting their entanglement. In Sec.~\ref{sec::trapped-ion}, we detail our proposed experimental implementation for trapped-ion platforms, before concluding in Sec.~\ref{sec::conclusion}. 

\section{Physical setting}
\label{sec::physical setting}
We consider a chain of $N$ spin-1/2 degrees of freedom described by an Ising Hamiltonian ($\hbar = 1$)
\begin{equation}
    H_S = - \sum_{i <j} J_{ij} \sigma^x_i\sigma^x_{j},
    \label{eq::HSIsingGeneral}
\end{equation}
with ferromagnetic coupling constants $J_{ij} >0$. To make the mechanism explicit, we will often specialize the results to the nearest-neighbor ferromagnetic chain, 
\begin{equation}
    H_S = -J \sum_j \sigma^x_j \sigma^x_{j+1}, \qquad J >0.
    \label{eq::IsingNearestNeighbour}
\end{equation}
This model provides a transparent setting for elucidating the role of local dissipation in selecting or generating correlations. Indeed, any operator diagonal in the $x$-basis trivially commutes with the Hamiltonian. In particular, $[H_S, \sigma_i^x \sigma^x_j] = 0$, which means that the coherent dynamics is neutral: it neither creates nor relaxes correlations along the Ising axis. The connected correlator $C^{xx}_{ij} = \braket{\sigma^x_i \sigma^x_j} - \braket{\sigma^x_i} \!\braket{\sigma^x_j}$ therefore provides a natural tool to diagnose the effect of the bath since any change in this observable must arise from the coupling to the environment. 

The Ising Hamiltonian~\eqref{eq::HSIsingGeneral} is classical in the $x$-basis and simply assigns an energy to each spin configuration. Because the energy cost of flipping one spin depends on the surrounding configuration, we introduce below structured reservoirs, that is, reservoirs whose response depends on the energy exchanged.

\subsection{Local and independent reservoirs}
Throughout this work, `local' refers to the support of the microscopic spin-reservoir coupling: each spin couples to its own reservoir, and distinct reservoirs are mutually independent. Thus, the bath is not allowed to provide a spatial channel between distinct spins: its adjustable resource is the temporal structure of each local reservoir.

We model this situation with an interaction Hamiltonian of the form 
\begin{equation}
    H_\mathrm{int} = \sum_j \left( L_j B_j^\dagger +  L_j^\dagger B_j \right),
    \label{eq::Hint}
\end{equation}
where $L_j$ acts only on spin $j$, while $B_j$ acts on the reservoir attached to that spin, as shown in Fig.~\ref{fig::1}. For concreteness, unless stated otherwise, we take identical reservoirs, each modeled as a collection of harmonic modes linearly coupled to their associated spin, so that the total environment Hamiltonian and reservoir operators $B_j$ take the form
\begin{equation}
 \begin{split}
    &H_E = \sum_{j,k} \omega_k b_{jk}^\dagger b_{jk}, \quad B_j = \sum_k g_k b_{jk}, \\
    \end{split}
\end{equation}
with $b_{jk}$ the annihilation operator for the $k^{\text{th}}$ bosonic mode of frequency $\omega_k$ of the $j^{\text{th}}$ reservoir, satisfying $[b_{jk}, b^\dagger_{j'k'}] = \delta_{jj'} \delta_{kk'}$, and with $g_k$ the coupling constant between each spin and the $k^{\text{th}}$ mode of its associated reservoir.

We take an initially factorized state $\rho_\mathrm{tot} = \rho_S(0) \otimes \rho_E$, with a product reservoir state $\rho_E = \otimes_j \rho_{E_j}$. In this context, mutual independence of the baths simply means 
\begin{equation}
    \langle B_i(t) B_j^\dagger (0) \rangle_E \equiv \mathrm{Tr}_E[B_i(t) B_j^\dagger(0) \rho_E]  = \alpha(t) \delta_{ij}.
\end{equation}
Here, $\alpha(t)$ is the temporal autocorrelation function of the reservoirs, $\rho_E$ is the reservoir state, assumed to be stationary with respect to the free bath Hamiltonian, and $B_i(t)$ is evolved under this free bath Hamiltonian $H_E$. This equation expresses the absence of cross-correlations between distinct reservoirs, while allowing correlations between two times within the same reservoir. For vacuum reservoirs, $\braket{B_i(t) B_j^\dagger (0)}_E$ completely characterizes the bath; at finite occupation, however, one must also retain $\braket{B_i^\dagger(t) B_j(0)}_E = \tilde{\alpha}(t) \delta_{ij}$, as discussed in Sec.~\ref{sec::finiteT}.

The function $\alpha(t)$ encodes the temporal memory of each local reservoir. Its frequency-domain counterpart is the dissipative bath correlation spectrum~\footnote{The imaginary part of the half Fourier transform in~\eqref{eq::def_S} produces the Lamb-shift contribution, which we will not need explicitly (see Section~\ref{sec::frombathmem}).}
\begin{equation}
    S(\omega) = \mathrm{Re} \int_0^{+\infty}dt~e^{i\omega t} \alpha(t),
    \label{eq::def_S}
\end{equation}
which determines the strength with which the reservoir induces system transitions at frequency $\omega$, and which reduces to the usual spectral density $J(\omega) = \pi \sum_k |g_k|^2 \delta(\omega - \omega_k)$ for zero-temperature reservoirs. If the bath correlations decay over a characteristic time $\tau_B$, its spectrum typically varies over a frequency range of order $1/\tau_B$. Hence, a memoryless correlation function $\alpha(t) \propto \delta(t)$ translates into a perfectly flat spectrum (i.e., an unstructured bath). This ideal white-noise limit assigns equal spectral weight to the relevant system transitions. By contrast, a structured spectrum can distinguish their energy costs, even when $\tau_B$ remains short compared with the spin relaxation time. Spectral selectivity --- the central environmental resource used in this work --- therefore does not necessarily require a breakdown of the Markov approximation for the reduced spin dynamics. 

\subsection{Symmetries and form of the coupling operators}
The Ising Hamiltonian~\eqref{eq::HSIsingGeneral} has two different global elementary symmetries that will play an important role in the following. The first is the unitary $\mathbb{Z}_2$ parity symmetry generated by 
\begin{equation}
    U_z = \prod_j \sigma^z_j.
\end{equation}
Because it flips the sign of each $\sigma_j^x$, it leaves every product $\sigma_i^x \sigma_j^x$ invariant. At the level of density matrices, it is represented by the superoperator 
\begin{equation}
    \mathcal{U}_z[\rho] = U_z \rho U_z^\dagger.
    \label{eq::weakSym}
\end{equation}
The second symmetry is an antiunitary, time-reversal-like symmetry $K_z$, implementing the complex conjugation in the $z$ basis. Since the Hamiltonian is real in this basis, it is invariant under $K_z$. At the superoperator level, we write 
\begin{equation}
    \mathcal{K}_z[\rho] = K_z \rho K_z^{-1}.
\end{equation}

We use these symmetries to define the class of coupling operators $L_j$ describing the spin-reservoir interaction~\eqref{eq::Hint}. This restriction is deliberate. Our goal is to isolate what can be achieved by reservoirs that are local and independent in space, and that do not themselves inject an explicit symmetry-breaking resource. In particular, breaking antiunitary or time-reversal symmetry at the superoperator level is known to play a key role in the generation of correlation and entanglement~\cite{Seif2022D}. We therefore require the coupling operators to respect, up to a phase, these two global symmetries of the Hamiltonian, i.e., 
\begin{equation}
    \mathcal{U}_z[L_j] = e^{i \phi_U} L_j, \qquad \mathcal{K}_z[L_j] = e^{i\phi_K} L_j.
\end{equation}
The phases $\phi_U$ and $\phi_K$ do not constitute an explicit symmetry-breaking resource, since they can be compensated by a corresponding transformation of the bath operators $B_j$. The transformation under $U_z$ nevertheless distinguishes two
physically different classes of system-reservoir couplings. Since
$U_z^2=\mathds{1}$, one necessarily has $e^{i\phi_U}=\pm1$. For $e^{i\phi_U}=+1$, the coupling operators commute with the parity, $[L_j,U_z]=0$. In this case, $U_z$ itself is a symmetry of the full system-environment Hamiltonian $H_{\mathrm{tot}}=H_S+H_E+H_{\mathrm{int}}$, namely $[H_{\mathrm{tot}},U_z]=0$, where $U_z$ is implicitly extended as $U_z\otimes\mathds{1}_E$. The system parity is therefore an exactly conserved quantity of the full dynamics, independently of any approximation used to derive a reduced description of the spin chain. For $e^{i\phi_U}=-1$, by contrast, the coupling operators anticommute with the parity, $\{L_j,U_z\}=0$. The system parity is then no longer conserved by the full system-environment Hamiltonian. Nevertheless, the total Hamiltonian still possesses a joint system-environment symmetry. Indeed, introducing the parity of the bosonic reservoirs, $P_E = \exp\left(i\pi\sum_{j,k} b_{jk}^\dagger b_{jk}\right)$, one readily verifies that $\widetilde U_z = U_z\otimes P_E$
commutes with $H_{\mathrm{tot}}$. Thus, in the first case, the system parity itself is conserved, in the second, only a joint system-environment parity is conserved. As we show below, these two classes of coupling operators give rise, respectively, to \textit{strong} and \textit{weak} $U_z$ symmetries of the reduced open-system dynamics.

For a single traceless one-site coupling operator, the requirements above select two families. The first,
\begin{equation}
    L_j^s = \sigma_j^z,
    \label{eq::Ljsz}
\end{equation}
commutes with $U_z$ and is invariant under $\mathcal{K}_z$. In the $x$-basis, it flips spin $j$, giving the most direct realization of the domain-wall process discussed above. This case will also play an important role in generating quantum correlations. The second family is
\begin{equation}
    L_j = \alpha \sigma_j^- + \beta \sigma_j^+, \qquad \alpha, \beta \in \mathbb{R},~|\alpha|\neq |\beta|,
    \label{eq::Ljsmsp}
\end{equation}
which instead anticommutes with $U_z$, while remaining invariant under $\mathcal{K}_z$. Here, the ladder operators $\sigma_j^\pm$ are defined as $\sigma_j^\pm = (\sigma_j^x \pm i \sigma_j^y)/2$. Notice that the restriction $|\alpha| \neq |\beta|$ excludes the trivial case $L_j = \sigma_j^x$ ($\alpha = \beta$), and the case $L_j = \sigma_j^y$ which is equivalent to the strong-symmetry class with $U_z$ replaced by $U_y = \prod_j \sigma_j^y$. For convenience, we will refer to them as the strong-symmetry and weak-symmetry classes of coupling operators, anticipating the symmetry structure they induce in the reduced dynamics.

\section{No-go theorem for flat baths}
\label{Section::no-go}
We first establish what local reservoirs cannot achieve when their spectrum is considered flat. For the strong-symmetric and weak symmetric classes of couplings introduced in Sec.~\ref{sec::physical setting}, we show that every connected Ising-axis correlator necessarily decays exponentially for arbitrary couplings $J_{ij}$. In particular, this means that the steady state cannot sustain Ising-axis correlations.
\subsection{Flat-bath or memoryless limit}
In the ideal white-noise limit $\alpha(t) \propto \delta(t)$, the bath is perfectly Markovian. The dynamics of the spin chain is then exactly captured by the Lindblad master equation~\cite{Lindblad1976b}
\begin{equation}
 \frac{d}{dt} \rho = -i [H_S, \rho] + \sum_j \gamma_j \mathcal{D}_{L_j}[\rho] \equiv \mathcal{L}[\rho], 
 \label{eq::LindbladLimit}
\end{equation}
with $\gamma_j >0$, the dissipator 
\begin{equation}
    \mathcal{D}_{L_j}[\rho] = 2L_j \rho L_j^\dagger - \{L_j^\dagger L_j, \rho\},
\end{equation}
 and $\mathcal{L}$ the Lindbladian, i.e., the generator of the dynamics. The jump operators are the local coupling operators $L_j$ introduced in the previous section [see Eq.~\eqref{eq::Ljsz} and Eq.~\eqref{eq::Ljsmsp}]. More details about this limit can be found in Appendix~\ref{memorylesslimit}.
\subsection{No-go theorem}
We now state the no-go theorem and the essence of its proof. A full demonstration is provided in Appendix~\ref{Proof}.

\textbf{No-go theorem.} \textit{Consider the local Lindblad dynamics given by Eq.~\eqref{eq::LindbladLimit}, with the Ising Hamiltonian $H_S$ defined in~\eqref{eq::HSIsingGeneral}, arbitrary couplings $J_{ij}$, and local jump operators $L_j$ from the strong-symmetric or weak-symmetric classes. Then, for any initial density matrix and any pair of sites $i \neq j$, the connected Ising correlator $C^{xx} (i, j) = \braket{\sigma^x_i \sigma^x_j} - \braket{\sigma^x_i} \!\braket{\sigma^x_j}$ ($i\neq j$) obeys 
\begin{equation}
    C^{xx}(i,j, t) = e^{-(A_i + A_j) t} C^{xx}(i, j, t=0),
\end{equation}
where the local damping rate $A_j$ is defined as 
\begin{equation}
A_j = \left\{ 
    \begin{array}{ll}
        4 \gamma_j& \mbox{if}~L_j = \sigma^z_j \\
        \gamma_j(\alpha-\beta)^2& \mbox{if}~~L_j = \alpha\sigma_j^- + \beta \sigma_j^+
    \end{array}
\right.
\end{equation}
Consequently, all the correlators $C^{xx}$ vanish in the steady state.}

The proof is straightforward. The key observation is that the Hamiltonian contribution vanishes identically (remember that is why we chose these correlators in the first place), while each local dissipator simply damps the $\sigma^x_j$ operator of each site $j$ at a rate that depends on the class of the coupling operator. In this memoryless limit, the bath only damps correlations and cannot select a steady state with nontrivial Ising-axis correlations.

\section{From bath memory to spatial correlations}
\label{sec::frombathmem}
The no-go theorem of the previous section demonstrates that a flat bath cannot stabilize Ising-axis correlations; this is because a given spin is damped at the same rate, regardless of what its neighbors are doing. We now show how the aforementioned obstruction can be bypassed by structured baths. Again, we stress that we do not introduce a spatially correlated bath, and we do not use a coupling operator $L_j$ that explicitly breaks the symmetries of the Ising Hamiltonian $H_S$. The only new ingredient is that the local reservoirs have a non-flat spectrum $S(\omega)$. 

As already briefly discussed in Sec.~\ref{sec::physical setting}, while a flat bath sees only a local spin flip, a structured bath perceives the energy cost of that flip. In an interacting Ising chain, however, this energy cost depends on the neighboring spins. Thus, a bath that is local in space but structured in time becomes sensitive to the many-body spin configurations. It is this simple physical mechanism that we exploit. Therefore, converting nontrivial temporal structures --- stored in the correlation function $\alpha(t)$ --- into spatial correlations does not require strong coupling, nor explicitly non-Markovian dynamics. Instead, the key ingredient is that the bath resolves the spectrum of $H_S$, and this already happens in the weak-coupling and secular regime.

In this section, we consider the generic weak-symmetry class of jump operators, Eq.~\eqref{eq::Ljsmsp}, namely $L_j = \alpha \sigma_j^- + \beta \sigma_j^+$ with $|\alpha| \neq |\beta|$. The strong-symmetry class will be discussed in Sec.~\ref{sec:sz}. Below, we will first discuss analytical results on the steady state properties of the system that can be obtained via the derivation of a standard weak-coupling master equation for the chain, highlighting in particular the emergence of a correlation length entirely controlled by the bath spectrum. Then, we will discuss how robust this universal behavior is beyond the weak-coupling and secularized regime by relying on numerically exact tensor-network calculations.

\subsection{Bohr frequencies and eigenoperators}
Deriving a secular master equation requires resolving the system coupling operators into system transitions with definite energy changes: the Bohr frequencies $\omega$. For the nearest-neighbor Ising chain, this decomposition has a particularly simple physical content since a local spin flip can only modify the two bonds adjacent to that spin. For the rest of this section, we thus consider the nearest-neighbor Ising chain. We will return to the general case in Sec.~\ref{sec:sz}.

We introduce raising and lowering operators along the Ising axis ($x$-basis)
\begin{equation}
    \tau_j^\pm = \frac{\sigma^z_j \mp i \sigma^y_j}{2},
\end{equation}
together with projectors onto the orientations --- in the $x$-basis --- of the neighboring spins
\begin{equation}
    \Pi_j^{ss'} = \frac{1+ s \sigma^x_{j-1}}{2} \frac{1+ s' \sigma^x_{j+1}}{2}, \quad s, s' = \pm1.
\end{equation}
Thus, $\Pi_j^{++}$, for instance, selects configurations in which both neighbors of site $j$ point along $+x$.

We denote by $\mathcal E_j(\omega)$ the subspace of local eigenoperators centered on site $j$ and associated with Bohr frequency $\omega$. Any operator
$O_j\in\mathcal E_j(\omega)$ satisfies
\begin{equation}
    [H_S,O_j(\omega)] = \omega O_j(\omega),
    \label{eq::conventionKj}
\end{equation}
so that $\omega$ is the energy change induced by $O_j(\omega)$. With this convention, $\omega >0$ corresponds to a process that raises the energy.

For the conditional spin flips introduced above, one readily finds
\begin{equation}
    [H_S, \Pi_j^{ss'} \tau_j^\eta ] = -2J \eta (s+s') \Pi_j^{ss'} \tau_j^\eta,\quad \eta = \pm1.
\end{equation}
The three possible Bohr-frequency sectors can therefore be identified
directly from the local spin configuration: when the two neighbors are aligned, $s=s'$, flipping the central spin changes the energy by $\pm4J$. The energy-raising subspace is thus
\begin{equation}
    \mathcal{E}_j(+4J) = \mathrm{span}\left\{\Pi_j^{++}\tau_j^-, \Pi_j^{--} \tau_j^+\right\}
\end{equation}
since $\Pi_j^{++} \tau_j^-$ and $\Pi_j^{--} \tau_j^+$ create two domain walls and raise the energy by $4J$. Conversely, the energy-lowering subspace is 
\begin{equation}
     \mathcal{E}_j(-4J) = \mathrm{span}\left\{\Pi_j^{++}\tau_j^+, \Pi_j^{--} \tau_j^-\right\},
\end{equation}
since $\Pi_j^{++}\tau_j^+$ and $\Pi_j^{--} \tau_j^-$ destroy two domain walls. Finally, when the two neighbors are antialigned, flipping the central spin only moves a domain wall from one adjacent bond to the other without modifying the system energy. Adding dephasing $\sigma_j^x$ that obviously belongs to this sector too, the energy-conserving subspace is 
\begin{equation}
    \mathcal{E}_j(0) = \mathrm{span}\left\{\Pi_j^{+-} \tau_j^+, \Pi_j^{+-} \tau_j^-, \Pi_j^{-+} \tau_j^+, \Pi_j^{-+} \tau_j^-, \sigma_j^x\right\}.
\end{equation}
The two nonzero-frequency subspaces are therefore two-dimensional, while the zero-frequency spin-flip and dephasing subspace is five-dimensional for every site $j$. Note also that, unless otherwise stated, we assume periodic boundary conditions (PBC) or open boundary conditions (OBC) in the thermodynamic limit $N \rightarrow +\infty$ throughout to ensure that every spin has two neighbors.

\begin{figure}[t!]
    \centering
    \includegraphics[width=0.9\linewidth]{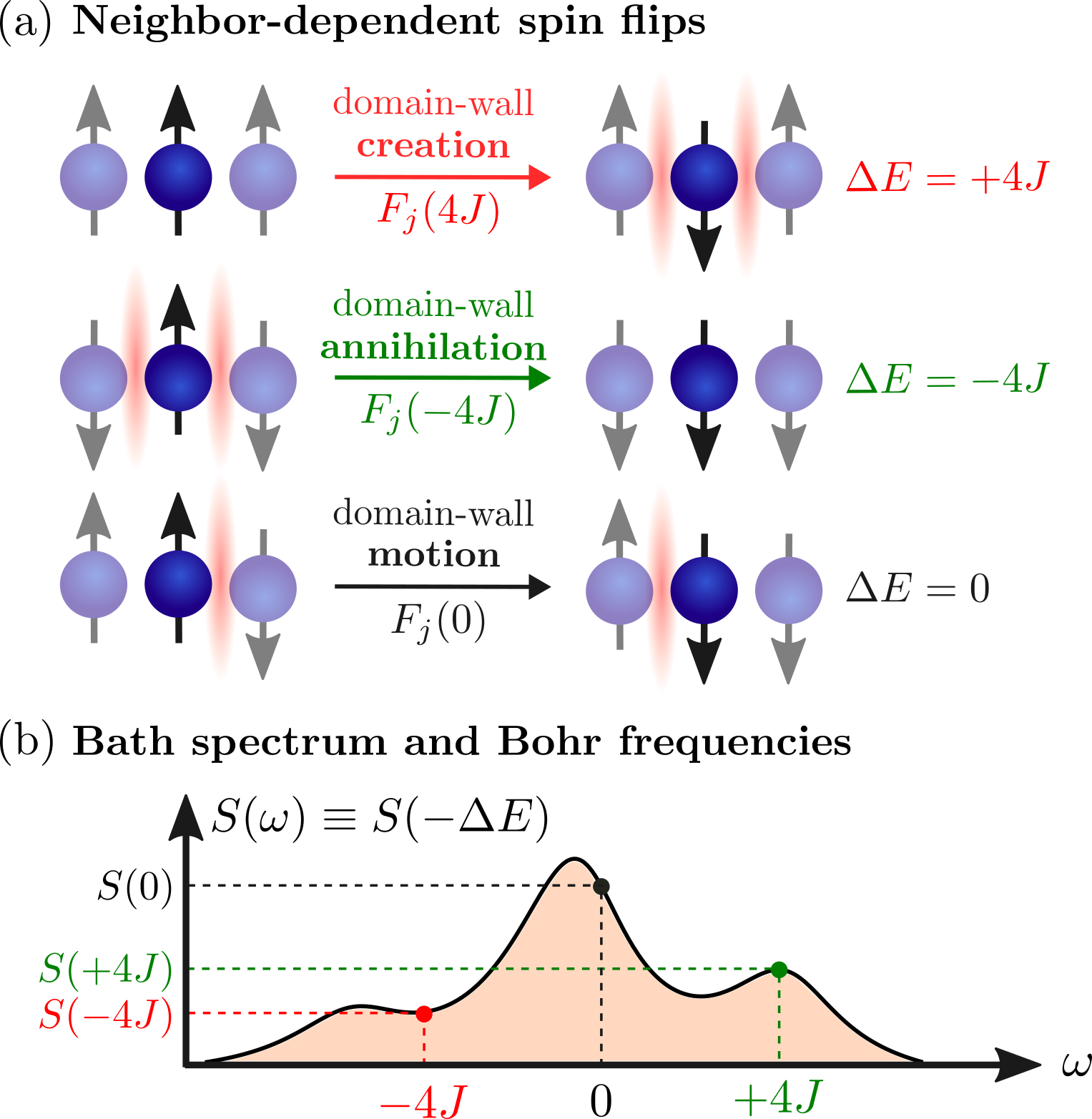}
    \caption{\textbf{Spectrally-resolved spin-flips:} (a) Sketch of the three possible neighbor-dependent spin flips associated with jump operators $F_j$ for the nearest-neighbor Ising chain. When the spins enclosing a spin $j$ are aligned with spin $j$, flipping spin $j$ creates two domain walls and increases the system energy by $+4J$ (red). When the spins enclosing a spin $j$ are aligned with each other but anti-aligned with spin $j$, flipping spin $j$ destroys the two domain walls and decreases the system energy by $-4J$ (green). When the spins enclosing a spin $j$ are anti-aligned, flipping spin $j$ induces domain wall motion and dephasing without modifying the system energy (black). (b) Bath spectrum $S(-\omega)$ sampled at the Bohr frequencies associated with the processes described above, highlighting how bath spectrum engineering can bias the competition between creation, annihilation, and motion of domain walls.}
    \label{fig::2}
\end{figure}

\subsection{Universal classical Ising correlations}
Having identified the possible energy-resolved spin flips, the remaining question is which linear combination inside each Bohr-frequency subspace is selected by the microscopic coupling $L_j$. This distinction is important because the secular approximation separates processes with different energy changes, but it does not destroy coherences \textit{inside} a given Bohr frequency sector. 

For the weak-symmetry class of coupling operators, that is $L_j = \alpha \sigma_j^- + \beta \sigma_j^+$ with $\alpha, \beta \in \mathbb{R}$ and $|\alpha| \neq |\beta|$, projecting $L_j$ onto the three local Bohr sectors gives 
\begin{equation}
    L_j = F_j(+4J) + F_j(-4J) + F_j(0),
\end{equation}
with
\begin{equation}
 \begin{split}
    F_j(+4J) &= \frac{(\alpha - \beta)}{2} \left(\Pi^{--}_j \tau_j^+ - \Pi_j^{++} \tau_j^-\right), \\
    F_j(-4J) &= \frac{(\alpha - \beta)}{2} \left(\Pi^{++}_j \tau_j^+ - \Pi_j^{--} \tau_j^-\right), \\
    F_j(0) &= \frac{(\alpha - \beta)}{2} \left(\Pi_j^{+-} + \Pi_j^{-+}\right)\left(\tau_j^+ - \tau_j^-\right) \\
    &\quad   + \frac{(\alpha + \beta)}{2} \sigma_j^x.
 \end{split}
\end{equation}
The action of these operators is sketched in Fig.~\ref{fig::2}.
A standard weak-coupling and secular derivation~\cite{BreuerPetruccione2002} leads then to the master equation 
\begin{equation}
 \frac{d\rho}{dt} = -i [H_S + H_\mathrm{LS}, \rho] + \sum_j \sum_{\omega = 0, \pm4J}  S(-\omega) \mathcal{D}_{F_j(\omega)}[\rho].
 \label{eq::secularME}
\end{equation}
Following standard practice, we assume that the Lamb-shift Hamiltonian $H_\mathrm{LS}$ is negligible, or coherently compensated. Note that the reversed sign in the argument of the bath spectrum $S(\omega)$ simply follows from our convention~\eqref{eq::conventionKj}, which implies that $F_j(+4J)$ raises the system energy by $4J$.

Crucially, taking diagonal matrix elements of the secular master equation yields a closed evolution for the Ising populations, even if the zero-frequency terms couple dephasing with domain-wall motion. Let $P_\mathbf{s}(t) = {}_x\!\braket{\boldsymbol{s}|\rho(t)|\boldsymbol{s}}_x $ be the population of the classical Ising configuration $\mathbf{s} = (s_1, ..., s_N)$ in the $x$-basis, where $s_j = \pm 1$, and let $\mathbf{s}^{(j)}$ denote the configuration obtained by flipping spin $j$. The corresponding energy change is
\begin{equation}
   \Delta E( \mathbf{s} \rightarrow \mathbf{s}^{(j)}) = E(\mathbf{s}^{(j)}) - E(\mathbf{s}) \in \{-4J, 0, +4J\}.
\end{equation}
We then obtain a closed Pauli master equation  
\begin{equation}
    \frac{d}{dt}P_\mathbf{s}(t) = \sum_j \left[W (\mathbf{s}^{(j)} \rightarrow \mathbf{s}) P_{\mathbf{s}^{(j)}}(t) - W(\mathbf{s} \rightarrow \mathbf{s}^{(j)}) P_\mathbf{s}(t)\right],
    \label{eq::Pauli_w}
\end{equation}
with the single-spin-flip rate 
\begin{equation}
 \begin{split}
    W(\mathbf{s} \rightarrow \mathbf{s}^{(j)}) 
    &= \frac{(\alpha - \beta)^2}{2}S\left(-\Delta E (\mathbf{s} \rightarrow \mathbf{s}^{(j)})\right).
 \end{split}
    \label{eq::single-flip_rate_weak}
\end{equation}
In Eq.~\eqref{eq::Pauli_w}, the first term is the incoming probability current from $\mathbf{s}^{(j)}$ to $\mathbf{s}$, while the second is the outgoing probability current from $\mathbf{s}$ to $\mathbf{s}^{(j)}$.

The structured reservoir, therefore, realizes a classical single-spin-flip process on the space of Ising configurations. Depending on the local energy change, that is $\Delta E = -4J, 0, +4J$, the corresponding event annihilates two domain walls, moves one domain wall, or creates two domain walls, respectively. The rates of these events are then directly set by the bath spectrum sampled at the right frequency $S(-\Delta E).$ Spatial correlations are thus generated by a purely classical kinetic mechanism: the bath spectrum biases domain-wall creation against domain-wall annihilation. Note that the closure is exact within the dissipative secular dynamics and holds for an arbitrary density matrix, not only for states initially diagonal in the Ising basis. 

The stationary distribution follows directly from local detailed balance. Defining an effective inverse temperature as 
\begin{equation}
    \beta_\mathrm{eff}   = \frac{1}{4J} \ln\left(\frac{S(4J)}{S(-4J)} \right),
\end{equation}
it follows that the ratio of the forward and reverse rates necessarily satisfies 
\begin{equation}
    \frac{W(\Delta E)}{W(-\Delta E)} = \frac{S(-\Delta E)}{S(\Delta E)} = e^{-\beta_\mathrm{eff} \Delta E}, \quad \Delta E = 0, \pm4J.
\end{equation}
Every elementary transition is therefore balanced by its reverse, and the exact stationary populations are 
\begin{equation}
      P_\mathbf{s}(t\rightarrow +\infty) \equiv P_\mathbf{s}^{\mathrm{ss}} = \frac{e^{-\beta_\mathrm{eff} E(\mathbf{s})}}{Z_\mathrm{eff}}, \quad Z_\mathrm{eff} = \sum_{\boldsymbol{s}} e^{-\beta_\mathrm{eff} E(\mathbf{s})}.
      \label{eq::Boltzmann}
\end{equation}
The reservoir itself need not be thermal, and we will explicitly consider nonthermal reservoirs in the next section. Rather, thermalization of the populations emerges from the sparse transition spectrum of the nearest-neighbor Ising chain. A generic nonthermal reservoir would assign different effective temperatures to different transition energies. Here, a single temperature is sufficient because only two nonzero frequencies are relevant. Note that the complete quantum steady state need not itself be Gibbsian since coherences may survive within degenerate energy subspaces. Its diagonal part is nevertheless fixed by the equation above independently of $\alpha$ and $\beta$ ($|\alpha| \neq |\beta|$). Hence, every observable diagonal in the Ising basis is universal across the entire family of coupling operators $L_j$ considered.

Introducing the spectral asymmetry ratio \begin{equation}
\chi = S(4J)/S(-4J),
\label{eq:analytic:correlation:lenght}
\end{equation}
the thermodynamic limit steady-state correlations are 
\begin{equation}
    \braket{\sigma^x_j \sigma^x_{j+r}} = \tanh(\beta_\mathrm{eff} J)^r,
         \label{eq::xxcorranal}
\end{equation}
with correlation length 
 \begin{equation}
 \xi^{-1} = -\ln|\tanh(\beta_\mathrm{eff}J)| = -\ln \left|\frac{\sqrt{\chi}-1}{\sqrt{\chi}+1} \right|
     \label{eq::xianal}.
 \end{equation}
Thus, an entire structured reservoir is reduced, as far as secular Ising correlations are concerned, to a single number $\chi$, that fixes the winner between domain-wall creation and annihilation. In other words, microscopically very different reservoirs can become indistinguishable at the level of steady-state correlations: an entire class of reservoirs sharing the same spectral asymmetry $\chi$ should yield the same correlation length.

\begin{figure}[t!]
    \centering
    \subfloat[]{
    \includegraphics[width=0.98\linewidth]{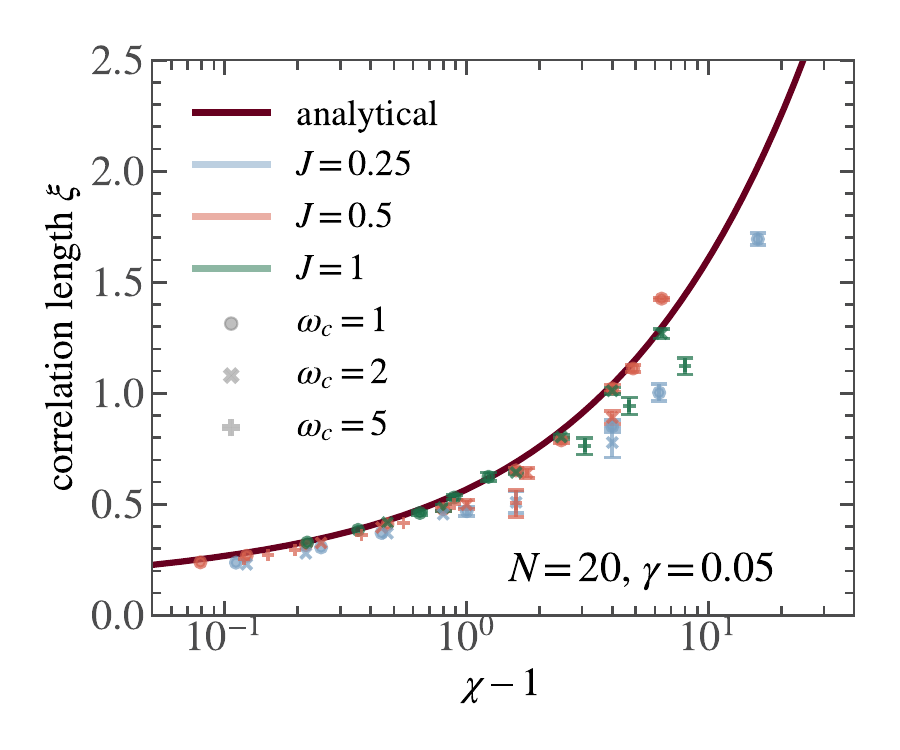}
    }
    \vspace{-0.8cm}
    \subfloat[]{
    \includegraphics[width=0.98\linewidth]{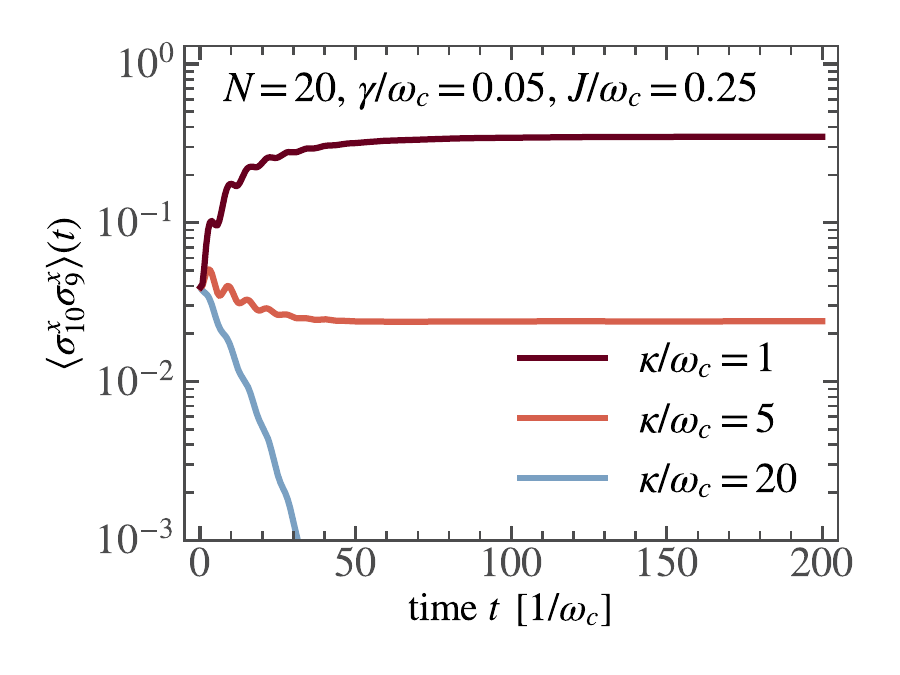}
    }
    \caption{\textbf{Universal behavior of the correlation length and buildup of correlations.} (a): Correlation length as a function of $\chi -1$ for different Lorentzian baths and compared with the analytical prediction of Eq.~\eqref{eq::xianal}. We work in units where $\gamma = 0.05$. (b): Example showing the $\kappa$-dependence in the buildup of correlations as a function of time. For a very short bath correlation time (large $\kappa$), the Ising-axis correlator $\braket{\sigma^x_{10} \sigma^x_9}$ decays almost exponentially in agreement with our no-go result. However, diminishing $\kappa$ increases the stabilized Ising-axis correlations. All points were computed with the exact Markovian embedding~\eqref{eq:markovian:embedding} with $L_j = \sigma_j^-$, for $N = 20$ sites using the CLIK-MPS method (see Appendix~\ref{app:sec:numerics} for details). }
    \label{fig::Universal}
\end{figure}

\subsection{Quantitative temporal-to-spatial conversion}
\label{sec::Quantitative}
The secular result makes a strong prediction: although a structured reservoir is characterized by its full spectrum $S(\omega)$, the secular steady-state Ising correlations depend only on the single ratio $\chi$. 

To test this prediction beyond the weak-coupling derivation, we consider that each spin interacts with its own independent Lorentzian reservoir, taken here at zero temperature for simplicity. In that case, the spectral function $S(\omega)$ reduces to the spectral density and is given by 
\begin{equation}
    S(\omega) = \frac{\gamma \kappa^2}{\kappa^2 + (\omega-\omega_c)^2}.
    \label{eq::lorentzian}
\end{equation}
This type of spectral density is both experimentally relevant and theoretically transparent. Such a Lorentzian spectrum is the effective environment generated by a single damped harmonic oscillator or pseudomode~\cite{Imamoglu1994Stochastic, Dalton2001Theory, Garraway1997Nonperturbative, Pleasance2020Generalized, Debecker2024C}, and therefore naturally describes superconducting qubits coupled to lossy microwave resonators~\cite{Blais2021Circuit}, atoms or artificial atoms in damped cavities~\cite{Garraway1997Nonperturbative, Dalton2001Theory}, and trapped-ion platforms in which cooled vibrational modes act as engineered reservoirs~\cite{Sun2025Q, So2025Q}. Theoretically, this choice is especially convenient because it admits a simple, exact Markovian embedding. Indeed, we can associate a local bosonic mode $a_j$ interacting with each spin and consider the enlarged Hamiltonian 
\begin{equation}
    H_\mathrm{emb} = H_S + \omega_c \sum_j a_j^\dagger a _j + g \sum_j (L_j a_j^\dagger + L_j^\dagger a_j),
    \label{eq::lindlabd_emb}
\end{equation}
with $g = \sqrt{\gamma \kappa}$. Each auxiliary mode is itself coupled to a structureless Markovian reservoir. The combined spin-pseudomode state $\rho_\mathrm{tot}$ then evolves according to the Lindblad equation 
\begin{equation}
    \frac{d\rho_\mathrm{tot}}{dt} = -i[H_\mathrm{emb}, \rho_\mathrm{tot}] + \kappa \sum_j \mathcal{D}_{a_j}[\rho_\mathrm{tot}],
    \label{eq:markovian:embedding}
\end{equation}
and the physical spin state is recovered by tracing out the auxiliary modes.

The same embedding also provides a direct time-domain description of the reservoir. Each damped pseudomode generates the local bath autocorrelation function 
\begin{equation}
    \alpha(t) = \gamma \kappa e^{-(\kappa+i\omega_c)t} \quad (t>0).
\end{equation}
Indeed, its one-sided Fourier transform, defined in Eq.~\eqref{eq::def_S}, yields exactly the Lorentzian spectrum in Eq.~\eqref{eq::lorentzian}. The parameter $\kappa$ therefore has a dual interpretation since it controls both the spectral linewidth and the inverse reservoir correlation time, i.e., $\tau_B = 1/\kappa$. Importantly, the secular steady-state correlations depend only on $\chi$, and are therefore insensitive to the overall spectral weight and to the particular parametrization of the spin-pseudomode coupling. Note, however, that we choose $g = \sqrt{\gamma \kappa}$ with $g^2/\kappa = \gamma $ fixed to ensure a proper large-$\kappa$ limit. As $\kappa \rightarrow +\infty$, the bath correlation time vanishes, and the spectral density becomes flat over all system frequencies. The reduced spin dynamics then obey the Lindblad limit master equation~\eqref{eq::LindbladLimit} with rate $\gamma$ and jump operator $L_j$~\cite{Debecker2026S, Debecker2024S} (see also Appendix~\ref{memorylesslimit}). Consequently, the no-go theorem applies and requires the steady-state correlations to vanish.

For the Lorentzian reservoir, the conversion of temporal bath structure into spatial correlations can be made fully explicit. Its spectral asymmetry ratio reads 
\begin{equation}
    \chi = \frac{\kappa^2 + (4J + \omega_c)^2}{\kappa^2 + (4J- \omega_c)^2}.
\end{equation}
Thus, this ratio decreases monotonically with the reservoir linewidth $\kappa$. Inserting this result in Eq.~\eqref{eq::xxcorranal} gives 
\begin{widetext}

\begin{equation}
    \xi^{-1}_L = \ln\left[\frac{\left(\sqrt{(1/\tau_B)^2 + (4J + \omega_c)^2} + \sqrt{(1/\tau_B)^2 + (4J - \omega_c)^2} \right)^2}{16 J \omega_c} \right].
\end{equation}
\end{widetext}
The spatial correlation length is consequently a monotonic function of the reservoir correlation time $\tau_B$. This expression makes quantitative the temporal-to-spatial conversion introduced earlier. At fixed spectral detuning, extending the lifetime of correlations in each local reservoir monotonically increases the spatial correlation length of the many-body steady state. In particular, on resonance, $\omega_c = 4J$, this mechanism reduces to a simple proportionality relation
\begin{equation}
    \xi_L = \frac{1}{2~\mathrm{arcsinh}(\kappa/8J)} \approx 4J \tau_B \quad (\kappa \ll J),
\end{equation}
and $\xi$ becomes theoretically unbounded.

Figure~\ref{fig::Universal} demonstrates that the universal prediction is not an artifact of the approximations. In panel~(a), we plot the steady-state correlation length obtained from the full spin-pseudomode dynamics~\eqref{eq::lindlabd_emb} against the spectral asymmetry $\chi$ for $L_j = \sigma_j^-$ (see Appendix~\ref{app:sec:numerics} for details about the tensor network simulations). Data obtained from different sets of parameters $\{J, \omega_c, \kappa \}$ ---corresponding to different reservoirs--- at fixed $\gamma$ quantitatively agree with the analytical curve~\eqref{eq::xianal}.  Thus, as far as $\xi$ is concerned, the system largely forgets the detailed shape of the reservoir spectrum, and only one quantity, i.e., $\chi$, is relevant. Deviations from the analytical prediction are more pronounced for the blue points, which correspond to the smallest $J$, where the secular approximation is less controlled. They accompany the emergence of antiferromagnetic correlations, as discussed in Appendix~\ref{app:beyondsecular}.

Panel~(b) provides a complementary dynamical picture. Starting from the same initial state, a narrow structured reservoir, corresponding to a small ratio $\kappa/\omega_c$, rapidly builds nearest-neighbor correlations. Decreasing the spectral resolution of the bath by increasing $\kappa$, however, leads to smaller selected correlations. For the broadest bath shown, the correlator is simply damped toward zero, thereby continuously recovering the flat-bath no-go result of Section~\ref{Section::no-go}.

\section{Global entanglement enabled by strong symmetry}
\label{sec:sz}
The previous section shows that a structured bath can select a classical distribution. We now show that strong symmetry has a striking dynamical consequence. The classical dynamics of the Ising populations is exactly mirrored by the set of coherences connecting globally reversed configurations. This, in turn, allows us to solve exactly for the steady state manifold despite the nonquadratic nature of the generator~\cite{Prosen2008T, Prosen2010Q, Barthel_2022}. Importantly, the result is valid for the general Ising model~(\ref{eq::HSIsingGeneral}) and at the Redfield level, that is, we do not have to assume the secular approximation. We then demonstrate that this steady manifold hosts globally entangled states. In particular, our protocol can stabilize highly pure GHZ-like states from strictly local baths without spatial correlations. 

\subsection{Mirroring the classical dynamics with strong symmetry}
Recall that the strong-symmetry coupling family $L_j^s = \sigma_j^z$ necessarily promotes the global parity symmetry $U_z$ to a strong symmetry since $[L_j^{s}, U_z] = 0$. This implies that the expectation value of $U_z$ is conserved by the dynamics~\cite{Buca2012}.

Interestingly, we will now show that the global strong symmetry $U_z$ creates a duplicate of the population dynamics in the globally reversed coherences sector $C_s(t) = {}_x\!\braket{\mathbf{s}|\rho|\mathbf{-s}}_x $. We start from the so-called Redfield master equation 
\begin{equation}
    \frac{d \rho}{dt} = -i[H_S, \rho] + \sum_j [\bar{L}_j^s\rho, L_j^s] + [L_j^s, \rho\bar{L}_j^{s\dagger}] = \mathcal{L}_\mathrm{R}[\rho],
    \label{eq::Redfield}
\end{equation}
where the effective coupling operator $\bar{L}_j^{s}$ is defined as 
\begin{equation}
    \bar{L}_j^{s} \equiv \int_0^{+\infty} d\tau \alpha(\tau) e^{-iH_S\tau} \sigma^z_j e^{+i H_S \tau}. 
\end{equation}
The Redfield master equation is more general than the secular master equation used in Sec.~\ref{sec::frombathmem} since it does not rely on the secular approximation. Equivalently, it retains the interferences between different Bohr sectors that the secular approximation neglects.

Crucially, the strong coupling operator implements a simple spin-flip in the $x$-basis, i.e., $\sigma^z_j \ket{\mathbf{s}}_x = \ket{\mathbf{s}^{(j)}}_x$. Therefore, the action of the effective coupling operator on $\ket{\mathbf{s}}_x$ becomes particularly simple 
\begin{equation}
 \begin{split}
    \bar{L}_j^{s}\ket{\mathbf{s}}_x &= \int_0^{+\infty} d\tau~\alpha(\tau)e^{-i [E(\mathbf{s}^{(j)}) - E(\mathbf{s)}] \tau} \ket{\mathbf{s}^{(j)}}_x \\
    &= \Gamma[-\Delta E(\mathbf{s}\rightarrow \mathbf{s}^{(j)})] \ket{\mathbf{s}^{(j)}}_x,
 \end{split}
\end{equation}
where 
\begin{equation}
    \Gamma(\omega) = \int_0^{+\infty}d\tau~\alpha(\tau)e^{i\omega\tau}.
\end{equation}
The effective coupling operator still does exactly the same local spin flip as $L_j^{s} = \sigma_j^z$, up to a constant factor that explicitly depends on the energy cost of that flip. 

Since both $L_j^s$ and $\bar{L}_j^s$ map a basis state $\ket{\mathbf{s}}_x$ onto one and only one other basis state $\ket{\mathbf{s}^{(j)}}_x$, projecting the Redfield master equation onto the population leads to the closed classical Pauli master equation
\begin{equation}
    \frac{d}{dt}P_\mathbf{s}(t) = \sum_j \left[W^s(\mathbf{s}^{(j)} \rightarrow \mathbf{s}) P_{\mathbf{s}^{(j)}}(t) - W^s(\mathbf{s} \rightarrow \mathbf{s}^{(j)}) P_\mathbf{s}(t)\right],
    \label{eq::Pauli_s}
\end{equation}
with the single-flip rate~$W^s(\mathbf{s}^{(j)} \rightarrow \mathbf{s}) = 2 S[- \Delta E(\mathbf{s}^{(j)} \rightarrow \mathbf{s})]$.
Comparing with the single-flip rate for the weak-symmetry coupling family~\eqref{eq::single-flip_rate_weak}, we get 
\begin{equation}
   W^s(\cdot ) = \frac{4}{(\alpha-\beta)^2} W(\cdot),
\end{equation}
which shows that the two classical generators $W^{s}$ and $W$ are simply proportional to each other. Thus, the weak and strong symmetry couplings generate the same classical dynamics, up to an overall rescaling of time. In particular, they share the same Boltzmann stationary distribution~\eqref{eq::Boltzmann}.

\begin{figure}
    \centering
    \includegraphics[width=0.75\linewidth]{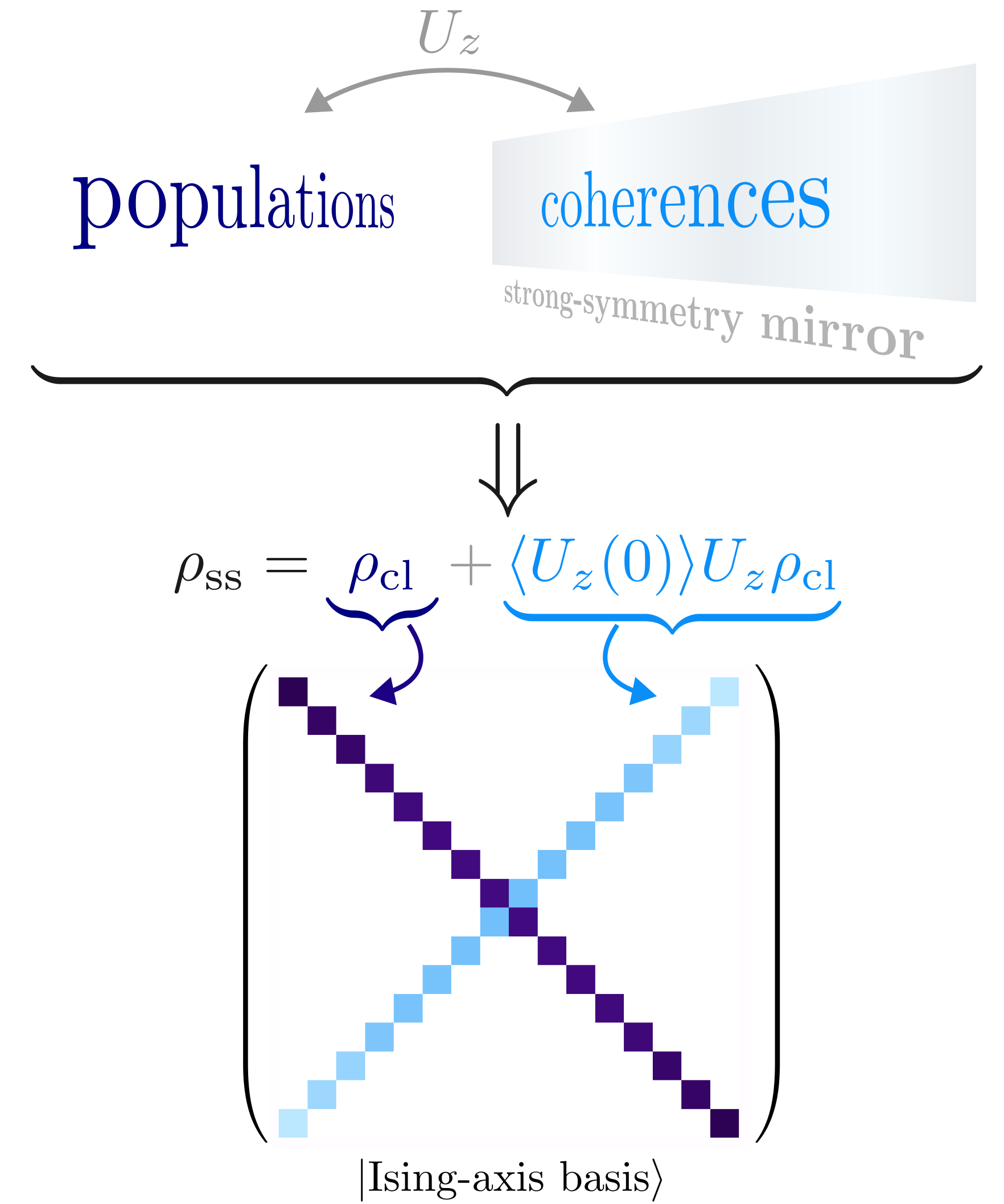}
    \caption{\textbf{Sketch of the role of strong symmetry.} The strong symmetry $U_z$ forces the coherences $C_\mathbf{s}$ between globally-reversed configurations to mirror the dynamics of the populations $P_\mathbf{s}$ in the Ising basis. This mechanism leads to the analytical steady-state manifold~\eqref{eq::analyticalSS_sz}. }
    \label{fig::mirror}
\end{figure}

Crucially, the strong symmetry mirrors the population's dynamics in the coherences $C_\mathbf{s} = {}_x\!\braket{\mathbf{s}|\rho|-\mathbf{s}}_x$ between the configuration $\mathbf{s}$ and its globally reversed partner $-\mathbf{s}$, as illustrated in Fig.~\ref{fig::mirror}. Indeed, since by definition of a strong symmetry $U_z$ commutes with both $L_j^s$ and $H_S$, we also have that $[U_z, \bar{L}_j^s] = 0$. The Redfield generator $\mathcal{L}_R$ then obeys the key identity
\begin{equation}
    \mathcal{L}_\mathrm{R}[A U_z] = \mathcal{L}_\mathrm{R}[A] U_z
\end{equation}
for any operator $A$. Suppose now that we have an arbitrary operator connecting globally reversed configurations $O = \sum_\mathbf{s} C_\mathbf{s} \ket{\mathbf{s}}_x\bra{-\mathbf{s}} = A U_z$, where $A$ is the diagonal operator $A = \sum_{\mathbf{s}} C_\mathbf{s} \ket{\mathbf{s}} \bra{\mathbf{s}}$. Then, the dynamics of $A U_z$ is 
$\frac{d}{dt} (AU_z) = \mathcal{L}_\mathrm{R}[AU_z] = \mathcal{L}_R[A] U_z$, and we obtain
\begin{equation}
   \begin{split}
       \frac{d}{dt} C_\mathbf{s} = \sum_j \left[W^s(\mathbf{s}^{(j)} \rightarrow \mathbf{s}) C_{\mathbf{s}^{(j)}}(t)  - W^s(\mathbf{s} \rightarrow \mathbf{s}^{(j)}) C_\mathbf{s}(t)\right],
   \end{split} 
\end{equation}
which is exactly the same evolution as the population~\eqref{eq::Pauli_s}. The only difference lies in the normalization. While $\sum_\mathbf{s} P_\mathbf{s} = 1$ , the global coherences $C_\mathbf{s}$ satisfy 
\begin{equation}
\begin{aligned}
    \sum_{\mathbf{s}} C_\mathbf{s}(t) &= \sum_{\mathbf{s}} {}_x \braket{\mathbf{s}|\rho(t)|-\mathbf{s}}_x  \\&= \mathrm{Tr}[\rho(t) U_z] 
    = \langle U_z(t) \rangle \equiv \braket{U_z(0)},
    \end{aligned}
\end{equation}
where the last line holds since parity is a conserved quantity.

Since they share the same dynamics, the globally reversed coherences  and the population must share the same steady state, up to the normalization factor $\braket{U_z(0)}$, yielding 
\begin{equation}
    C_\mathbf{s}^\mathrm{ss} = \braket{U_z(0)} P_\mathbf{s}^\mathrm{ss}.
\end{equation}
The linearity of the Redfield generator guarantees that the following remarkably simple manifold is steady
\begin{equation}
    \rho_\mathrm{ss} = \rho_\mathrm{cl} \left[\mathds{1} + \braket{U_z(0)}U_z\right] 
    \label{eq::analyticalSS_sz}, 
\end{equation}
with $\rho_\mathrm{cl} = \sum_\mathbf{s} P_\mathbf{s}^\mathrm{ss} \ket{\mathbf{s}}\bra{\mathbf{s}} $, as sketched in Fig.~\ref{fig::mirror}. Equation~\eqref{eq::analyticalSS_sz} is the central consequence of the strong symmetry in the infinite-time limit, and we show in Appendix~\ref{appendix::complteness_ss} that this manifold is complete in the secular limit, and generically complete in the Redfield case. Once the stationary distribution of the classical master equation~\eqref{eq::Pauli_s} is known, the associated quantum stationary state directly follows from Eq.~\eqref{eq::analyticalSS_sz}.
This equation cleanly separates reservoir structure and symmetry.

For the nearest-neighbor chain, the classical stationary distribution was found in Sec.~\ref{sec::frombathmem} to be thermal at the effective inverse temperature $\beta_\mathrm{eff}$. We thus obtain 
\begin{equation}
 \rho_\mathrm{ss} = \frac{e^{-\beta_\mathrm{eff}H_S}}{Z_\mathrm{eff}}[\mathds{1} + \braket{U_z(0)} U_z].
\end{equation}
From this equation, the different roles played by the spectral structure of the bath and the strong symmetry are clearly delineated. Indeed, the spectral structure fixes $\beta_\mathrm{eff}$, and thereby the classical distribution of Ising configurations. The initial value $\braket{U_z(0)}$, conserved throughout the dynamics, determines how strong the coherent link is between each configuration and its globally reversed partner. One may thus view the strong symmetry as a quantum glue that coherently binds globally reversed copies of the same purely classical process.

\subsection{Global entanglement and classical distinguishability} \label{sec::globalentanglement}
Surprisingly, despite the strictly local nature of the dissipative dynamics, the quantum correlations encoded in the steady states belonging to the manifold~\eqref{eq::analyticalSS_sz} are exclusively global. To see this, consider any proper subsystem $A \subset \{1, ..., N \}$ with a  complement $\bar{A}$. Because the coherence term in Eq.~\eqref{eq::analyticalSS_sz} connects configurations that are globally reversed (thus, they differ on every spin), the reduced steady state  $\rho_\mathrm{ss, A} \equiv  \mathrm{Tr}_{\bar A}[\rho_\mathrm{ss}] $ obtained by tracing out $\bar{A}$ reads
\begin{equation}
   \rho_\mathrm{ss, A} = 
   \mathrm{Tr}_{\bar{A}}[\rho_\mathrm{cl}] + \braket{U_z(0)}\mathrm{Tr}_{\bar{A}}[\rho_\mathrm{cl} U_z]  = \mathrm{Tr}_{\bar{A}}[\rho_\mathrm{cl}].
\end{equation}
Thus, every proper reduced density matrix is diagonal in a product basis and is consequently separable. Any entanglement contained in $\rho_\mathrm{ss}$ is therefore genuinely global, and completely invisible in every proper subsystem.

Interestingly, we now show that the global entanglement  stored in the steady state manifold  and detected by the negativity can be linked exactly to a purely classical notion of distinguishability. Indeed, consider again a bipartition $A|\bar{A}$ of the full chain and define $\mathbf{s}^{(A)}$ as the configuration where all spins belonging to $A$ have been flipped, i.e., $\mathbf{s}^{(A)} = (-\mathbf{s}_A, \mathbf{s}_{\bar{A}})$. The negativity quantifying the entanglement associated with the bipartition $A|\bar{A}$ then reads  
\begin{equation}
    \mathcal{N}_{A|\bar{A}} = \frac14 \left[ \sum_\mathbf{s} \left|P_\mathbf{s}^\mathrm{ss} - |\braket{U_z(0)}| P_{\mathbf{s}^{(A)}}^{\mathrm{ss}}\right| - (1- |\braket{U_z(0)}|)\right],
    \label{eq::neg_general}
\end{equation}
and simplifies to
\begin{equation}
    \mathcal{N}_{A|\bar{A}}= \frac14 \left[ \sum_\mathbf{s} |P_\mathbf{s}^\mathrm{ss} -  P_{\mathbf{s}^{(A)}}^{\mathrm{ss}} |\right] = \frac12  \delta (P^\mathrm{ss}, P^{\mathrm{ss}}_{(A)} ),
    \label{eq::neg_symmetry_sector}
\end{equation}
for an initial state in a given symmetry sector ($\braket{U_z(0)} = \pm 1$). Here, we have introduced the total variation distance
\begin{equation}
    \delta(P, Q) \equiv \frac12 \sum_\mathbf{s}|P_\mathbf{s} - Q_\mathbf{s}|,
    \label{eq::total_variation_distance}
\end{equation}
defined for any pair $(P, Q)$ of probability distributions. We also introduced $P^{\mathrm{ss}}_{(A)}$ as the probability distribution that associates the probability $P^\mathrm{ss}_{\mathbf{s}^{(A)}}$ with the configuration $\mathbf{s}$. Explicit derivations of Eq.~\eqref{eq::neg_general} and Eq.~\eqref{eq::neg_symmetry_sector} can be found in Appendix~\ref{app:negativity}. 

Equations~\eqref{eq::neg_general} and \eqref{eq::neg_symmetry_sector} expose the two ingredients required for entanglement in this steady-state manifold. The first is the coherence fixed by the conserved strong-symmetry operator $\braket{U_z(0)}$. If the two parity sectors carry equal weight, that is $\braket{U_z(0)} = 0$, then the negativity vanishes for all bipartitions, as already expected from Eq.~\eqref{eq::analyticalSS_sz} since the steady state is then entirely classical. The second ingredient is a nontrivial classical distribution selected by the bath. Indeed, if $P_\mathbf{s}^\mathrm{ss}$ is uniform--- as for a structureless bath--- then $P^\mathrm{ss} = P^\mathrm{ss}_{(A)}$ for any bipartition, and the negativity again vanishes. Equivalently, the flat-bath dynamics is unital. This is consistent with recent results showing that Abelian strong symmetries, such as the $\mathbb{Z}_2$ symmetry considered here, cannot by themselves enforce entanglement under unital dynamics~\cite{Moharramipour2024S, Li2025H} (in sharp contrast with the non-Abelian case). This explains why, in our setting, strong symmetry alone is not sufficient: bath structure is needed here to shape the classical distribution and make the dynamics non-unital.

The role of the structure of the bath in generating entanglement becomes even more transparent once a parity sector is fixed. In this case, the expression~\eqref{eq::neg_symmetry_sector} reveals that the negativity associated with the bipartition $A|\bar{A}$ is given by the classical distinguishability of the probability distributions $P^\mathrm{ss}$ and $P^\mathrm{ss}_{(A)}$. Maximizing the negativity thus amounts to making the two classical distributions $P^\mathrm{ss}$ and $P_{(A)}^\mathrm{ss}$ as distinguishable as possible. The bath structure can then be optimized to guarantee the largest distinguishability between these two distributions. In the limiting case in which their support becomes disjoint, $P^\mathrm{ss}$ and $P^\mathrm{ss}_{(A)}$ are perfectly distinguishable $\delta(P^\mathrm{ss}, P_{(A)}^\mathrm{ss}) = 1$, and the negativity reaches its maximal value $\mathcal{N}_{A|\bar{A}} = \frac12$ within the steady-state manifold~\eqref{eq::analyticalSS_sz}. 

\begin{figure}[t!]
    \centering
    \includegraphics[width=0.9\linewidth]{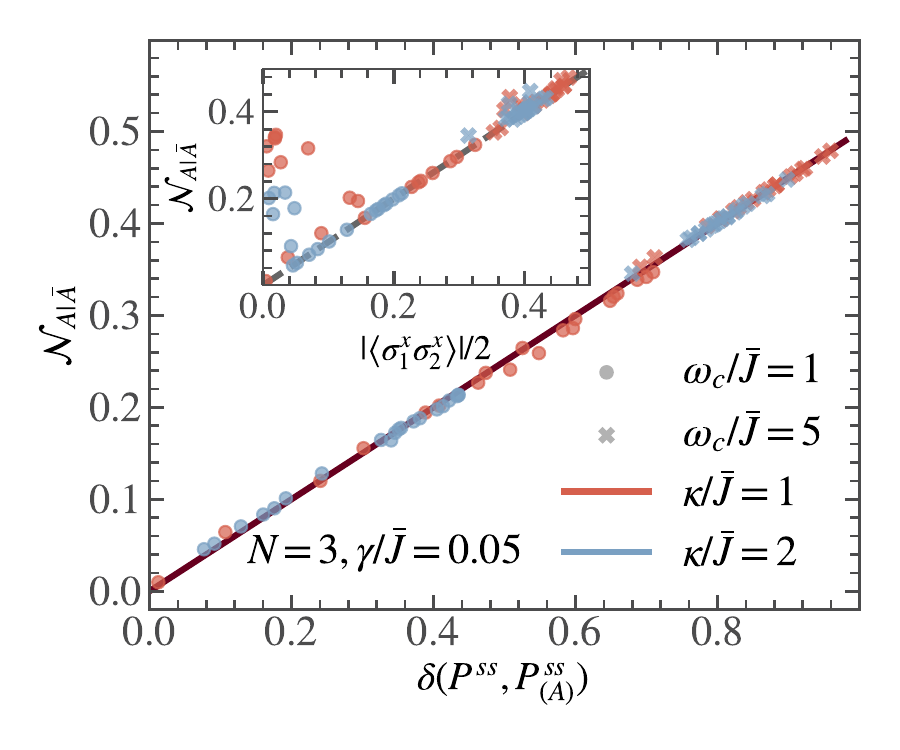}
    \caption{\textbf{Connecting entanglement and classical distinguishability.} Main panel: Negativity for the bipartition $1|(2, 3)$ of the steady state of the full Markovian embedding~\eqref{eq:markovian:embedding} compared to the distinguishability result~\eqref{eq::neg_symmetry_sector} . The elements of the coupling matrix $J_{ij}$ are drawn randomly from a lognormal distribution with $\mu=0$ and $\sigma=0.5$ (with median $\bar J= \exp(\mu)=1$), and $\braket{U_z(0)} = +1$.
    Inset: the simple bound~\eqref{eq::lower_bound_neg} is tight and holds in the case of the Markovian embedding.
    }
    \label{fig::negativity:vs:distinguishability}
\end{figure}

The conversion of classical distinguishability into entanglement is illustrated in Fig.~\ref{fig::negativity:vs:distinguishability}. For three spins with randomly chosen Ising couplings, we show the negativity across a representative bipartition ($1|(2, 3)$)---obtained from the exact Markovian embedding~\eqref{eq:markovian:embedding}---as a function of the classical distributions $P^\mathrm{ss}$ and $P^\mathrm{ss}_{(1)}$. The numerical points closely follow a straight line with slope $1/2$, demonstrating the accuracy of Eq.~\eqref{eq::neg_symmetry_sector} for general Ising couplings $J_{ij}$.

\subsection{Detection of negativity and lower bound}
Having determined that the negativity cannot exceed $1/2$, we now show that, once a symmetry sector is fixed, it can also be lower bounded by an Ising correlator. Specifically, starting from Eq.~\eqref{eq::neg_symmetry_sector}, one can show (see Appendix~\ref{app:negativity} for details) that the negativity associated with the bipartition $A|\bar{A}$ reads 
\begin{equation}
    \mathcal{N}_{A|\bar{A}} \geq \frac12 |\braket{\sigma^x_i \sigma^x_j}|,
    \label{eq::lower_bound_neg}
\end{equation}
where $i \in A$ and $j \in \bar{A}$. Optimizing over all pairs $(i\in A, j \in \bar{A})$ naturally yields the tightest bound of this form. Importantly, all members of the steady-state manifold~\eqref{eq::analyticalSS_sz} share the same Ising correlator $\braket{\sigma^x_i \sigma^x_j}$ once $\rho_\mathrm{cl}$ is fixed. Equation~\eqref{eq::lower_bound_neg} should therefore not be interpreted as a generic entanglement witness. Rather, it becomes valid because the initial state is known to belong to a definite strong-symmetry sector (which fixes the global coherences). The inset of Fig.~\ref{fig::negativity:vs:distinguishability} shows that the bound~\eqref{eq::lower_bound_neg} is valid beyond Redfield, captures entanglement, and can be saturated for many realizations in a small chain (three sites).

Detecting such global entanglement is a very challenging task~\cite{Guhne2009E, Lu2007E, Wu2026E} because it typically requires addressing all particles simultaneously or performing a global measurement of the state. By contrast, our bound~\eqref{eq::lower_bound_neg} only involves a two-site correlator that can be easily measured. Notice that the value of this correlator solely depends on the reduced state $\rho_{ij}$, which is purely separable. Nevertheless, once the symmetry sector is known, it provides quantitative information about entanglement across the full bipartition $A|\bar{A}$. The strong symmetry supplies here the missing global information that allows a simple two-site measurement to constrain an intrinsically global quantum property. As we show in the next Sec.~\ref{sec::nn}, this lower bound becomes an exact equality for nearest-neighbor interactions.

\subsection{Genuine multipartite entanglement}
Global entanglement should be distinguished from genuine multipartite entanglement (GME), which is a stronger notion. Here, global entanglement means that any proper reduced state is separable. In contrast, a mixed state is said to be genuinely multipartite entangled if it cannot be written as a convex combination of biseparable states, i.e., states that are separable across some bipartition of the system. Note that the bipartition may differ between the different terms of the mixture. Thus, a mixed state that is entangled across every bipartition need not necessarily be GME. 

The special structure of our steady-state manifold~\eqref{eq::analyticalSS_sz} makes it possible to characterize exactly GME. We first perform a change of basis to a GHZ basis that pairs coherently $\ket{\mathbf{s}}$ and $\ket{-\mathbf{s}}$
\begin{equation}
    \ket{\mathrm{GHZ}_\mathbf{s}^\pm} \equiv \frac{\ket{\mathbf{s}}_x \pm \ket{-\mathbf{s}}_x}{\sqrt2},
\end{equation}
in terms of which our analytical steady states read 
\begin{equation}
    \rho_\mathrm{ss} = \frac12 \sum_{\mathbf{s}}[p^+_\mathbf{s} \ket{\mathrm{GHZ}^+_\mathbf{s}} \bra{\mathrm{GHZ}^+_\mathbf{s}} + p_\mathbf{s}^- \ket{\mathrm{GHZ}^-_\mathbf{s}} \bra{\mathrm{GHZ}^-_\mathbf{s}}],
\end{equation}
with $p_\mathbf{s}^\pm = (1 \pm \braket{U_z(0)}) P_\mathbf{s}^\mathrm{ss}$ and the factor $1/2$ avoids double counting. For this class of states diagonal in this GHZ basis, GME is characterized exactly by the largest GHZ-basis population. Specifically, $\rho_{\mathrm{ss}}$ is GME if and only if there is a population $p_\mathrm{max}$ greater than $1/2$~\cite{Guhne2010S}. Maximizing $p_\mathbf{s}^\pm$ then yields 
\begin{equation}
    p_\mathrm{max} = (1+|\braket{U_z(0)}|) \max_\mathbf{s} P_\mathbf{s}^\mathrm{ss} > \frac12. 
    \label{eq::GME_necessaryAndSufficient}
\end{equation}
This last expression reduces to the particularly simple condition 
\begin{equation}
    \max_\mathbf{s} P_\mathbf{s}^\mathrm{ss} > \frac14, 
    \label{eq::GME_symmetry_sector}
\end{equation}
once a symmetry sector is fixed. GME emerges naturally once the structure of the bath concentrates sufficient classical probability on a single GHZ basis state. 

We now show that GME can be certified using only two-spin correlators.  The idea is to particularize Eq.~\eqref{eq::GME_symmetry_sector} to the case where the maximum $P_\mathbf{s}$ is realized for the pure ferromagnetic configurations ($\mathbf{s} = +...+$ and $\mathbf{s} = -...-$). This is because the Ising correlator $\braket{\sigma^x_i \sigma^x_j}$ directly gives the probability $P(E_{ij})$ that spins $i$ and $j$ are antialigned (in the $x$ basis) through 
\begin{equation}
    P(E_{ij}) = [1- \braket{\sigma^x_i \sigma^x_j}]/2,
\end{equation}
where $E_{ij}$ is the event that spins $i$ and $j$ are antialigned. Let now $\Lambda$ be any set of $N-1$ links connecting all $N$ spins (e.g., $\Lambda = \{ (1, 2), (2, 3), ..., (N-1, N)\}$). Since a configuration is either fully polarized or there exists at least one mismatch, the conservation of probability gives 
\begin{equation}
    2 P_{+...+}^\mathrm{ss} + P\left(\bigcup_{(i, j) \in \Lambda} E_{ij} \right) = 1.
\end{equation}
The events $E_{ij}$ generally overlap, so the probability of finding at least one mismatch is bounded by the sum of the individual mismatch probabilities $P(E_{ij})$. Hence,
\begin{equation}
    2 P_{+...+}^\mathrm{ss} \geq 1 - \frac12 \sum_{(i, j) \in \Lambda} [1 - \braket{\sigma^x_i \sigma^x_j}].
\end{equation}
This last result, combined with the GME condition~\eqref{eq::GME_symmetry_sector}, yields the simple sufficient criterion
\begin{equation}
    \sum_{(i, j) \in \Lambda}[1 - \braket{\sigma^x_i \sigma^x_j}] < 1 \implies \mathrm{GME}. 
    \label{eq::witness_GME_sym_sector}
\end{equation}
Thus, once the symmetry sector is fixed, GME can be certified from a connected set of only $N-1$ two-spin correlators. Since all Ising correlators commute, they can furthermore be extracted all at once from the same measurement in the $x$ basis. Moreover, we curiously have that every corresponding two-spin reduced state remains separable. At first sight, our result may seem impossible as it is well known that no criterion based solely on two-point correlators (like $\braket{\sigma^x_i \sigma^x_j}$) can detect the entanglement of the graph-state family that contains the GHZ state~\cite{Gittsovich2010M}. This apparent tension is resolved as soon as one remembers that Eq.~\eqref{eq::witness_GME_sym_sector} relies on the knowledge that the initial state satisfies $|\braket{U_z(0)}| = 1$. The strong symmetry, therefore, supplies precisely the global information that is absent from the two-spin reduced state. Compared to efficient GME certification present in the literature~\cite{Toth2005D, Knips2016M, Wu2026E, Li2026}, our setting provides a distinct route. Indeed, instead of extracting the missing many-body information from additional observables, it is fixed dynamically by the strong symmetry. In the following, we will mainly work in a regime where the two fully polarized configurations acquire appreciable stationary weight.

\subsection{Nearest-neighbor interactions} \label{sec::nn}
For nearest-neighbor interactions, we obtain simple expressions for the negativity and the GME witness because the stationary populations are described by the effective Gibbs distribution $P_\mathbf{s}^\mathrm{ss} = e^{-\beta_\mathrm{eff} E(\mathbf{s})}/ Z_\mathrm{eff}$. Indeed, using Eq.~\eqref{eq::neg_general}, the negativity can then be recast, for OBC in the thermodynamic limit, in the following form
\begin{equation}
    \mathcal{N}_{A|\bar{A}} = \frac14 \left[(1+ |\braket{U_z(0)}|) |\braket{\sigma^x_i \sigma^x_{j}}| - (1-|\braket{U_z(0)}|) \right]_+,
    \label{eq::neg_NN_general}
\end{equation}
as shown in Appendix~\ref{app:negativity}, where an exact expression for PBC and finite $N$ is also provided. Here, $[A]_+ \equiv \mathrm{max}(A, 0)$, and $(i, j)$ labels the spins living at the border between $A$ and $\bar{A}$. In the following, we assume that every local bath has the same spectral asymmetry ratio $\chi$, which implies that the relevant Ising-axis correlator is $\braket{\sigma^x_j \sigma^x_{j+1}}$. Consequently, the negativity becomes nonzero only when 
\begin{equation}
    |\braket{\sigma^x_j \sigma^x_{j+1}}| > \frac{1 - |\braket{U_z(0)}|}{1+ |\braket{U_z(0)}|}.
\end{equation}
In order to stabilize an entangled steady state, the Ising-axis correlator---which solely depends on $\beta_\mathrm{eff} J$---needs to overcome a threshold that decreases monotonically with the initial parity value. For its maximal value $|\braket{U_z(0)}| = 1$, the threshold becomes trivial, and we obtain
\begin{equation}
    \mathcal{N}_{A|\bar{A}} = \frac{1}{2} |\braket{\sigma^x_j \sigma^x_{j+1}} | = \frac12 |\tanh(\beta_\mathrm{eff} J)|.
    \label{eq::neg_NN_symsector}
\end{equation}
Nearest-neighbor interactions then provide an example where the general bound~\eqref{eq::lower_bound_neg} is saturated. Interestingly, in our context, a two-spin Ising correlator measured anywhere in the chain gives the exact negativity across any contiguous bipartition. Moreover, this negativity is nonzero as soon as there is some spectral asymmetry $\chi \neq 1$, that is $\beta_\mathrm{eff} \neq 0$. 

\begin{figure}[ht!]
    \centering
    \subfloat[]{
    \includegraphics[width=0.9\columnwidth]{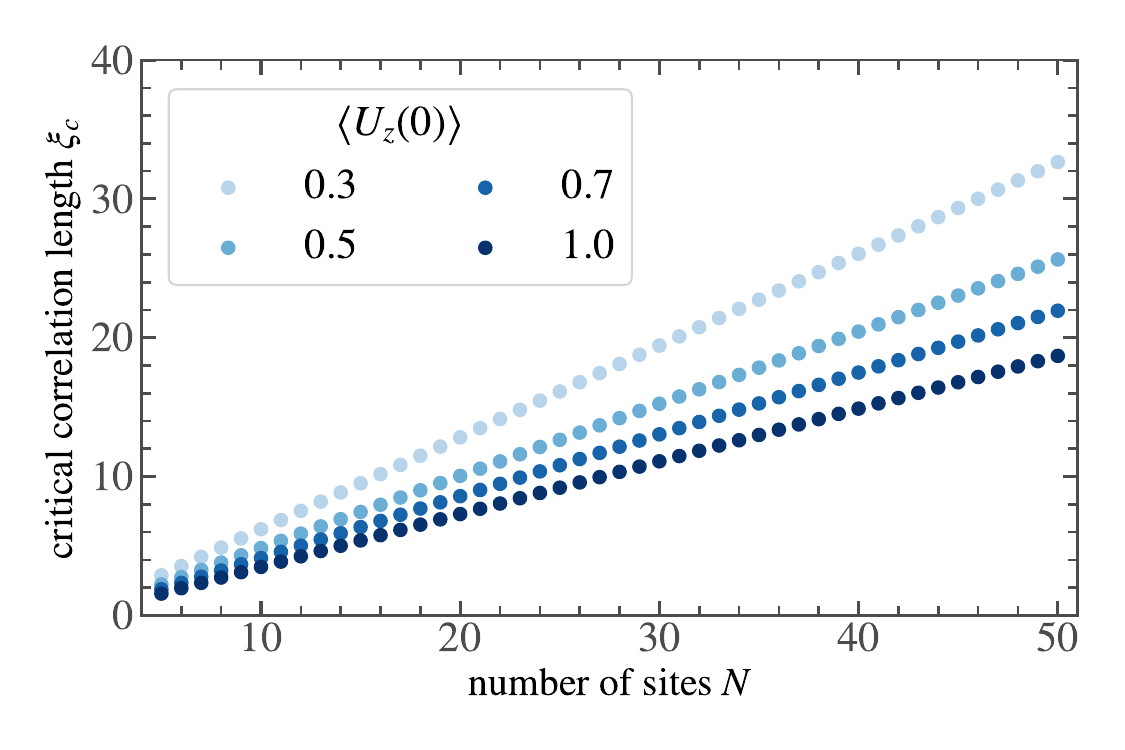}
    }
    \vspace{-0.8cm}
    \subfloat[]{
    \includegraphics[width=0.9\linewidth]{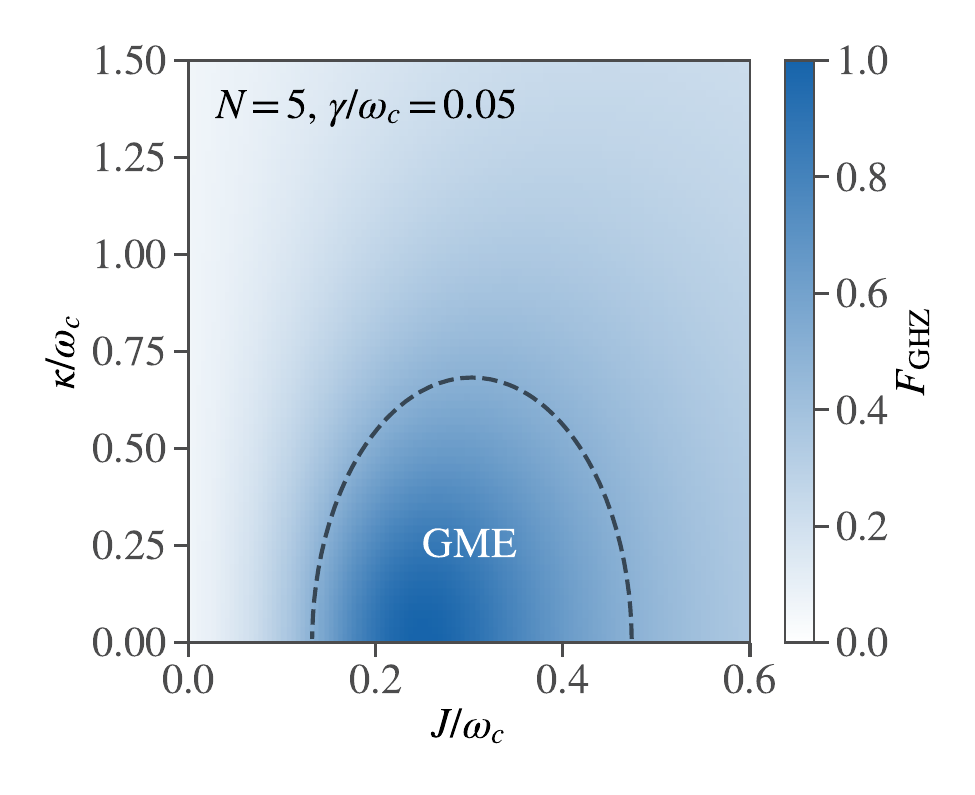}
    }
    \vspace{-0.4cm}
    \caption{\textbf{Conditions for genuine multipartite entanglement in the steady state.}
    (a): Critical correlation length $\xi_c$ (obtained from Eq.~\eqref{eq::GME_analytical_NN}) required for GME steady states shown for different values of the $z$ parity $\braket{U_z(0)}$ (see legend), and in the nearest-neighbor case. The dots show the exact threshold, while solid lines demonstrate that the $\xi(N) \approx N / (2 \mathrm{arcosh}(1+|\!\braket{U_z(0)}\!|)$, for $N$ large enough. Increasing the initial parity reduces the correlation length required for GME.
    (b): Analytical GHZ fidelity $F_{\rm GHZ}$ of Eq.~\eqref{eq::analytical_fid_GHZ} for the steady state in the parity-even subspace. The black line separates the two regions where $F_{\rm GHZ} > 0.5$ and $F_{\rm GHZ} < 0.5$. 
    }
    \label{fig::GHZ_Fidelity}
\end{figure}

For nearest-neighbor interactions, characterizing GME is vastly simplified. This comes from the fact that the Gibbs distribution for the population (with $\beta_\mathrm{eff} >0$) necessarily implies that $\max_\mathbf{s} P_\mathbf{s}^\mathrm{ss} = P^\mathrm{ss}_{+...+}$. The necessary and sufficient criterion~\eqref{eq::GME_necessaryAndSufficient} becomes 
\begin{equation}
    (1+|\braket{U_z(0)}|) P_{+...+}^\mathrm{ss}> \frac12.
\end{equation}
A straightforward calculation, however, reveals that 
\begin{equation}
    P_{+...+} ^\mathrm{ss}= \frac{\cosh^N \left(\frac{1}{2\xi}\right)}{2 \cosh\left(\frac{N}{2\xi}\right)},
\end{equation}
from which we obtain that the steady state is GME if and only if 
\begin{equation}
    (1+ |\braket{U_z(0)}|) \cosh^N\left(\frac{1}{2\xi}\right) > \cosh\left( \frac{N}{2\xi}\right).
    \label{eq::GME_analytical_NN}
\end{equation}
Once the initial parity is known, the correlation length alone determines whether the steady state is GME. Since $\xi$ is itself determined by the spectral asymmetry $\chi$, this condition directly connects the steady-state GME with the spectral structure of the baths. Denoting by $\xi_c$ the critical correlation length that saturates the inequality~\eqref{eq::GME_analytical_NN}, we find that the ansatz $\xi_c = c N$ yields 
\begin{equation}
    c \rightarrow \frac{1}{2 \mathrm{arcosh}(1+|\braket{U_z(0)}|)} \quad \mathrm{as}~N\rightarrow+\infty.
\end{equation}
As shown in Fig.~\ref{fig::GHZ_Fidelity} (a), this linear growth is already evident for small system sizes and preparing the initial state with $|\braket{U_z(0)}| = 1$ minimizes the required correlation length, with $\xi_c \approx 0.38N$ for large N. 
Being GME and maximizing the negativity across all bipartitions (as given by Eq.~\eqref{eq::neg_NN_symsector}) requires both that $\braket{U_z(0)}= \pm 1$, and $\beta_\mathrm{eff} \rightarrow +\infty$. In this limit, we are effectively cooling down the system, and as configurations containing domain walls lose weight, the steady state approaches
\begin{equation}
    \rho_\mathrm{ss} \approx \ket{\mathrm{GHZ}_\pm} \bra{\mathrm{GHZ}_\pm }.
\end{equation}
To quantify the resemblance between the steady states that we prepare and the asymptotic pure steady states, we evaluate the fidelity 
\begin{equation}
    F_\mathrm{GHZ} = \braket{\mathrm{GHZ}_+|\rho_\mathrm{ss}|\mathrm{GHZ}_+} = 2P_{+\cdots +}^\mathrm{ss}, 
    \label{eq::analytical_fid_GHZ}
\end{equation}
for an initial state in the even parity sector. Fig.~\ref{fig::GHZ_Fidelity} (b) examines how this fidelity behaves by tuning the parameters $J/\omega_c$ and $\kappa/\omega_c$. For five sites, the fidelity, which is in excellent agreement with the exact Markovian embedding (see Appendix~\ref{app:sec:numerics}), becomes close to unity near $4J = \omega_c $ for sufficiently narrow reservoirs. Note that these parameters are natural as they correspond to a very low effective temperature $1/\beta_\mathrm{eff}$. The black dashed contour is defined by $F_\mathrm{GHZ} = 1/2$, and any state with a higher fidelity is GME. The region supporting this certification is substantially broader than the region where the overlap approaches unity: moderate spectral selectivity already produces GME, while a narrower, resonant reservoir brings the stationary state close to the pure GHZ state. 

\section{Trapped-Ion Implementation}
\label{sec::trapped-ion}
In this last section, we propose a concrete experimental trapped-ion implementation to test our theoretical predictions. Our focus is the buildup of GME from effectively independent reservoirs, with their spectra providing a control knob.  After discussing the impact of finite temperature for the reservoirs, we close this discussion by showing numerical simulations for typical experimentally relevant parameters, which confirm the theoretical results.
\subsection{Proposed experimental implementation}
\begin{figure}[t!]
    \centering
    \includegraphics[width=1\linewidth]{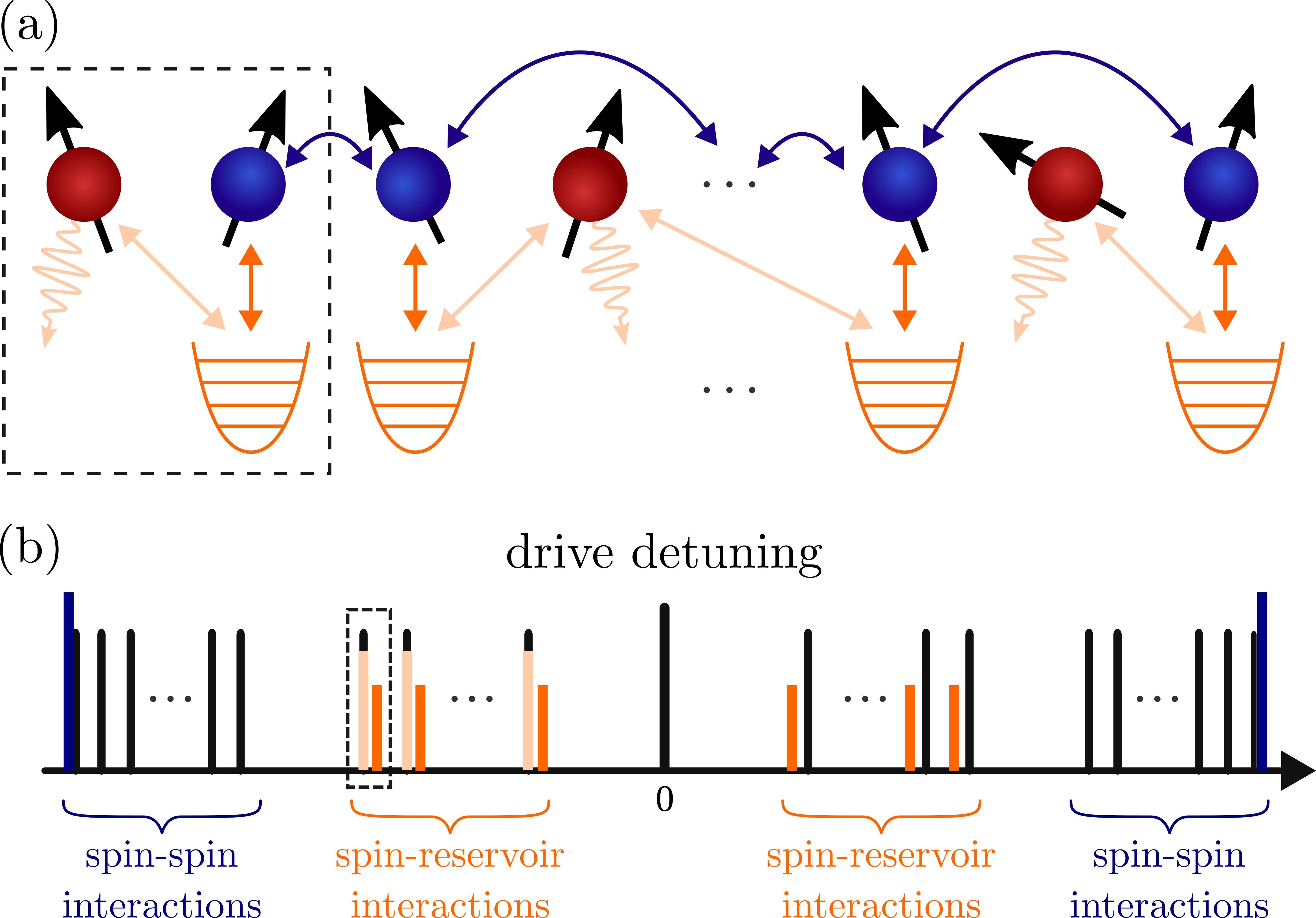}
    \caption{\textbf{Conceptual experimental scheme for implementing engineered-dissipation protocols in a trapped-ion chain.} (a) One-dimensional ion chain comprising system spins (blue) and ancillary ions (red). Blue curves denote nearest-neighbor spin interactions, while dark orange lines indicate coupling to local reservoir modes represented by harmonic wells. These modes are damped via red-sideband cooling of the ancillary ions, illustrated by light orange lines and dissipation arrows. (b) Multifrequency driving scheme addressing two motional sideband branches associated with orthogonal trapping directions. Blue bichromatic tones generate nearest-neighbor spin interactions under global illumination of the system ions, while orange tones engineer programmable reservoir frequencies and spin-reservoir couplings. Here, we show the GHZ-state generation protocol, where dark orange bichromatic pairs are used. In the classical protocol (not shown), these tones are replaced by single off-resonant red-sideband tones applied to the system ions, similar to the resonant light-orange tones used for sub-Doppler cooling of selected motional modes via the ancillary ions. Dashed boxes highlight a representative spin-reservoir-ancilla configuration in (a) and the corresponding driving tones in (b).}
    \label{fig::exp}
\end{figure}

Trapped ions are well-suited to implementing the engineered-dissipation protocols discussed here. The spins can be encoded in individually addressed hyperfine qubits~\cite{pagano2025varenna,monroe2021programmable}, while local reservoirs can be realized using distinct collective normal modes of the ion chain through a synthetic-space mapping~\cite{lemmer2018engineering, Sun2025Q}. This mapping establishes an effective one-to-one spin-mode coupling structure through ion- and mode-resolved sideband drives, such that locality is defined by the engineered couplings rather than the spatial confinement of the collective vibrational modes \cite{Marijanovic2026}. The proposed ion-chain architecture and multifrequency driving scheme are schematically illustrated in Fig.~\ref{fig::exp}.

The two protocols for engineering the local reservoir modes and their couplings to the spins are distinguished by the choice of local spin operators $L_j$, realized through appropriate sideband drives. For the classical protocol, single-tone red-sideband drives applied independently to each system ion realize Jaynes-Cummings-type interactions with $L_j=\sigma_j^-$, whereas GHZ-state generation employs balanced bichromatic drives with suitable phases to implement Hermitian couplings $L_j=\sigma_j^y$.  Following Refs.~\cite{So2024T, So2025Q}, we formulate the latter protocol in a rotated spin basis, with the $y$- and $z$-axes interchanged ($y\leftrightarrow z$), and initialize the system in the fully polarized product state $\ket{\downarrow,\downarrow,\ldots,\downarrow}_y$. Selective addressing of the spin-mode pairs defined by the prescribed synthetic-space mapping yields spin-reservoir interactions of the form $\sigma_j^y(a_j+a_j^\dagger)$. Residual off-target spin-phonon and spin-spin couplings are analyzed in detail in Appendix~\ref{app:one_to_one}, where we show that their secular contributions vanish and the remaining time-dependent contributions are bounded in the Magnus expansion. Under resolved-frequency averaging, their standalone secular contribution generates only a global phase, which can be neglected.

In both protocols, the effective reservoir frequencies $\omega_j$ are controlled by the drive detunings relative to the corresponding motional sidebands. Thus, they can be made uniform in a programmable way, $\omega_j=\omega_c$. Similarly, uniform spin-reservoir coupling strengths, $g_j=g$, can be achieved by adjusting the drive amplitudes to compensate for variations in the corresponding spin-mode Lamb-Dicke parameters, which reflect each ion's participation in its addressed mode~\cite{pagano2025varenna,monroe2021programmable}. The one-to-one assignment of collective modes to system ions can also be optimized to minimize the laser power required for the prescribed couplings. This optimization favors spin-mode pairs with larger Lamb-Dicke magnitudes, which require smaller drive amplitudes at fixed coupling strength.

Moreover, spin-spin interactions can be generated by off-resonantly coupling the system ions to a motional branch orthogonal to that used for spin-reservoir coupling, employing bichromatic laser tones similar to those used for spin-reservoir interactions in the GHZ-state generation protocol~\cite{monroe2021programmable}. In the dispersive regime, where the detunings from the motional sidebands are much larger than the corresponding spin-phonon coupling strengths, virtual phonon excitations mediate effective Ising-type interactions, with coupling strengths and spatial profiles determined by the normal-mode structure, drive detunings, and mode-dependent spin-phonon couplings~\cite{Kyprianidis2024interaction}. By tailoring these parameters, the interaction range of $J_{ij}$ can be tuned from long-range couplings~\cite{De2025NonEquilibrium} $J_{ij}\sim J_0/|i-j|^{\alpha_J}$ with power law exponent $0.5\lesssim\alpha_J\lesssim1.5$ to predominantly short-range interactions of the form $J_{ij}\sim J_0 e^{-\beta_J |i-j|}$ ~\cite{De2026StringBreaking}. Here, we propose using orthogonal motional branches to facilitate independent control of the coherent spin-spin interactions and the spin-phonon couplings employed to engineer the local reservoirs.

Controlled dissipation of multiple reservoir modes can be achieved through sympathetic cooling using individually addressed ancillary ions subjected to sub-Doppler cooling~\cite{So2025Q, so2026finitetemp}. Since the additional ancillary ions participate in the collective vibrational modes, their cooling enables phonon removal from the designated reservoir modes without directly affecting the system qubits. Importantly, a single ancillary ion can cool multiple reservoir modes by addressing their respective red sidebands, provided it has nonzero participation in each mode. This eliminates the need for a dedicated ancillary ion for each reservoir. By tuning the cooling parameters, multiple modes can be damped with independently adjustable rates $\kappa_j$, allowing uniform dissipation, $\kappa_j=\kappa$, or mode-dependent damping profiles. Furthermore, controlled heating may be introduced to engineer finite-temperature reservoirs with tunable effective thermal occupations~\cite{so2026finitetemp}.

\begin{figure}[t!]
    \centering
    \subfloat[]{
    \includegraphics[width=0.98\linewidth]{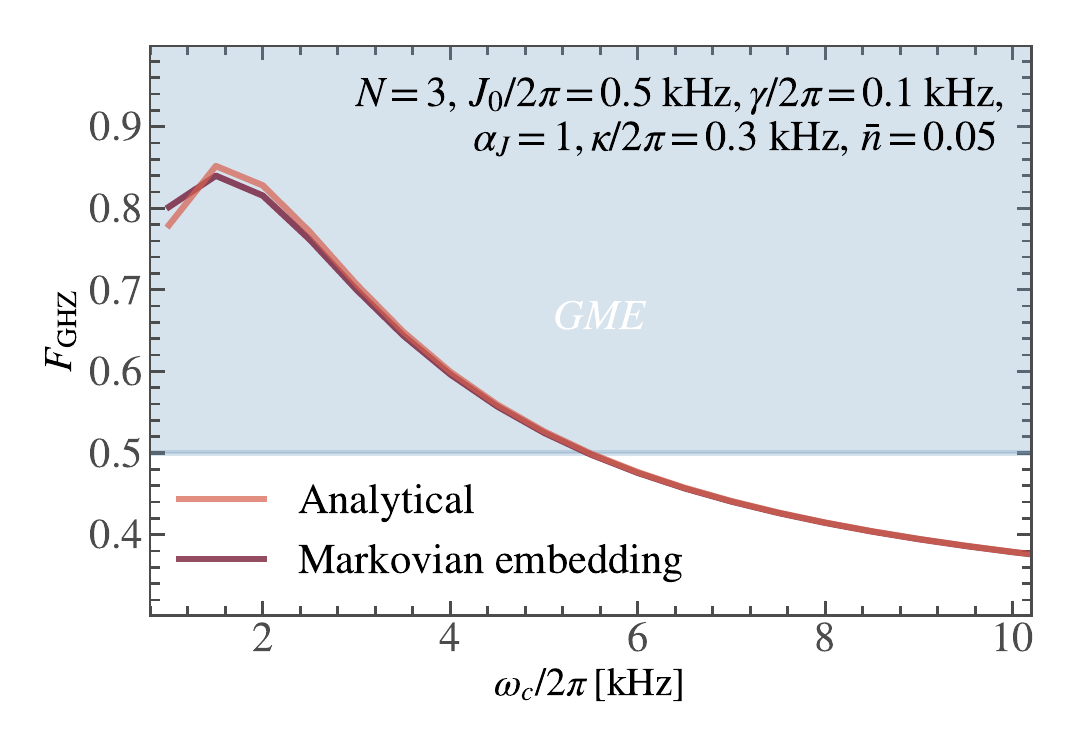}
    }
    \label{fig::exp:params}
    \vspace{-0.8cm}
    \subfloat[]{
    \includegraphics[width=0.98\linewidth]{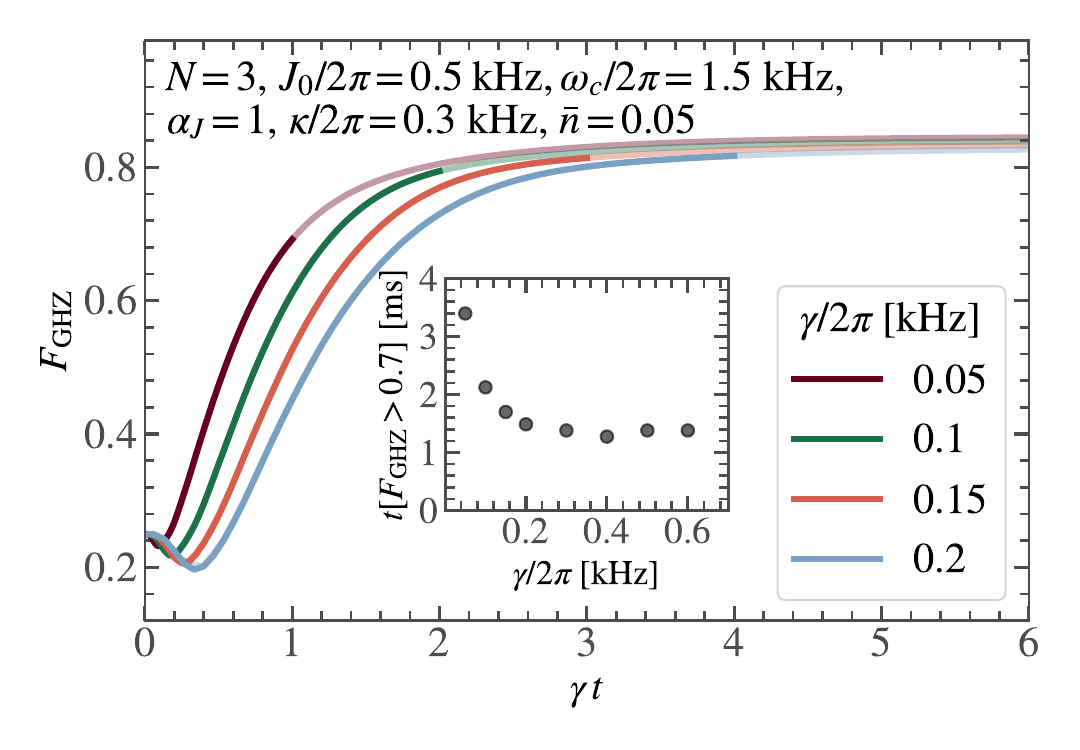}
    }
    \vspace{0.4cm}
    \caption{\textbf{Fidelity with the GHZ state for typical experimental parameters.} (a) Fidelity  of the steady state reached from an initial state with $\braket{U_z(0)} = +1$. Tuning the spectral function by changing $\omega_c$ allows to see the emergence of GME. (b) Role of $\gamma$ to converge more rapidly to the steady state. Varying $\gamma$ within the range $\gamma/2\pi = 0.05-0.2~\mathrm{kHz}$ does not affect the steady state reached, as shown by the collapsing of all curves at long times. Increasing $\gamma$, however, speeds up the preparation. For a fixed final time $t_f = 3.75~\mathrm{ms}$, the achieved fidelity is indicated by the boundary between the opaque and lighter segments of each curve. The initial state is always taken to be the state $|+\cdots  +\rangle_z$. The inset shows the time needed in $\mathrm{ms}$ to reach $F_\mathrm{GHZ} > 0.7$ as a function of $\gamma$. For both panels, we consider $\bar{n}=0.05$, $N = 3$ and $\kappa/2\pi = 0.3~\mathrm{kHz} $.}
    \label{fig::experimental:simulation}
\end{figure}

\subsection{Finite-temperature reservoirs and preparation dynamics}
\label{sec::finiteT}
The trapped-ion implementation naturally raises the question of finite reservoir occupation, $\bar{n}$, that was previously assumed to be zero (see Sec.~\ref{sec::physical setting}). Even in the absence of deliberate heating, intrinsic motional heating prevents the reservoir modes from remaining exactly in their vacuum states: the average bosonic occupation can be as low as $\bar{n}=0.05$, but not zero. Here, we discuss how this finite occupation enters our description and affects the preparation of entangled steady states. In the following, the Ising couplings are left arbitrary, and we focus on the strong-symmetry class $L_j^s = \sigma^z_j$ since it generates GME. 

Considering this finite occupation for the motional modes simply amounts to extending the Markovian embedding construction as follows 
\begin{equation}
 \begin{split}
 \frac{d \rho_\mathrm{tot}}{dt} &= -i [H_\mathrm{emb}, \rho_\mathrm{tot}]  \\
 &+ \kappa \left[\sum_j (\bar{n}+1) \mathcal{D}_{a_j}[\rho_\mathrm{tot}] + \bar{n}\mathcal{D}_{a_j^\dagger}[\rho_\mathrm{tot}]\right].
 \end{split}
\end{equation}
The only difference with Eq.~\eqref{eq:markovian:embedding} is the presence of a small heating term $\mathcal{D}_{a_j^\dagger}$. For modes initially in their stationary thermal states, the two correlation functions associated with each reservoir (see Sec.~\ref{sec::physical setting}) read 
\begin{equation}
 \begin{split}
    \alpha_{\bar{n}}(t) &= |g|^2 (\bar{n} +1) e^{-(\kappa + i \omega_c) t}, \\
    \tilde{\alpha}_{\bar{n}}(t) &= |g|^2 \bar{n} e^{-(\kappa - i \omega_c) t}
 \end{split}
\end{equation}
for all times $t >0$. Because of the form of the coupling operator, only their sum is relevant, and the associated spectrum $S_{\bar{n}}(\omega)$ is 
\begin{equation}
    S_{\bar{n}}(\omega) = (\bar{n}+1) S(\omega) + \bar{n} S(-\omega),
\end{equation}
with $S(\omega)$ the zero temperature spectral function. Nonzero occupation, therefore, slightly enhances the spectrum at $\omega$, and more importantly, degrades the bias by adding a contribution at $-\omega$. In other words, $\bar{n}$ limits the spectral asymmetry ratio $\chi$ for all $\omega$
\begin{equation}
 \begin{split}
    \chi_{\bar{n}} &= \frac{(\bar{n}+1) S(\omega) + \bar{n} S(-\omega)}{(\bar{n} +1) S(-\omega) +  \bar{n} S(\omega)} \\
    &= \frac{(\bar{n}+1) \chi + \bar{n}}{(\bar{n} +1) + \bar{n} \chi} \leq \frac{\bar{n} +1}{\bar{n}},
    \end{split}
\end{equation}
where the bound is reached when the zero-temperature asymmetry ratio $\chi$ goes to infinity. 

Beyond this limitation, the analytical construction extends naturally and directly with the substitution $S(\omega) \rightarrow S_{\bar{n}}(\omega)$. Thus, the stationary state remains 
\begin{equation}
    \rho_\mathrm{ss} = \rho_\mathrm{cl, \bar{n}} [\mathds{1} + \braket{U_z(0)} U_z],
    \label{eq::exact:ss:finiten}
\end{equation}
where the classical state $\rho_\mathrm{cl, \bar{n}}$ is the steady state of the classical Pauli master equation~\eqref{eq::Pauli_s} with the replacement $S(\omega) \rightarrow S_{\bar{n}}(\omega)$. 

We close this section by studying the behavior of the GHZ fidelity for experimentally-relevant parameters, as depicted in Fig.~\ref{fig::experimental:simulation}. This includes a finite reservoir occupation $\bar{n} = 0.05$, and power-law Ising couplings $J_{ij} = J_0/|i-j|^{\alpha_J}$ with $\alpha_J = 1$.  Panel (a) highlights the decisive role of the reservoir spectral function. With all other parameters fixed, tuning its center $\omega_c/2\pi$ in the range $1-10~\mathrm{kHz}$ changes the steady state and can drive the GHZ fidelity largely across the $0.5$ fidelity threshold for GME. Furthermore, the close agreement between the analytical prediction of Eq.~\eqref{eq::exact:ss:finiten} and the Markovian embedding supports the applicability of our analytical solution in this parameter regime.

According to the Redfield description, varying $\gamma$ leaves the steady state unchanged since it simply multiplies all rates of the classical Pauli master equation~\eqref{eq::Pauli_s} by the same factor. Rather, $\gamma$ is effectively rescaling the dynamics. Consistent with this prediction, Fig.~\ref{fig::experimental:simulation}~(b) shows that the Markovian embedding leads to nearly identical steady-state fidelities. Increasing $\gamma$ speeds up the dynamics as shown by the fidelity reached at the boundary between the opaque and lighter segments, corresponding to a fixed final time $t_f = 3.75~\mathrm{ms}$. As further illustrated in the inset, the time to reach $F_\mathrm{GHZ} > 0.7$ can be decreased from $3.4~\mathrm{ms}$ to $1.3~\mathrm{ms}$. The subsequent saturation of this improvement signals that increasing $\gamma$ leads to a departure from the simple time rescaling predicted by Redfield and does not indefinitely accelerate the dynamics.
\section{Conclusion and outlook}
\label{sec::conclusion}
We have shown that mutually independent local reservoirs can generate spatial correlations and (genuine) multipartite entanglement when their temporal structure is suitably engineered. For a broad class of open Ising chains, we first established a no-go theorem showing that local and frequency-independent  dissipation exponentially suppresses Ising-axis correlations. Structured reservoirs evade this obstruction by resolving the energy changes associated with local spin flips and selectively biasing domain-wall creation and annihilation. In the weak-coupling and secular regime, this mechanism leads to an exact classical dynamics for the Ising populations and, for nearest-neighbor chains, to steady-state correlations governed by a single effective temperature, independently of the microscopic form of the local system-bath couplings considered. The interaction, therefore, converts the temporal structure stored separately in each reservoir into spatial order in the many-body system.

Strong symmetry provides the additional ingredient required to transform bath-induced spatial correlations into multipartite entanglement. Although weak- and strong-symmetry couplings generate the same classical population dynamics, they lead to fundamentally different quantum steady states. When the global $\mathds{Z}_2$ symmetry is strong, coherences between globally reversed spin configurations obey the same dynamics as the populations and remain tied to the conserved parity of the initial state. We derived the resulting steady-state manifold exactly and showed that initially separable states within a fixed symmetry sector can reach entangled steady states whose every proper reduced state is separable. Our analytical solution establishes a general relation between the distinguishability of classical probability distributions and the negativity, valid for arbitrary Ising couplings and initial states confined to a single symmetry sector. Within such a symmetry sector, we further showed that negativity and GME can be certified using only two-point correlators. Particularizing these results to nearest-neighbor chains leads to especially simple necessary and sufficient relations between two-point correlations and GME. Moreover, we showed that, for initial states with a given parity, the steady state approaches a pure GHZ state as the effective temperature tends to zero. All the aforementioned exact results expose the central mechanism: spectral structure determines which classical configurations are populated, while strong symmetry can convert this population imbalance into quantum entanglement. 

Combining Lorentzian reservoirs, an exact Markovian embedding, and tensor-network simulations, we numerically demonstrate that the universal correlation mechanism is not an artifact of the approximations. Furthermore, using exact diagonalization for small chains, we also numerically confirm the building of GME from initial separable states, in excellent agreement with the analytical predictions.

Finally, we proposed a concrete trapped-ion implementation in which each spin couples selectively to a distinct collective vibrational mode, with dissipation realized by sympathetic cooling of ancillary ions. Drive detunings, cooling rates, and drive amplitudes control the effective reservoir frequencies, linewidths, and spin-reservoir coupling strengths, respectively. An orthogonally polarized motional branch mediates the Ising interactions, facilitating independent control of interactions and dissipation. Our correlation-based entanglement criteria also offer an accompanying detection strategy without full state tomography, provided the initial state belongs to a definite symmetry sector.

This work raises several research questions, and we detail a few below. A first natural question concerns extending the mechanism developed here to non-Abelian strong symmetries. Indeed, non-Abelian symmetries are known to play a key role in generating highly entangled stationary states even under unital dynamics~\cite{Moharramipour2024S, Li2025H}, while our mechanism generates entanglement through spectral selectivity with an Abelian strong symmetry (parity). How do these two routes combine, and how much additional control over the steady-state entanglement can we gain when bath spectral engineering is considered? Another question concerns spectral functions with gaps or finite support. In the weak-coupling description, suppressing selected transition frequencies forbids spin flips for particular neighboring configurations, thereby imposing kinetic constraints. This suggests a connection with constrained models such as PXP, whose coherent quantum many-body dynamics supports quantum many-body scars~\cite{Turner2018W}. Can spectral engineering, combined with suitable coherent dynamics, be used to prepare scarred states or protect their characteristic revivals? A third research direction could be to study the complex interplay between coherent long-range interactions and local dissipative dynamics in higher dimensions. Indeed, our approach applies to arbitrary Ising couplings and therefore provides a concrete route for analyzing such questions. Lastly, we mention more pragmatic questions about the dynamical cost of preparing such states. This could involve analyzing how the preparation time and dissipative gap scale with system size, and their relation to general time-entanglement trade-offs in dissipative state preparation~\cite{Pocklington2024}.

\begin{acknowledgments}

MM acknowledges funding from the Royal Society, Research Ireland, and the European Union’s Horizon research and innovation programme under the Marie Skłodowska-Curie grant agreement No 101264565. 
PW and SP acknowledge support by the Deutsche Forschungsgemeinschaft (DFG, German Research Foundation) under Germany’s Excellence Strategy-426 EXC-2111-390814868.
GP acknowledges support from the Welch Foundation Award (grant no. C-2154), the NSF CAREER Award (grant no. PHY-2144910), and the NSF NRT Award (grant no. 2346014). GP acknowledges that this material is based on work supported by the U.S Department of Energy, Office of Science, Office of Nuclear Physics under the Early Career Award (grant no. DE-SC0023806). Computational resources were provided by the Consortium des Equipements de Calcul Intensif (CECI), funded by the Fonds de la Recherche Scientifique de Belgique (F.R.S.-FNRS) under Grant No.\ 2.5020.11.
\end{acknowledgments}

\FloatBarrier

\appendix 

\section{Memoryless or white noise limit}
\label{memorylesslimit}
In this section, we provide some details about the white noise limit---focusing on its realizations with a Lorentzian bath. First, recall that the dynamics of the reduced spin system (i.e., the Ising chain)  is determined by the bath correlation function (assumed identical for each site)
\begin{equation}
    \alpha(t-s) = |g|^2 e^{-(\kappa+i\omega_c)(t-s)}, \quad t\geq s
\end{equation}
with $g$ the coupling constant appearing in Eq.~\eqref{eq::lindlabd_emb}.
Now, notice that by setting $g^2 = \gamma\kappa$ and taking the limit $\kappa \rightarrow +\infty $, the correlation function becomes 
\begin{equation}
    \alpha(t-s) = \gamma \kappa e^{-i\omega_c (t-s)-\kappa (t-s) } \longrightarrow \gamma  \delta(t-s)
\end{equation}
in the limit of ultrafast damping $\kappa \rightarrow + \infty$. More precisely, this memoryless limit is defined by $\kappa \rightarrow +\infty$ while keeping $g^2/\kappa = \gamma$ constant. In terms of the spectral function, which reads explicitly,
\begin{equation}
    S(\omega) = \mathrm{Re}\int_0^{+\infty}d\tau\alpha(\tau)e^{i\omega \tau} = \frac{\gamma \kappa^2}{\kappa^2 + (\omega-\omega_c)^2},
\end{equation}
the limit $\kappa \rightarrow +\infty$ destroys all spectral selectivity:
\begin{equation}
    \lim_{\kappa \rightarrow +\infty}S(\omega) = \gamma = \mathrm{const.}
\end{equation}
In other words, the bath becomes completely flat in this limit.
Since the correlation function acts as a memory kernel of the reduced dynamics of the spin chain, $\alpha(t-s) \propto \delta(t-s)$ implies that only times $s=t$ will contribute to its dynamics. Consequently, one can show that the dynamics of the Ising chain is exactly captured by the Lindblad master equation (see Ref.~\cite{Debecker2024S, Debecker2024C})
\begin{equation}
    \frac{d\rho}{dt} = -i [H_S, \rho] + \gamma \sum_j(2L_j  \rho L_j^\dagger - L_j^\dagger L_j \rho - \rho L_j^\dagger L_j ). 
\end{equation}
with $\rho$ the reduced density matrix of the spin degrees of freedom. 
\section{Proof of the no-go theorem for flat baths}
In this section, we provide a proof of the no-go theorem presented in the main text. Specifically, we want to demonstrate that for the local jump operators $L_j^s = \sigma_j^z$ and $L_j = \alpha \sigma_j^- + \beta \sigma_j^+$ with $|\alpha| \neq |\beta|,~\alpha, \beta \in \mathbb{R}$ and any Ising Hamiltonian $H_S$ and rate $\gamma_j>0$, the correlators $C_{xx} (i, j) = \braket{\sigma^x_i \sigma^x_j} - \braket{\sigma^x_i} \braket{\sigma^x_j}$ ($i\neq j$) vanish identically in the steady state. \\

Let us show that $\braket{\sigma^x_i \sigma^x_j} \rightarrow 0$ as $t \rightarrow +\infty$. A direct computation gives 
\begin{equation}
 \begin{split}
         \dot{\sigma^x_i \sigma^x_{j}} &= \mathcal{L}^\dagger [\sigma^x_i \sigma^x_j] = i [H_S, \sigma^x_i \sigma^x_j] \\
         &+ \sum_k \gamma_k (2L_k^\dagger \sigma^x_i \sigma^x_j L_k - \{L^\dagger_k L_k, \sigma^x_i \sigma^x_j \} \\
    &= \sum_k \gamma_k (2L_k^\dagger \sigma^x_i \sigma^x_j L_k - \{L^\dagger_k L_k, \sigma^x_i \sigma^x_j \}. \\
    \end{split}
\end{equation}
In this last sum, only terms $k=i$ and $k=j$ give nontrivial contributions. Thus, 
\begin{equation}
 \begin{split}
         \dot{\sigma^x_i \sigma^x_{j}} &= 
    \sum_k \gamma_k (2L_k^\dagger \sigma^x_i \sigma^x_j L_k - \{L^\dagger_k L_k, \sigma^x_i \sigma^x_j \}. \\
    &= \gamma_i (2L_i^\dagger \sigma^x_i L_i - \{L_i^\dagger L_i, \sigma^x_i\}) \sigma^x_j \\
    &+ \gamma_j (2L_j^\dagger \sigma^x_jL_j - \{L_j^\dagger L_j, \sigma^x_j\})\sigma^x_i, \\
    &= \gamma_i \mathcal{D}^\dagger_{L_i}[\sigma^x_i] \sigma^x_j + \gamma_j \mathcal{D}^\dagger_{L_j}[\sigma^x_j] \sigma^x_i.
    \end{split}
\end{equation}
Now, since 
\begin{equation}
 \begin{split}
    \mathcal{D}^\dagger_{L}[\sigma^x] &= \left\{
    \begin{array}{ll}
        -(\alpha-\beta)^2\sigma^x & \mbox{if } L = \alpha\sigma^-+\beta\sigma^+ \\
     -4 \sigma^x &  \mbox{if } L = \sigma^z,
    \end{array}
\right.
 \end{split}
\end{equation}
we have that $\braket{\sigma^x_j} \rightarrow 0$ as $t \rightarrow +\infty$, and
the equations of motion for $C_{xx}(i, j)$ close
\begin{equation}
    \dot{\sigma^x_i \sigma^x_j} = - r \gamma \sigma^x_i \sigma^x_j,
\end{equation}
with $r>0$ fixed by the choice of the local jump operators at sites $i$ and $j$. This proves that the type of dynamics considered here exponentially suppresses $xx$ correlations and $\braket{\sigma^x_i \sigma^x_j} = C_{xx}(i, j) = 0$ in the steady state, as claimed in the main text. 
\label{Proof}
\section{Beyond secular effects and crossover}
\label{app:beyondsecular}
In this section, we provide additional details about the deviation from the secular limit presented in Sec.~\ref{sec::frombathmem}.

According to the analytical result~\eqref{eq::xxcorranal}, positive effective temperature $\beta_\mathrm{eff}$ implies that the Ising-axis correlations are necessarily ferromagnetic. One can work at negative $\beta_\mathrm{eff}$ by choosing the spectral function $S(\omega)$ such that $S(-4J) > S(4J)$. For these negative temperatures, the physics is essentially the same, with the Ising-axis correlations being antiferromagnetic.

\begin{figure}[ht!]
    \centering
    \includegraphics[width=0.98\linewidth]{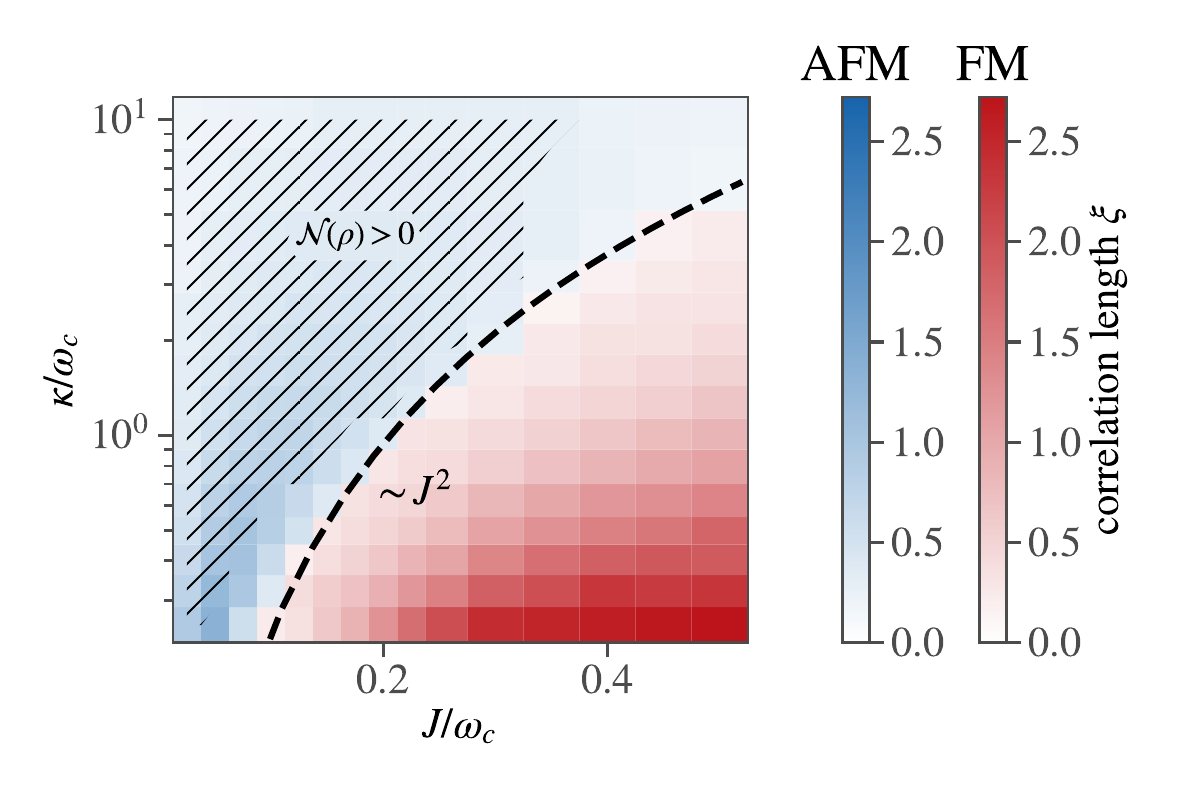}
    \vspace{-0.4cm}
    \caption{\textbf{Behavior of the Ising-axis correlations beyond the secular approximation.} We consider the full Markovian embedding with $N=20$ and $\gamma/\omega_c=0.5$. 
    We show the correlation length $\xi$ in the steady state. While the secularized approximation predicts ferromagnetic (FM) behavior at $J>0$, we see a pronounced crossover between a FM and AFM region, where the crossover line shows a $\kappa\sim J^2$ dependence. 
    Additionally, we highlight the region of small but finite two-spin negativity $\mathcal N(\rho)$, which lies in the AFM region. 
    }
    \label{fig:crossover}
\end{figure}

Interestingly, we find that this crossover from ferromagnetic to antiferromagnetic (AFM) Ising-axis correlations can happen even if $\beta_\mathrm{eff}$ is kept strictly positive, as shown in Fig.~\ref{fig:crossover}. These AFM correlations appear for parameters that break the secular approximation, and we also observe a small but nonzero two-spin negativity in this AFM region that is $\kappa$-independent. More importantly, these AFM correlations may explain why the blue points in Fig.~\ref{fig::Universal} (corresponding to the smallest value of $J$) fall further away from the analytical curve. Indeed, recall that by reducing the Hamiltonian frequency $J$, we necessarily degrade the secular approximation. 
\section{Completeness of the steady-state manifold with strong-symmetric couplings}
\label{appendix::complteness_ss}
\subsection{Secular master equation}
Let us derive the explicit form of the secular master equation with the strong-symmetric coupling operators $L_j^{s} = \sigma_j^z$. Projecting these operators onto Bohr sectors gives 
\begin{equation}
    L_j^{s} = \sum_\omega F_j^{s}(\omega),
    \label{eq::app_bohr_sz}
\end{equation}
where, the operators $F_j^s(\omega)$ explicitly reads
\begin{equation}
    F_j^{s}(\omega) = \sum_{\mathbf{s}: E(\mathbf{s}^{(j)}) -E(\mathbf{s}) = \omega } \ket{\mathbf{s}^{(j)}}_x \bra{\mathbf{s}}.
\end{equation}
The secular generator $\mathcal{L}_\mathrm{sec}^{s}$ is then given by 
\begin{equation}
    \mathcal{L}_\mathrm{sec}^{s}[\cdot] = -i[H_{S} + H_\mathrm{LS}, \cdot] + \sum_{j, \omega} S(-\omega) \mathcal{D}_{F_j^s(\omega)}[\cdot],
\end{equation}
with the Lamb-shift Hamiltonian 
\begin{equation}
    H_\mathrm{LS} = \sum_{j, \omega} \mathrm{Im}\left[\Gamma(-\omega)\right] F_j^{s \dagger }(\omega) F_j^s(\omega).
\end{equation}
While the secular generator is different from the Redfield one as it neglects all terms connecting different Bohr sectors $\omega \neq \omega'$, it is straightforward to see that both $\rho_\mathrm{cl}$ and $U_z \rho_\mathrm{cl}$ are also steady states of $\mathcal{L}_\mathrm{sec}$. 
\subsection{Completeness in the secular regime}
While we have argued that the manifold~\eqref{eq::analyticalSS_sz} is also steady for the secular generator, it need not be complete, that is the long time dynamics could lead to steady states that do not belong to this manifold. Here, we demonstrate that it is not the case---in the secular regime---under the general assumption that the spectral function is strictly positive $S(\omega ) >0$ for all $\omega$, and that the (otherwise arbitrary) Ising couplings $J_{i> j}$ define a connected graph.

Proving completeness of this manifold can be done in three simple steps. First, we notice that 
\begin{equation}
    \rho_\mathrm{cl} = \sum_\mathbf{s} P_\mathbf{s}^\mathrm{ss}\ket{\mathbf{s}}_x \bra{\mathbf{s}}, 
\end{equation}
is unique and full rank. This comes from the fact that $S(\omega) >0$ for all $\omega$ directly implies that it is possible to go from any given configuration to any other configuration with positive probability and in a finite amount of time. In other words, the Markov chain defined by the Pauli master equation~\eqref{eq::Pauli_s} is irreducible. This, in turn, implies that $\rho_\mathrm{cl}$ is unique and $P_\mathbf{s}^\mathrm{ss} >0 ~\forall \mathbf{s}$~\cite{Norris1997}.

The next step is to realize that since $\rho_\mathrm{cl}$ is full rank, then it is sufficient to show that the following commutant
\begin{equation}
    \{H_{S}, F_j^{s}(\omega), F_j^{s\dagger}(\omega)\}', 
\end{equation}
which is the set of operators that commute with $H_{S}$ and all $F_j(\omega)$, $F_j^\dagger (\omega)$ for all $j$ and $\omega$, is of dimension two by virtue of Prop.~8 of Ref.~\cite{Deschamps2016S}. Specifically, we will show that 
\begin{equation}
    \{H_S, F_j^s(\omega), F_j^{s\dagger}(\omega)\}' = \mathrm{span} \{\mathds{1}, U_z\},
    \label{eq::app_commutant}
\end{equation}
which, physically, means that the trace and the parity of $\rho$ are the only two conserved quantities.

The third and last step of this proof consists thus in demonstrating the commutant equality~\eqref{eq::app_commutant}. 
Let $O$ be in this commutant.  Then, $[O, F_j(\omega)] = [O, F_j^\dagger (\omega)] = 0$ for all $j$ and $\omega$. Since $F_j(\omega)$ are the Bohr components of the coupling operator $L_j^s = \sigma_j^z$, we have $\sum_\omega F_j(\omega) = \sigma^z_j$ (see Eq.~\eqref{eq::app_bohr_sz}). Hence, $[O, \sigma^z_j] = 0$ for all $j$. Therefore, $O$ must be diagonal in the $z$ basis 
\begin{equation}
    O = \sum_{\mathbf{s}_z} c_{\mathbf{s}_z} \ket{\mathbf{s}}_z\bra{\mathbf{s}}.
\end{equation}
For $O$ lives in the commutant, we also have $[O, H_S] = 0$ with $H_S = - \sum_{i < j} J_{ij} \sigma^x_i \sigma^x_j$. This gives successively
\begin{equation}
 \begin{split}
 &\sum_{\mathbf{s}_z}\sum_{i < j} J_{ij} c_{\mathbf{s}_z} \left(\ket{\mathbf{s}^{(i, j)}}_z\bra{\mathbf{s}} - \ket{\mathbf{s}}_z \bra{\mathbf{s}^{(i, j)}}\right) = 0 \\
 &\iff J_{ij} (c_{\mathbf{s}_z} -c_{\mathbf{s}_z^{(i, j)}}) = 0 \quad \forall \mathbf{s}_z,~ i< j,
    \end{split}
\end{equation}
Thus, whenever $J_{ij} \neq 0$, the coefficients associated with  $\ket{\mathbf{s}}_z$ and $\ket{\mathbf{s}^{(i, j})}_z$---obtained from flipping sites $i$ and $j$ in the $z$ basis---are identical. Now, because the couplings $J_{ij}$ define a connected graph by hypothesis, there exists a path with only nonzero $J_{ij}$ that connects any two spins $i_0$ and $j_0$. Let $(i_0, k_1, k_2, ..., j_0)$ be such a path. Successively flipping pairs of spins along the path then leads to 
\begin{equation}
    c_{\mathbf{s}_z} = c_{\mathbf{s}_z^{(i_0, k_1)}} = c_{\mathbf{s}_z^{(i_0, k_2)}} = c_{\mathbf{s}_z^{(i_0, k_3)}} ... = c_{\mathbf{s}_z^{(i_0, j_0)}},
\end{equation}
where for the first equality we flipped $(i_0, k_1)$, then $(k_1, k_2)$, $(k_2, k_3), ...$. This means that all the coefficients $c_{\mathbf{s}_z}$ are the same in a given class of configurations that are related by the flipping of an even or odd number of spins. In other words, the coefficient $c_{\mathbf{s}_z}$ can only take two values: one for each parity, that is, 
\begin{equation}
    O = c_+ \Pi_+ + c_- \Pi_-,
\end{equation}
with the parity projectors $\Pi_\pm = (\mathds{1} \pm U_z)/2$. This proves that $O \in \mathrm{span}\left\{\mathds{1}, U_z\right\}$. Conversely, if $O \in \mathrm{span}\left\{\mathds{1}, U_z\right\}$, then it commutes with $H_S$ and $F_j^s(\omega), F_j^{s \dagger}(\omega)$ for all $j$ and $\omega$ since $U_z$ trivially commutes with $H_S$ and $\sigma^z_j$ (and hence also with the spectral projectors of $H_S$---the operators $F_j^{s}(\omega)$. We therefore conclude that the kernel of $\mathcal{L}_\mathrm{sec}$ is of dimension two, and the manifold~\eqref{eq::analyticalSS_sz} is complete at the secular level.

\subsection{Completeness for the Redfield generator}
Having shown that the secular generator has a two-dimensional kernel, we now prove that it is still the case---except for some fine-tuned parameters---for the Redfield generator. More precisely, we show that its kernel is generically two-dimensional, and these accidental degeneracies can be lifted by adding an arbitrarily small admixture of white noise.  

Let us start with a counterexample. Taking two sites, $H_S = -J \sigma^x_1 \sigma^x_2$ and $L_j^s = \sigma^z_j$ as usual, the operators $\bar{L}_j^{s}$ can be readily computed 
\begin{equation}
 \begin{split}
    \bar{L}_1^{s} &= \frac{\Gamma(2J) + \Gamma(-2J)}{2} \sigma^z_1 + \frac{\Gamma(2J) - \Gamma(-2J)}{2i} \sigma^y_1 \sigma^x_2, \\
    \bar{L}_2^{s} &= \frac{\Gamma(2J) + \Gamma(-2J)}{2} \sigma^z_2 + \frac{\Gamma(2J) - \Gamma(-2J)}{2i} \sigma^y_2 \sigma^x_1.
 \end{split}
\end{equation}
Now, straightforward algebra leads to the conclusion that the following four-dimensional subspace
\begin{equation}
    \mathrm{span}\left\{\rho_\mathrm{cl}, U_z\rho_\mathrm{cl}, \sigma^z_1, \sigma^z_2 \right\}, \quad \rho_\mathrm{cl} = \frac14 \left(\mathds{1} + \frac23 \sigma^x_1 \sigma^x_2\right)
\end{equation}
is steady for $\kappa = 2J$, $\omega_c= 4J$, and $g^2 = 5J^2$. Note that these parameters lie outside the domain of validity of the Redfield master equation; the example establishes an algebraic possibility of the Redfield generator.

The counterexample above is not the norm; rather, it is the exception. In the parameter regimes explored numerically, the stationary states of the Redfield dynamics agree with our analytical prediction~\eqref{eq::analyticalSS_sz}. More importantly, we now show that these extra stationary directions are unstable under the addition of an arbitrary slight white-noise background. In other words, exceptional generators can be approached arbitrarily closely with generators that have a two-dimensional kernel. Throughout, we keep our usual assumptions, that is $S(\omega) >0$ everywhere, and we assume a connected interaction graph.

Let us first consider the flat bath limit $\Gamma(\omega) = \gamma >0$, which leads to the generator (see Appendix~\ref{memorylesslimit})
\begin{equation}
   \mathcal{L}_\mathrm{flat} = -i[H_S, \cdot] + \gamma \sum_j \mathcal{D}_{\sigma^z_j}[\cdot].
\end{equation}
We note that $\dim \ker \mathcal{L}_\mathrm{flat} = 2$. Indeed, the full-rank state $\mathds{1}/2^N$ is stationary. Moreover, since, from the previous section we know that $\{H_S, \sigma^z_j \}'$ is two-dimensional, we directly conclude that $\dim \ker \mathcal{L}_\mathrm{flat} = 2$. 

The next step consists in constructing the new generator 
\begin{equation}
    \mathcal{L}(\lambda) = (1-\lambda)\mathcal{L}_\mathrm{R} + \lambda \mathcal{L}_\mathrm{flat},
\end{equation}
and realizing that the kernel of $\mathcal{L}(\lambda)$ is generically of dimension two, i.e., it is two-dimensional for all values of $\lambda$ except for finitely many. Indeed, this comes from the fact that 
\begin{equation}
    \mathrm{dim}\left( \mathrm{ker} \mathcal{L}(\lambda)\right) = \mathrm{dim}\left( \mathrm{ker} \mathcal{L^\dagger}(\lambda)\right) = 2 + \mathrm{dim}\left(\mathrm{ker}\mathcal{L}(\lambda)^\dagger|_\mathcal{A}\right),
    \label{eq::app_dimension_ker_red}
\end{equation}
where, in the last equality, $\mathcal{L}(\lambda)^\dagger|_\mathcal{A}$ denotes the restriction of  $\mathcal{L}(\lambda)^\dagger$ to the space of operators orthogonal to $\mathds{1}$ and $U_z$
\begin{equation}
    \mathcal{A} = \left\{A: \mathrm{Tr}[A] = 0~\mathrm{and}~\mathrm{Tr}[AU_z] = 0\right\}.
\end{equation}
Eq.~\eqref{eq::app_dimension_ker_red} comes from the fact that for any positive $S(\omega)$ and connected graph $J_{ij}$, we have shown in the previous section that both $\mathds{1}$ and $U_z$ lie in the kernel of $\mathcal{L}^\dagger$.

The map $\mathcal{T}(\lambda) = \mathcal{L}^\dagger(\lambda)|_\mathcal{A}$ need not be an endormorphism; it is thus generically represented by a rectangular matrix. The operator $\mathcal{T}^\dagger (\lambda) \mathcal{T}(\lambda)$, however, is represented by a square matrix and $\mathrm{ker} T(\lambda) = \ker(\mathcal{T}(\lambda)^\dagger  \mathcal{T}(\lambda))$. Now, since  $P(\lambda) = \mathrm{det}(\mathcal{T}(\lambda)^\dagger \mathcal{T}(\lambda))$ is a polynomial in $\lambda$, it can only have finitely many roots that would increase the dimension of the kernel. Furthermore, $P(\lambda)$ is not the zero polynomial since for $\lambda = 1$, $\mathcal{L}(\lambda) = \mathcal{L}_\mathrm{flat}$, which implies $P(1) \neq 0$.
\section{Exact computation of the negativity} \label{app:negativity}
In this Appendix, we derive the bipartite negativity of the steady states
\begin{equation}
    \rho_\mathrm{ss} = \rho_\mathrm{cl} (\mathds{1} + \braket{U_z(0)} U_z), \quad \rho_\mathrm{cl} = \sum_{\mathbf{s}} P_\mathbf{s}^{\mathrm{ss}} \ket{\mathbf{s}}_x\bra{\mathbf{s}},
    \label{app:ss_strong}
\end{equation}

\subsection{General couplings $J_{ij}$}
We start by noting that the classical dynamics is invariant under a global spin flip, and we always work in conditions that ensure the uniqueness of the stationary distribution $P_\mathbf{s}^\mathrm{ss}$ (i.e., $S(\omega) >0$; see Appendix~\ref{appendix::complteness_ss}). We therefore have $P_\mathbf{s}^\mathrm{ss} = P_{-\mathbf{s}}^\mathrm{ss}$, and the steady state family~\eqref{app:ss_strong} can be written as 
\begin{equation}
    \rho_\mathrm{ss} = \sum_\mathbf{s} P_\mathbf{s}^\mathrm{ss} (\ket{\mathbf{s}}_x \bra{\mathbf{s}} + \braket{U_z(0)} \ket{\mathbf{s}}_x\bra{-\mathbf{s}}).
\end{equation}
Let us now consider a given proper global bipartition $A|\bar{A}$ of the spin chain, and let $\mathbf{s}_A$ and $\mathbf{s}_{\bar{A}}$ be the configurations associated with sets $A$ and $\bar{A}$, respectively. The negativity across the bipartition $A|\bar{A}$ is defined as 
\begin{equation}
    \mathcal{N}_{A|\bar{A}} \equiv \frac{\norm{\rho_{ss}^{T_A}}_1 - 1}{2},
    \label{eq:def_neg}
\end{equation}
where $\cdot ^{T_A}$ indicates the partial transpose with respect to $A$, and $\norm{\cdot}_1$ is the trace norm. The partial transpose $\rho_\mathrm{ss}^{T_A}$ with respect to system $A$ then reads:

\begin{widetext}
\begin{equation}
\begin{aligned}
\rho_{\mathrm{ss}}^{T_A}
&= \sum_{\mathbf{s}=(\mathbf{s}_A,\mathbf{s}_{\bar A})}
P_{\mathbf{s}}^{\mathrm{ss}}
\Bigl[
    \ket{\mathbf{s}}_x \bra{\mathbf{s}}
    + \braket{U_z(0)}
    \bigl(
        \ket{\mathbf{s}_A,\mathbf{s}_{\bar A}}_x
        \bra{-\mathbf{s}_A,-\mathbf{s}_{\bar A}}
    \bigr)^{T_A}
\Bigr] \\
&= \sum_{\mathbf{s}=(\mathbf{s}_A,\mathbf{s}_{\bar A})}
P_{\mathbf{s}}^{\mathrm{ss}}
\Bigl[
    \ket{\mathbf{s}}_x \bra{\mathbf{s}}
    + \braket{U_z(0)}
    \ket{-\mathbf{s}_A,\mathbf{s}_{\bar A}}_x
    \bra{\mathbf{s}_A,-\mathbf{s}_{\bar A}}
\Bigr],
\end{aligned}
\end{equation}
\end{widetext}
where we used the fact that the diagonal term is unaffected by the
partial transpose. Now, in the second term, the dyadic products can be conveniently written as $\ket{-\mathbf{s}_A, \mathbf{s}_{\bar{A}}} \bra{\mathbf{s}_A, -\mathbf{s}_{\bar{A}}} = \ket{\mathbf{s}^{(A)}} \bra{-\mathbf{s}^{(A)}}$ by defining $\ket{\mathbf{s}^{(A)}} \equiv \ket{-\mathbf{s}_A, \mathbf{s}_{\bar{A}}}$, i.e., $\ket{\mathbf{s}^{(A)}}$ corresponds to the configuration $\mathbf{s}$ obtained by reversing all spins belonging to $A$. Relabeling $\mathbf{s}^{(A)} \rightarrow \mathbf{s}$ yields
\begin{equation}
    \rho_\mathrm{ss}^{T_A} = \sum_\mathbf{s} P_\mathbf{s}^\mathrm{ss} \ket{\mathbf{s}}\bra{\mathbf{s}} + \braket{U_z(0)} \sum_\mathbf{s} P^\mathrm{ss}_{\mathbf{s}^{(A)}} \ket{\mathbf{s}}{\bra{-\mathbf{s}}}.
\end{equation}
This last equation shows that, by working in the subspaces spanned by $\ket{\mathbf{s}}$ and $\ket{-\mathbf{s}}$, $\rho_\mathrm{ss}^{T_A}$ can be written as the direct sum of simple two-dimensional matrices 
\begin{equation}
    R_\mathbf{s} = 
    \begin{pmatrix}
    P_\mathbf{s} & \braket{U_z(0)}P_{\mathbf{s^{(A)}}} \\
    \braket{U_z(0)}P_{\mathbf{s^{(A)}}} & P_\mathbf{s}
    \end{pmatrix},
\end{equation}
with eigenvalues $\lambda_\mathbf{s}^\pm = P_\mathbf{s} \pm |\braket{U_z(0)} |P_\mathbf{s}^{(A)}$. Consequently, the trace norm of $\rho_\mathrm{ss}^{T_A}$ is
\begin{equation}
    \begin{split}
    \norm{\rho_\mathrm{ss}^{T_A}}_1 &= \frac12 \sum_{\mathbf{s}}|\lambda_\mathbf{s}^+| + |\lambda_\mathbf{s}^-| \\
    &= \frac12(1+ |\braket{U_z(0)}|) + \frac12 \sum_\mathbf{s}|P_\mathbf{s}- |\braket{U_z(0)}| P_{\mathbf{s}^{(A)}}|,
    \end{split}
\end{equation}
where the factor $1/2$ prevents double counting, and we used $\sum_{\mathbf{s}} P_\mathbf{s} = 1$. Using the definition of the negativity~\eqref{eq:def_neg}, we finally obtain the result stated in the main text:
\begin{equation}
    \mathcal{N}_{A|\bar{A}} = \frac14 \left[ \sum_\mathbf{s} |P_\mathbf{s}^\mathrm{ss} - |\braket{U_z}_0| P_{\mathbf{s}^{(A)}}^{\mathrm{ss}}| - (1- |\braket{U_z}_0|)\right].
\end{equation}

In a fixed parity symmetry sector, we simply have $\braket{U_z(0)} = \pm 1$, and the negativity simplifies to
\begin{equation}
    \mathcal{N}_{A|\bar{A}} = \frac{1}{4} \sum_\mathbf{s}|P^\mathrm{ss}_\mathbf{s} - P^\mathrm{ss}_{\mathbf{s}^{(A)}}| = \frac12 \delta (P^\mathrm{ss}, P^{\mathrm{ss}(A)}),
\end{equation}
with $\delta (\cdot, \cdot)$ the total variation distance introduced in the main text. This relation has an important consequence. Indeed, consider a spin $i$ belonging to $A$ and $j$ in $\bar{A}$. The mean value of the correlator $\sigma^x_i \sigma^x_j$ can be written as 
\begin{equation}
    \braket{\sigma^x_i \sigma^x_j} = \sum_{\mathbf{s}} s_i s_j P_\mathbf{s} =  \sum_{\mathbf{s}} (-s_i) s_j P_{\mathbf{s}^{(A)}},
\end{equation}
or, equivalently, 
\begin{equation}
    2 \braket{\sigma^x_i \sigma^x_j} = \sum_{\mathbf{s}} s_i s_j (P_\mathbf{s} - P_{\mathbf{s}^{(A)}}). 
\end{equation}
Therefore, the absolute value of this Ising correlator directly gives a lower bound for the negativity 
\begin{equation}
    |\braket{\sigma^x_i \sigma^x_j}| \leq \frac12 \sum_\mathbf{s} |P_\mathbf{s} - P_{\mathbf{s}^{(A)}}| \iff \mathcal{N}_{A|\bar{A}} \geq \frac12 |\braket{\sigma^x_i \sigma^x_j}|.
\end{equation}
\subsection{Nearest-neighbor case }
For the nearest-neighbor case and for a contiguous bipartition $A = \{1, ..., \ell\}$ (with $\bar{A} = \{\ell+1, ..., N\}$), the negativity only depends on the correlator $\braket{\sigma^x_\ell \sigma^x_{\ell+1}}$ and the initial condition. Below, we demonstrate Eq.~\eqref{eq::neg_NN_general} valid in the thermodynamic limit, and we provide an exact finite-size expression for PBC. 

We first notice that Eq.~\eqref{eq::neg_general} can be conveniently written as 
\begin{equation}
    \mathcal{N}_{A|\bar{A}} = \frac12 \sum_\mathbf{s} \left[|\braket{U_z(0)}| P_\mathbf{s} - P_{\mathbf{s}^{(A)}} \right]_+,
\end{equation}
by rewriting $1- |\braket{U_z(0)}| = \sum_\mathbf{s} (P_\mathbf{s} - |\braket{U_z(0)}| P_{\mathbf{s}^{(A)}})$, and exploiting the elementary identity $|x| -x  = 2 \max(-x, 0) \equiv 2 [-x]_+$. The further simplifications for nearest-neighbor interactions and OBC in thermodynamic limit or PBC at finite $N$ come from using $P_{\mathbf{s}} = e^{-\beta_{\mathrm{eff}} E(\mathbf{s})/Z_{\mathrm{eff}}}$. 

For OBC in the thermodynamic limit $N \rightarrow +\infty$, flipping all spins in $A$ only affects the energy at the cut between $A$ and $\bar{A}$. We therefore have $E(\mathbf{s}^{(A)}) - E(\mathbf{s}) = 2J s_{\ell} s_{\ell +1}$. This implies
\begin{equation}
    \mathcal{N}_{A|\bar A} = \frac12 \sum_\mathbf{s} P_\mathbf{s} \left[|\braket{U_z(0)}| -e^{-2\beta_\mathrm{eff} J s_\ell s_{\ell +1}}\right]_+.
\end{equation}
Now, we split this sum in two: all the configurations for which $s_\ell s_{\ell +1} = +1 $ and $s_\ell s_{\ell +1} = -1$:
\begin{equation}
    \begin{split}
        \mathcal{N}_{A|\bar A} &= \frac12 \left( P_\mathrm{a} [|\braket{U_z(0)}| - e^{-2 \beta_\mathrm{eff} J}]_+ \right. \\
        &\left. + P_\mathrm{aa} [|\braket{U_z(0)}| - e^{2 \beta_\mathrm{eff} J}]_+\right) \\
        &= \frac12 P_\mathrm{a} [|\braket{U_z(0)}| - e^{-2 \beta_\mathrm{eff} J}]_+,
    \end{split}
\end{equation}
where $P_\mathrm{a}$ (resp. $P_\mathrm{aa}$) is the probability that spins $\ell$ and $\ell+1$ are aligned (resp. antialigned). Here, the second line vanishes because $e^{2 \beta _\mathrm{eff}J}  > 1 $ for $\beta_\mathrm{eff}, J >0$ and $|\braket{U_z(0)}|$ is always smaller than one. The probability $P_a$, however, adopts a simple form in terms of the Ising-axis correlator 
\begin{equation}
    P_\mathrm{a} = \frac{1 + \braket{\sigma^x_\ell \sigma^x_{\ell+1}}}{2}.
\end{equation}
This yields, with $e^{-2x} = \frac{1- \tanh(x)}{1+ \tanh(x)}~\forall x\in \mathbb{R}$, 
\begin{equation}
    \mathcal{N}_{A|\bar{A}} = \frac14 \left[(1+ |\braket{U_z(0)}|) |\braket{\sigma^x_\ell \sigma^x_{\ell+1}}| - (1-|\braket{U_z(0)}|) \right]_+,
\end{equation}
as claimed in the main text. For $\beta_\mathrm{eff} J < 0$, a similar reasoning applies. Only the antialigned contribution survives, and the absolute value of the correlator is then needed.

For PBC and finite $N$ (with $N \geq 3$), the main difference is that reversing all spins in $A$ will now affect the two cuts between $A$ and $\bar{A}$. Consequently, we have $E(\mathbf{s}^{(A)}) - E(\mathbf{s}) = 2J (s_{\ell} s_{\ell + 1} + s_{N} s_1).$ Thus 
\begin{equation}
    \mathcal{N}_{A|\bar{A}} = \frac12 \sum_\mathbf{s} P_\mathbf{s}\left[|\braket{U_z(0)}| - e^{-2 \beta_\mathrm{eff} J(s_\ell s_{\ell +1} + s_{N} s_1)} \right]_+.
\end{equation}
Now, we split again the sum according to whether the spins at each cut are aligned or antialigned. For $\beta_\mathrm{eff}, J >0$, only configurations in which both $s_1 s_N$ and $s_\ell s_{\ell +1} $ are aligned contribute, and 
\begin{equation}
    \mathcal{N}_{A|\bar{A}} = \frac12 P_{a, a}\left[|\braket{U_z(0)}| - e^{-4 \beta_\mathrm{eff} J}\right]_+.
\end{equation}
Here, we introduced $P_{a, a}$, the probability that both $s_1 s_N$ and $s_\ell s_{\ell +1} $ are aligned. Computing $P_{a, a}$ from the Gibbs weight is a standard exercise and gives 
\begin{equation}
    P_{a, a} = \frac{(1+ \tanh(\beta_\mathrm{eff} J))^2 (1 + \tanh^{N-2}(\beta_\mathrm{eff}J))}{4(1 + \tanh^{N}(\beta_\mathrm{eff}J))}.
\end{equation}
Introducing the shorthand $t \equiv \tanh(\beta_\mathrm{eff}J)$, we obtain 
\begin{equation}
    \mathcal{N}_{A|\bar{A}} = \frac{1+ t^{N-2}}{8 (1+ t^N)} \left[|\braket{U_z(0)}| (1+t)^2- (1- t)^2\right]_+.
\end{equation}
In a symmetry sector, this last expression simplifies to 
\begin{equation}
    \mathcal{N}_{A|\bar{A}} = \frac{t + t^{N-1}}{2(1+ t^N)} = \frac12 \braket{\sigma^x_{\ell} \sigma^x_{\ell +1}},
\end{equation}
where we identified the finite-size Ising-axis correlator for PBC. Finally, for $\beta_\mathrm{eff} J < 0$, a perfectly analogous reasoning gives the same expression with the correlator replaced by its absolute value.
\section{Numerical details}
\label{app:sec:numerics}

\begin{figure}
    \centering
    \includegraphics[width=\linewidth]{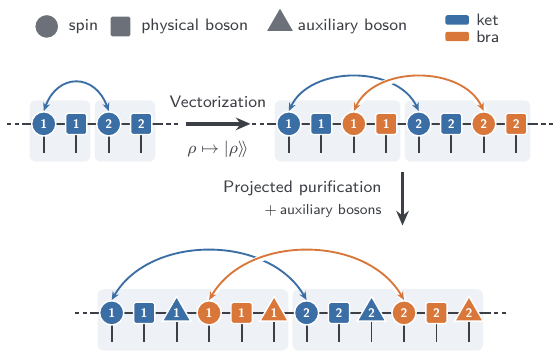}
    \caption{\textbf{Matrix-product-state (MPS) representation of the Markovian-embedded system }(Eq.~\ref{eq::lindlabd_emb}). In each schematic, we show two minimal units and for clarity only the longest-range couplings are displayed. Top-left: physical system consisting of one spin and one boson per unit. Top-right: to simulate the dissipative dynamics, we employ vectorization~\cite{Casagrande2021}, which doubles the Hilbert space. Bottom: to artificially enforce particle-number conservation on the bosonic sites, we use projected purification~\cite{Koehler_2021}: an auxiliary boson is introduced for each physical boson, and the enlarged Hilbert space is subsequently projected onto a subspace isomorphic to the original physical one.}
    \label{fig:app:mps:cartoon}
\end{figure}

To numerically study the Markovian embedded system (Eq.~\ref{eq::lindlabd_emb}), we employ the CLIK-MPS (Complex-time Lindbladian Krylov subspace Matrix Product States) framework~\cite{Westhoff2025A}, which allows one to systematically compute the low-lying
eigenmodes of the Lindbladian superoperator $\hat{\mathcal{L}}$.
Here we are interested in the steady state only, i.e.\ the right eigenvector of the vectorized Lindbladian corresponding to the eigenvalue $\lambda = 0$.
The main idea, based on~\cite{Minganti2022}, is that instead of running a time evolution until brute-force convergence to the steady state, which requires times $t \gg 1/|\mathrm{Re}\lambda_2|$ set by the dissipative gap, one can perform a \emph{short} time evolution, use the time-evolved state to span a Krylov subspace, and diagonalize the Lindbladian within that subspace.
This reduces the required evolution time substantially and yields the steady state with high accuracy.

Concretely, CLIK-MPS builds this subspace from two separate time evolutions.
The first is a short real-time evolution starting from a traceful initial state, which generates a set of time-evolved states with a non-vanishing overlap with the steady state.
The second evolution starts from a traceless initial state and is performed along a linearly inclined contour in the complex-time plane~\cite{Grundner2024},
parametrized by $z(t) = t\, e^{-i\alpha}$, with a small positive angle $\alpha$.
Evolving along this contour modifies the damping rate of the $k$-th eigenmode from $\xi_k = -\mathrm{Re}(\lambda_k)$
to $\xi_k = -\mathrm{Re}(\lambda_k)\cos\alpha -
\mathrm{Im}(\lambda_k)\sin\alpha$, thereby enhancing the contribution of
eigenmodes with $\mathrm{Im}(\lambda_k) > 0$ relative to a purely real-time
evolution and improving the resolution of low-lying excited states. Moreover, the use of two time evolutions (one with $\alpha=0$ and the other with $\alpha>0$), allows one to obtain many linearly-independent time-evolved states within a relatively short time evolution interval.
Once all time-evolved states are collected, the matrix elements of the Lindbladian $\mathcal{L}_{ij} =
\langle\!\langle\rho(t_i)|\hat{\mathcal{L}}|\rho(t_j)\rangle\!\rangle$ and the overlaps $M_{ij} = \langle\!\langle\rho(t_i)|\rho(t_j)\rangle\!\rangle$
(Gram matrix) are computed~\cite{Paeckel2026NS}.
The Gram matrix is then diagonalized, $M = USU^\dagger$, and used to construct
the transformation $X = US^{-1/2}$ into an orthonormal basis, discarding eigenvalues below a numerical threshold $\epsilon=10^{-14}$ to remove linear
dependencies.
This avoids Gram-Schmidt orthogonalization, which is numerically unstable for MPS.
Finally, the effective Lindbladian $\mathcal{L}_\mathrm{eff} =
X^\dagger \mathcal{L} X$ is diagonalized in the orthonormalized Krylov
subspace and the eigenstate corresponding to $\lambda = 0$ is identified as
the steady state.
 
\begin{figure}
    \centering
    \subfloat[]{
    \includegraphics[width=0.9\columnwidth]{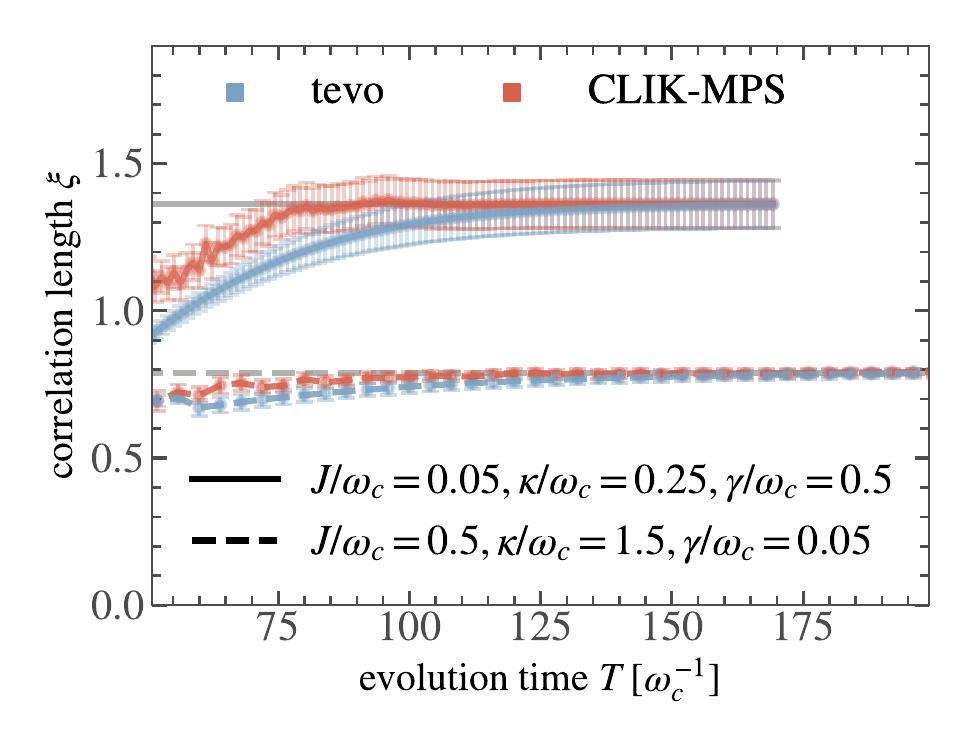}
    }
    \vspace{-0.8cm}
    \subfloat[]{
    \includegraphics[width=0.9\linewidth]{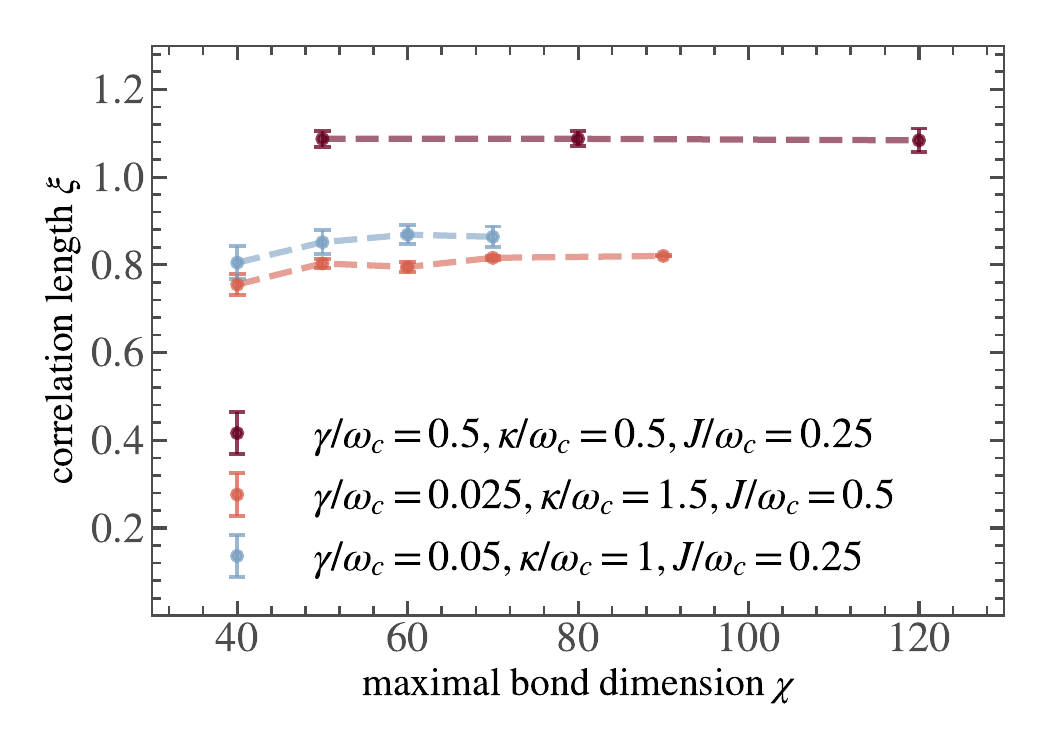}
    }
    \caption{\textbf{Convergence of CLIK-MPS.} (a) Comparing the correlation length calculated from the time evolved state (blue) with the result from CLIK-MPS with a Krylov space spanned from time evolved states up until the same evolution time (orange), measured in $\omega_c^{-1}$. 
    (b) Convergence of $\xi$ with the maximal bond dimension chosen for the time evolution used in CLIK-MPS for multiple parameter sets.
    }
    \label{fig:clik:vs:time}
\end{figure}

States and operators are represented in the matrix product state (MPS) and
matrix product operator (MPO) formalism, respectively~\cite{Schollwoeck2011}.
The density matrix $\hat{\rho}$ is vectorized~\cite{Amshallem2015} (and indicated as $|\rho\rangle\!\rangle$) by doubling the system size,
alternating physical and auxiliary sites on the enlarged lattice, so that the
Lindblad equation maps onto a Schr\"{o}dinger equation with a non-Hermitian
generator on the doubled Hilbert space $\partial_t |\rho\rangle\!\rangle = \hat{\mathcal{L}}\, |\rho\rangle\!\rangle$.
This vectorization scheme keeps the Lindbladian as local as possible in the MPO
representation, avoiding the introduction of long-range couplings~\cite{Casagrande2021}.
The time evolution is carried out using the local subspace expansion
time-dependent variational principle (LSE-TDVP)~\cite{Yang2020}, which is well-suited for bosonic degrees of freedom with large local Hilbert space dimensions.
For the bosonic sites, we additionally employ projected purification (PP)~\cite{Koehler_2021}, which
enforces particle number conservation, allowing us to exploit the $U(1)$ symmetry of the system exactly and to represent the vectorized density matrix in the appropriate particle-number sector. The MPS representation of a minimal system composed of two spins and two bosons employing vectorization and projected purification is shown schematically in Fig.~\ref{fig:app:mps:cartoon}. 
The implementation of CLIK-MPS used in this work is semi-automated. 
While the angle $\alpha$, timestep $\delta t$ and maximal bond dimension $\chi$ need to be chosen manually, the time evolution stops automatically as soon as the Krylov space is converged. 
This is made possible by calculating the Krylov space on-the-fly: Whenever a new timestep has been calculated for both time evolutions, its overlaps and expectation values with the other states are calculated in parallel, and the Lindbladian is diagonalized. 
This allows us to track the convergence of the lowest eigenvalue at each timestep. 
As soon as the eigenvalue is smaller than a threshold $|\lambda|<\mu$ for multiple timesteps in a row (here we choose $\mu=6\times 10^{-3}$ throughout), we are guaranteed that the steady state is sufficiently converged. 
To curtail problems due to a low number of Krylov vectors, we set a minimal evolution time $T_\mathrm{min}$, after which the above stopping criterion is used. $T_\mathrm{min}$ is parameter dependent, and we increase it whenever $\kappa$ or $\gamma$ decrease.
Additionally, in order to include only time-evolved states having enough overlap with the low lying eigenvalues, we track the Lindbladian residual $\Vert\mathcal L\rho(t)\Vert$ and only keep states into the Krylov space which feature a Lindbladian residual smaller than $9 \times 10^{-2}$. In Fig. \ref{fig:clik:vs:time} a) we show that CLIK-MPS converges substantially faster to the steady state value of the correlation length $\xi$ than a simple time evolution. Note that the exact advantage provided by CLIK-MPS depends on the considered system parameters.
We always calculate correlation lengths $\xi$ by fitting with an exponential ansatz $C(i-j) = \langle \sigma^x_i\sigma^x_j \rangle \sim e^{-|i-j|}/\xi$. To ensure that all results are comparable, we consider only values of the correlator $C(i-j)>\epsilon$, where we set the threshold to $\epsilon=10^{-2}$ for the data in Fig.~\ref{fig::Universal}. For Fig.~\ref{fig:crossover}, we need to lower the threshold to $\epsilon=0.8\times 10^{-3}$, since we find very small correlation lengths in the crossover region which we could not resolve otherwise . This threshold is needed because some steady states feature small deviations due to truncation and orthogonalization errors, which get visible at small values of $C$. Consequently, for parameters yielding very small correlation lengths ($\xi\ll 1$), a trustworthy fit can only be done with a few data points, leading to no estimate for the fitting error of $\xi$. These data points therefore do not feature an error bar in Fig.~\ref{fig::Universal}. 
To establish whether the correlation length $\xi$ fitted from the Krylov space is sufficiently converged, we calculate the deviation from the mean $\bar \xi$ when successively adding more states into the Krylov space. We always calculate this deviation for the last five $\xi$-s, and do not terminate CLIK-MPS before the deviation is less than $1\%$ of the mean value.
We also performed scaling analysis of the correlation length with the system size to check for finite size effects. We found that the correlation length for the tested parameters does not change upon increasing the system size up to $N=40$ sites.
Additionally, we performed convergence analysis with the maximal bond dimension allowed during time evolution, as shown in Fig.~\ref{fig:clik:vs:time} b). For most calculations we find that a bond dimension of $\chi=50$ suffices. For the calculations in Fig.~\ref{fig:crossover} that feature $\kappa/\omega_c<0.4$ we increase the bond dimension to $\chi=80$.
In Fig. \ref{fig::Universal} (b), we initialize with a pure product state that has the local representation $\ket{\psi} = \cos\theta \, \ket{\uparrow} + \sin \theta \, \ket{\downarrow}$, where we can control the initial correlator precisely, as $\langle \sigma^x_i\sigma^x_{i+1} \rangle = 4\cos^2 \theta\, \sin^2\theta$, and we chose $\theta=0.1$.

\begin{figure}
    \centering
    \subfloat[]{
    \includegraphics[width=0.9\columnwidth]{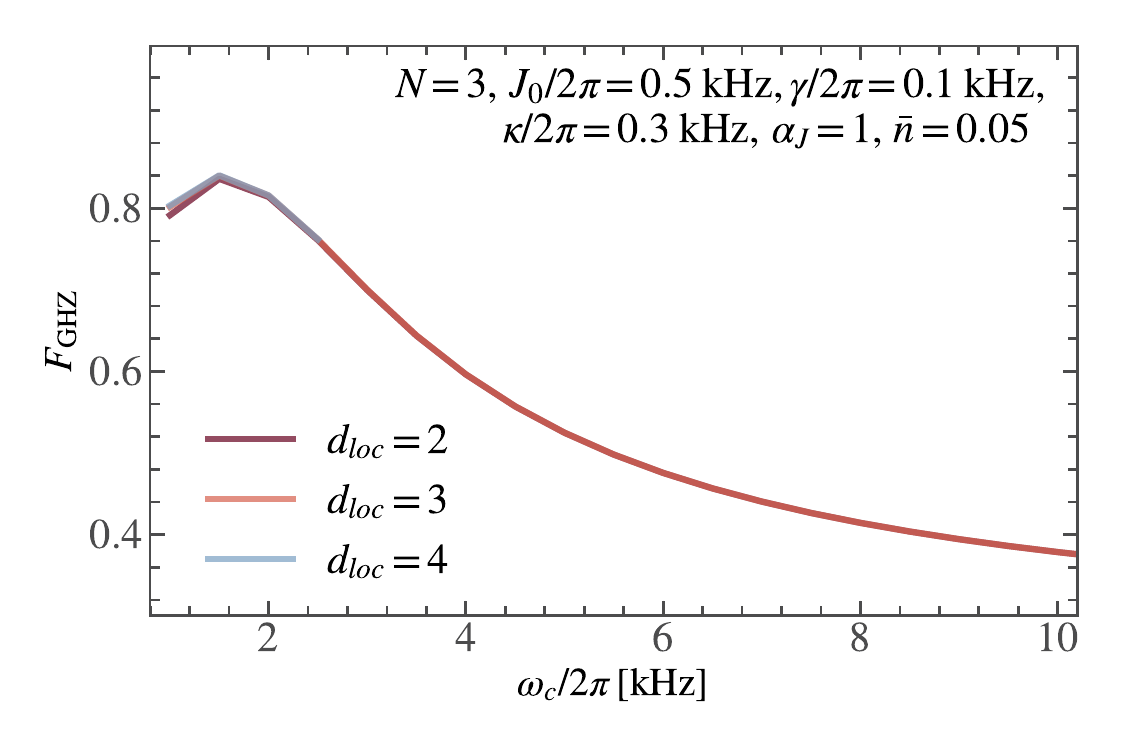}
    }
    \vspace{-0.8cm}
    \subfloat[]{
    \includegraphics[width=0.9\columnwidth]{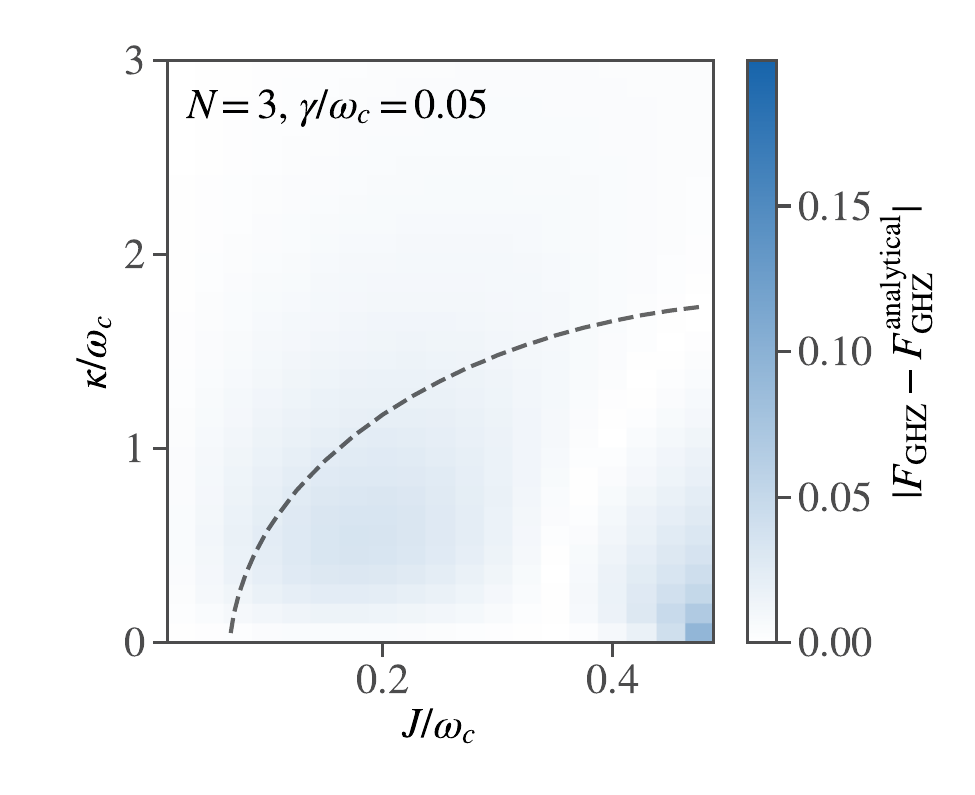}
    }
    \caption{\textbf{Comparison of Markovian embedding and analytical results}. (a) GHZ fidelity for different local bosonic dimensions. We find that for small system sizes, it typically suffices to cut the bosonic Hilbert space at a local dimension of $d_{loc}=2$. 
    (b) Comparison of the GHZ fidelity calculated for the full Markovian embedding with the analytical result Eq.~\eqref{eq::analytical_fid_GHZ}.
    }
    \label{fig:bosonic:dimension:comp}
\end{figure}

For some small scale calculations, we use exact diagonalization; more precisely, we represent the Liouvillian as a sparse matrix and evolve with it to find the steady state.
For small system sizes, we find that the bosonic Hilbert space dimension can be cut off quite low at $d_{loc}=2$, which makes exact diagonalization possible. A comparison between steady states calculated for two different bosonic dimensions can be found in Fig.~\ref{fig:bosonic:dimension:comp} (a). 
For large scale calculations with CLIK-MPS, we use a local bosonic dimension of 11.
Interestingly, we find that for small values of $\gamma/\omega_c$, the analytical results derived for Redfield match the full Markovian embedding quite well, as shown in Fig.~\ref{fig:bosonic:dimension:comp} (b).
\section{Residual interactions from spin-reservoir coupling in trapped-ion simulators}
\label{app:one_to_one}

In Sec.~\ref{sec::trapped-ion}, we consider an experimental scheme in which each ion receives a distinct bichromatic Raman drive tuned near its designated motional reservoir mode. Specializing the multitone construction of Ref.~\cite{Marijanovic2026} to this addressing pattern, we derive the residual off-target spin-phonon couplings and show that, when considered alone, their secular contribution generates only a global phase under resolved-frequency averaging.

Formally, we begin with $N$ driven ions in a chain supporting $M\geq N$ collective motional modes and set $\hbar=1$. Each ion $i$ receives a single bichromatic Raman drive with beatnotes $\omega_0\pm\mu_i$, equal Rabi frequency $\Omega_i$ per beatnote, and optical phases $\varphi_i^{B,R}$. As shown in Fig.~\ref{fig::exp}(b), the drive detuning magnitudes $|\mu_i|$ relative to the qubit resonance $\omega_0$ are distinct and chosen so that drive $i$ predominantly addresses mode $i$ in the set $\mathcal Q=\{1,\ldots,N\}$. Denoting the physical frequency of mode $p$ by $\nu_p$, we choose $\delta_i=\mu_i-\nu_i<0$, which sets the positive effective phonon frequency $\omega_i=-\delta_i$~\cite{So2026C}. The spin-phonon coupling strengths are $g_{ip}=\eta_{ip}\Omega_i$, where $\eta_{ip}$ is the Lamb-Dicke parameter. The optical phases are set to give a common spin phase $\phi=(\varphi_i^B+\varphi_i^R)/2-\pi/2$, fixing the spin axis $\sigma_i^\phi=\sigma_i^x\cos\phi+\sigma_i^y\sin\phi$. The phase difference $\psi_i=(\varphi_i^B-\varphi_i^R)/2$ determines the motional phase of drive $i$.

After the optical rotating-wave approximation, first-order Lamb-Dicke expansion, and neglect of the carrier and target counter-rotating terms ($|\Omega_i/\mu_i|\ll1$), the resonant-frame Hamiltonian is $H^{(\mathrm{res})}(t)\simeq H_0+H_{\rm cr}(t)$ with
\begin{equation}
H_0=\sum_{i=1}^{N}\left[\omega_i a_i^\dagger a_i+\frac{g_{ii}}2 \sigma_i^\phi(a_ie^{i\psi_i}+a_i^\dagger e^{-i\psi_i})\right].
\label{eq:app_H0}
\end{equation}
This frame rotates addressed modes at $\mu_p$ and spectators at $\nu_p$. Writing $f_i(t)=\cos(\mu_i t+\psi_i)$ and $\lambda_p=\mu_p$ for $p\in\mathcal Q$ and $\lambda_p=\nu_p$ otherwise, the residual interaction is
\begin{equation}
H_{\rm cr}(t)=\sum_{i=1}^{N}\sum_{\substack{p=1\\p\ne i}}^{M}g_{ip}\sigma_i^\phi f_i(t)(a_p e^{-i\lambda_p t}+a_p^\dagger e^{i\lambda_p t}).
\label{eq:app_Hcr}
\end{equation}

Suppose that $|\mu_i-\lambda_p|\gg|g_{ip}|$ for $p\ne i$ and that all sum-frequency terms are also far off resonance. We perform secular averaging over times satisfying $|\mu_i-\lambda_p|t\gg1$ for $p\ne i$ and $|\mu_i-\mu_j|t\gg1$ for $i\ne j$. The Magnus expansion for evolution under $H_{\rm cr}(t)$ gives $U_{\rm cr}(t)=\exp(\overline\Omega_1+\overline\Omega_2)$, where
\begin{align}
\overline\Omega_1(t)&=\sum_{i=1}^{N}\sum_{\substack{p=1\\p\ne i}}^{M}\sigma_i^\phi\bigl[d_{ip}(t)a_p^\dagger-d_{ip}^*(t)a_p\bigr],
\label{eq:app_displacement}\\
\overline\Omega_2(t)&=i\sum_{i,j=1}^{N}\sum_{\substack{p=1\\p\ne i,j}}^{M}g_{ip}g_{jp}\sigma_i^\phi \sigma_j^\phi I_{ij,p}(t),
\label{eq:app_Omega2}
\end{align}
with
\begin{align}
d_{ip}(t)&=-\frac{g_{ip}}{2}\left[e^{i\psi_i}\frac{e^{i(\lambda_p+\mu_i)t}-1}{\lambda_p+\mu_i}\right.\nonumber\\
&\qquad\qquad\qquad\left.+\,e^{-i\psi_i}\frac{e^{i(\lambda_p-\mu_i)t}-1}{\lambda_p-\mu_i}\right],
\end{align}
\begin{align}
I_{ij,p}(t)&=\frac{1}{\lambda_p^2-\mu_j^2}\int_0^t ds\,f_i(s)\Bigl[\lambda_p f_j(s)\nonumber\\&\quad-\lambda_p\cos\psi_j\cos(\lambda_p s)+\mu_j\sin\psi_j\sin(\lambda_p s)\Bigr].
\end{align}
Under the off-resonance conditions considered here, the terms proportional to $\cos(\lambda_p s)$ and $\sin(\lambda_p s)$ lead to bounded contributions after integration against $f_i(s)$. For the remaining term, distinct positive drive frequencies imply $\int_0^t ds\,f_i(s)f_j(s) =\frac{t}{2}\delta_{ij}+\mathcal O(1)$, where $\mathcal O(1)$ denotes terms bounded in $t$ at fixed nonzero frequencies and detunings. Thus, only the self terms contribute secularly. Using $(\sigma_i^\phi)^2=\mathbf1$, we obtain
\begin{equation}
\overline\Omega_2^{(\mathrm{sec})}(t)=-\frac{it}{2}\mathbf1\sum_{i=1}^{N}\sum_{\substack{p=1\\p\ne i}}^{M}\frac{g_{ip}^{\,2}\lambda_p}{\mu_i^2-\lambda_p^2},
\label{appeq::residual}
\end{equation}
which generates only a global phase. After discarding this phase, the secular effective contribution of $H_{\rm cr}(t)$ vanishes.

\bibliographystyle{apsrev4-2}
\bibliography{bib}

\begin{thebibliography}{80}%
\makeatletter
\providecommand \@ifxundefined [1]{%
 \@ifx{#1\undefined}
}%
\providecommand \@ifnum [1]{%
 \ifnum #1\expandafter \@firstoftwo
 \else \expandafter \@secondoftwo
 \fi
}%
\providecommand \@ifx [1]{%
 \ifx #1\expandafter \@firstoftwo
 \else \expandafter \@secondoftwo
 \fi
}%
\providecommand \natexlab [1]{#1}%
\providecommand \enquote  [1]{``#1''}%
\providecommand \bibnamefont  [1]{#1}%
\providecommand \bibfnamefont [1]{#1}%
\providecommand \citenamefont [1]{#1}%
\providecommand \href@noop [0]{\@secondoftwo}%
\providecommand \href [0]{\begingroup \@sanitize@url \@href}%
\providecommand \@href[1]{\@@startlink{#1}\@@href}%
\providecommand \@@href[1]{\endgroup#1\@@endlink}%
\providecommand \@sanitize@url [0]{\catcode `\\12\catcode `\$12\catcode `\&12\catcode `\#12\catcode `\^12\catcode `\_12\catcode `\%12\relax}%
\providecommand \@@startlink[1]{}%
\providecommand \@@endlink[0]{}%
\providecommand \url  [0]{\begingroup\@sanitize@url \@url }%
\providecommand \@url [1]{\endgroup\@href {#1}{\urlprefix }}%
\providecommand \urlprefix  [0]{URL }%
\providecommand \Eprint [0]{\href }%
\providecommand \doibase [0]{https://doi.org/}%
\providecommand \selectlanguage [0]{\@gobble}%
\providecommand \bibinfo  [0]{\@secondoftwo}%
\providecommand \bibfield  [0]{\@secondoftwo}%
\providecommand \translation [1]{[#1]}%
\providecommand \BibitemOpen [0]{}%
\providecommand \bibitemStop [0]{}%
\providecommand \bibitemNoStop [0]{.\EOS\space}%
\providecommand \EOS [0]{\spacefactor3000\relax}%
\providecommand \BibitemShut  [1]{\csname bibitem#1\endcsname}%
\let\auto@bib@innerbib\@empty
%</preamble>
\bibitem [{\citenamefont {Diehl}\ \emph {et~al.}(2008)\citenamefont {Diehl}, \citenamefont {Micheli}, \citenamefont {Kantian}, \citenamefont {Kraus}, \citenamefont {B{\"u}chler},\ and\ \citenamefont {Zoller}}]{Diehl2008Q}%
  \BibitemOpen
  \bibfield  {author} {\bibinfo {author} {\bibfnamefont {S.}~\bibnamefont {Diehl}}, \bibinfo {author} {\bibfnamefont {A.}~\bibnamefont {Micheli}}, \bibinfo {author} {\bibfnamefont {A.}~\bibnamefont {Kantian}}, \bibinfo {author} {\bibfnamefont {B.}~\bibnamefont {Kraus}}, \bibinfo {author} {\bibfnamefont {H.~P.}\ \bibnamefont {B{\"u}chler}},\ and\ \bibinfo {author} {\bibfnamefont {P.}~\bibnamefont {Zoller}},\ }\href {https://doi.org/10.1038/nphys1073} {\bibfield  {journal} {\bibinfo  {journal} {Nature Physics}\ }\textbf {\bibinfo {volume} {4}},\ \bibinfo {pages} {878} (\bibinfo {year} {2008})}\BibitemShut {NoStop}%
\bibitem [{\citenamefont {Verstraete}\ \emph {et~al.}(2009)\citenamefont {Verstraete}, \citenamefont {Wolf},\ and\ \citenamefont {Ignacio~Cirac}}]{Verstraete2009Q}%
  \BibitemOpen
  \bibfield  {author} {\bibinfo {author} {\bibfnamefont {F.}~\bibnamefont {Verstraete}}, \bibinfo {author} {\bibfnamefont {M.~M.}\ \bibnamefont {Wolf}},\ and\ \bibinfo {author} {\bibfnamefont {J.}~\bibnamefont {Ignacio~Cirac}},\ }\href {https://doi.org/10.1038/nphys1342} {\bibfield  {journal} {\bibinfo  {journal} {Nature Physics}\ }\textbf {\bibinfo {volume} {5}},\ \bibinfo {pages} {633} (\bibinfo {year} {2009})}\BibitemShut {NoStop}%
\bibitem [{\citenamefont {Tindall}\ \emph {et~al.}(2019)\citenamefont {Tindall}, \citenamefont {Bu\ifmmode~\check{c}\else \v{c}\fi{}a}, \citenamefont {Coulthard},\ and\ \citenamefont {Jaksch}}]{Tindall2019}%
  \BibitemOpen
  \bibfield  {author} {\bibinfo {author} {\bibfnamefont {J.}~\bibnamefont {Tindall}}, \bibinfo {author} {\bibfnamefont {B.}~\bibnamefont {Bu\ifmmode~\check{c}\else \v{c}\fi{}a}}, \bibinfo {author} {\bibfnamefont {J.~R.}\ \bibnamefont {Coulthard}},\ and\ \bibinfo {author} {\bibfnamefont {D.}~\bibnamefont {Jaksch}},\ }\href {https://doi.org/10.1103/PhysRevLett.123.030603} {\bibfield  {journal} {\bibinfo  {journal} {Phys. Rev. Lett.}\ }\textbf {\bibinfo {volume} {123}},\ \bibinfo {pages} {030603} (\bibinfo {year} {2019})}\BibitemShut {NoStop}%
\bibitem [{\citenamefont {Zhan}\ \emph {et~al.}(2026)\citenamefont {Zhan}, \citenamefont {Ding}, \citenamefont {Huhn}, \citenamefont {Gray}, \citenamefont {Preskill}, \citenamefont {Chan},\ and\ \citenamefont {Lin}}]{Zhan2026rapid}%
  \BibitemOpen
  \bibfield  {author} {\bibinfo {author} {\bibfnamefont {Y.}~\bibnamefont {Zhan}}, \bibinfo {author} {\bibfnamefont {Z.}~\bibnamefont {Ding}}, \bibinfo {author} {\bibfnamefont {J.}~\bibnamefont {Huhn}}, \bibinfo {author} {\bibfnamefont {J.}~\bibnamefont {Gray}}, \bibinfo {author} {\bibfnamefont {J.}~\bibnamefont {Preskill}}, \bibinfo {author} {\bibfnamefont {G.~K.-L.}\ \bibnamefont {Chan}},\ and\ \bibinfo {author} {\bibfnamefont {L.}~\bibnamefont {Lin}},\ }\href {https://doi.org/10.1103/wzb3-dbg9} {\bibfield  {journal} {\bibinfo  {journal} {Phys. Rev. X}\ }\textbf {\bibinfo {volume} {16}},\ \bibinfo {pages} {011004} (\bibinfo {year} {2026})}\BibitemShut {NoStop}%
\bibitem [{\citenamefont {Pocklington}\ and\ \citenamefont {Clerk}(2025)}]{Pocklington2025A}%
  \BibitemOpen
  \bibfield  {author} {\bibinfo {author} {\bibfnamefont {A.}~\bibnamefont {Pocklington}}\ and\ \bibinfo {author} {\bibfnamefont {A.~A.}\ \bibnamefont {Clerk}},\ }\href {https://doi.org/10.1103/PhysRevLett.134.050603} {\bibfield  {journal} {\bibinfo  {journal} {Phys. Rev. Lett.}\ }\textbf {\bibinfo {volume} {134}},\ \bibinfo {pages} {050603} (\bibinfo {year} {2025})}\BibitemShut {NoStop}%
\bibitem [{\citenamefont {Bao}\ and\ \citenamefont {Hou}(2025)}]{Bao2025A}%
  \BibitemOpen
  \bibfield  {author} {\bibinfo {author} {\bibfnamefont {R.}~\bibnamefont {Bao}}\ and\ \bibinfo {author} {\bibfnamefont {Z.}~\bibnamefont {Hou}},\ }\href {https://doi.org/10.1103/g94p-7421} {\bibfield  {journal} {\bibinfo  {journal} {Phys. Rev. Lett.}\ }\textbf {\bibinfo {volume} {135}},\ \bibinfo {pages} {150403} (\bibinfo {year} {2025})}\BibitemShut {NoStop}%
\bibitem [{\citenamefont {Westhoff}\ \emph {et~al.}(2025)\citenamefont {Westhoff}, \citenamefont {Paeckel},\ and\ \citenamefont {Moroder}}]{Westhoff2025bec}%
  \BibitemOpen
  \bibfield  {author} {\bibinfo {author} {\bibfnamefont {P.}~\bibnamefont {Westhoff}}, \bibinfo {author} {\bibfnamefont {S.}~\bibnamefont {Paeckel}},\ and\ \bibinfo {author} {\bibfnamefont {M.}~\bibnamefont {Moroder}},\ }\href {https://doi.org/10.1103/nm4b-9w5c} {\bibfield  {journal} {\bibinfo  {journal} {Phys. Rev. A}\ }\textbf {\bibinfo {volume} {112}},\ \bibinfo {pages} {L061304} (\bibinfo {year} {2025})}\BibitemShut {NoStop}%
\bibitem [{\citenamefont {Lee}\ \emph {et~al.}(2013)\citenamefont {Lee}, \citenamefont {Gopalakrishnan},\ and\ \citenamefont {Lukin}}]{Lee2013Unconventional}%
  \BibitemOpen
  \bibfield  {author} {\bibinfo {author} {\bibfnamefont {T.~E.}\ \bibnamefont {Lee}}, \bibinfo {author} {\bibfnamefont {S.}~\bibnamefont {Gopalakrishnan}},\ and\ \bibinfo {author} {\bibfnamefont {M.~D.}\ \bibnamefont {Lukin}},\ }\href {https://doi.org/10.1103/PhysRevLett.110.257204} {\bibfield  {journal} {\bibinfo  {journal} {Phys. Rev. Lett.}\ }\textbf {\bibinfo {volume} {110}},\ \bibinfo {pages} {257204} (\bibinfo {year} {2013})}\BibitemShut {NoStop}%
\bibitem [{\citenamefont {Jin}\ \emph {et~al.}(2016)\citenamefont {Jin}, \citenamefont {Biella}, \citenamefont {Viyuela} \emph {et~al.}}]{Jin2016Cluster}%
  \BibitemOpen
  \bibfield  {author} {\bibinfo {author} {\bibfnamefont {J.}~\bibnamefont {Jin}}, \bibinfo {author} {\bibfnamefont {A.}~\bibnamefont {Biella}}, \bibinfo {author} {\bibfnamefont {O.}~\bibnamefont {Viyuela}}, \emph {et~al.},\ }\href {https://doi.org/10.1103/PhysRevX.6.031011} {\bibfield  {journal} {\bibinfo  {journal} {Phys. Rev. X}\ }\textbf {\bibinfo {volume} {6}},\ \bibinfo {pages} {031011} (\bibinfo {year} {2016})}\BibitemShut {NoStop}%
\bibitem [{\citenamefont {Syassen}\ \emph {et~al.}(2008)\citenamefont {Syassen}, \citenamefont {Bauer}, \citenamefont {Lettner} \emph {et~al.}}]{Syassen2008S}%
  \BibitemOpen
  \bibfield  {author} {\bibinfo {author} {\bibfnamefont {N.}~\bibnamefont {Syassen}}, \bibinfo {author} {\bibfnamefont {D.~M.}\ \bibnamefont {Bauer}}, \bibinfo {author} {\bibfnamefont {M.}~\bibnamefont {Lettner}}, \emph {et~al.},\ }\href {https://doi.org/10.1126/science.1155309} {\bibfield  {journal} {\bibinfo  {journal} {Science}\ }\textbf {\bibinfo {volume} {320}},\ \bibinfo {pages} {1329} (\bibinfo {year} {2008})}\BibitemShut {NoStop}%
\bibitem [{\citenamefont {Mi}\ \emph {et~al.}(2024)\citenamefont {Mi}, \citenamefont {Michailidis}, \citenamefont {Shabani} \emph {et~al.}}]{Mi2024E}%
  \BibitemOpen
  \bibfield  {author} {\bibinfo {author} {\bibfnamefont {X.}~\bibnamefont {Mi}}, \bibinfo {author} {\bibfnamefont {A.~A.}\ \bibnamefont {Michailidis}}, \bibinfo {author} {\bibfnamefont {S.}~\bibnamefont {Shabani}}, \emph {et~al.},\ }\href {https://doi.org/10.1126/science.adh9932} {\bibfield  {journal} {\bibinfo  {journal} {Science}\ }\textbf {\bibinfo {volume} {383}},\ \bibinfo {pages} {1332} (\bibinfo {year} {2024})}\BibitemShut {NoStop}%
\bibitem [{\citenamefont {So}\ \emph {et~al.}(2024)\citenamefont {So}, \citenamefont {Suganthi}, \citenamefont {Menon} \emph {et~al.}}]{So2024T}%
  \BibitemOpen
  \bibfield  {author} {\bibinfo {author} {\bibfnamefont {V.}~\bibnamefont {So}}, \bibinfo {author} {\bibfnamefont {M.~D.}\ \bibnamefont {Suganthi}}, \bibinfo {author} {\bibfnamefont {A.}~\bibnamefont {Menon}}, \emph {et~al.},\ }\href {https://doi.org/10.1126/sciadv.ads8011} {\bibfield  {journal} {\bibinfo  {journal} {Science Advances}\ }\textbf {\bibinfo {volume} {10}},\ \bibinfo {pages} {eads8011} (\bibinfo {year} {2024})}\BibitemShut {NoStop}%
\bibitem [{\citenamefont {So}\ \emph {et~al.}(2026{\natexlab{a}})\citenamefont {So}, \citenamefont {Duraisamy~Suganthi}, \citenamefont {Zhu} \emph {et~al.}}]{So2025Q}%
  \BibitemOpen
  \bibfield  {author} {\bibinfo {author} {\bibfnamefont {V.}~\bibnamefont {So}}, \bibinfo {author} {\bibfnamefont {M.}~\bibnamefont {Duraisamy~Suganthi}}, \bibinfo {author} {\bibfnamefont {M.}~\bibnamefont {Zhu}}, \emph {et~al.},\ }\href {https://doi.org/10.1038/s41467-025-67116-6} {\bibfield  {journal} {\bibinfo  {journal} {Nature Communications}\ }\textbf {\bibinfo {volume} {17}},\ \bibinfo {pages} {438} (\bibinfo {year} {2026}{\natexlab{a}})}\BibitemShut {NoStop}%
\bibitem [{\citenamefont {Sun}\ \emph {et~al.}(2025)\citenamefont {Sun}, \citenamefont {Kang}, \citenamefont {Nuomin} \emph {et~al.}}]{Sun2025Q}%
  \BibitemOpen
  \bibfield  {author} {\bibinfo {author} {\bibfnamefont {K.}~\bibnamefont {Sun}}, \bibinfo {author} {\bibfnamefont {M.}~\bibnamefont {Kang}}, \bibinfo {author} {\bibfnamefont {H.}~\bibnamefont {Nuomin}}, \emph {et~al.},\ }\href {https://doi.org/10.1038/s41467-025-59296-y} {\bibfield  {journal} {\bibinfo  {journal} {Nature Communications}\ }\textbf {\bibinfo {volume} {16}},\ \bibinfo {pages} {4042} (\bibinfo {year} {2025})}\BibitemShut {NoStop}%
\bibitem [{\citenamefont {Greenberger}\ \emph {et~al.}(1989)\citenamefont {Greenberger}, \citenamefont {Horne},\ and\ \citenamefont {Zeilinger}}]{Greenberger1989}%
  \BibitemOpen
  \bibfield  {author} {\bibinfo {author} {\bibfnamefont {D.~M.}\ \bibnamefont {Greenberger}}, \bibinfo {author} {\bibfnamefont {M.~A.}\ \bibnamefont {Horne}},\ and\ \bibinfo {author} {\bibfnamefont {A.}~\bibnamefont {Zeilinger}},\ }\bibinfo {title} {Going beyond bell's theorem},\ in\ \href {https://doi.org/10.1007/978-94-017-0849-4_10} {\emph {\bibinfo {booktitle} {Bell's Theorem, Quantum Theory and Conceptions of the Universe}}},\ \bibinfo {editor} {edited by\ \bibinfo {editor} {\bibfnamefont {M.}~\bibnamefont {Kafatos}}}\ (\bibinfo  {publisher} {Springer Netherlands},\ \bibinfo {address} {Dordrecht},\ \bibinfo {year} {1989})\ pp.\ \bibinfo {pages} {69--72}\BibitemShut {NoStop}%
\bibitem [{\citenamefont {Monz}\ \emph {et~al.}(2011)\citenamefont {Monz}, \citenamefont {Schindler}, \citenamefont {Barreiro}, \citenamefont {Chwalla}, \citenamefont {Nigg}, \citenamefont {Coish}, \citenamefont {Harlander}, \citenamefont {H\"ansel}, \citenamefont {Hennrich},\ and\ \citenamefont {Blatt}}]{PhysRevLett.106.130506}%
  \BibitemOpen
  \bibfield  {author} {\bibinfo {author} {\bibfnamefont {T.}~\bibnamefont {Monz}}, \bibinfo {author} {\bibfnamefont {P.}~\bibnamefont {Schindler}}, \bibinfo {author} {\bibfnamefont {J.~T.}\ \bibnamefont {Barreiro}}, \bibinfo {author} {\bibfnamefont {M.}~\bibnamefont {Chwalla}}, \bibinfo {author} {\bibfnamefont {D.}~\bibnamefont {Nigg}}, \bibinfo {author} {\bibfnamefont {W.~A.}\ \bibnamefont {Coish}}, \bibinfo {author} {\bibfnamefont {M.}~\bibnamefont {Harlander}}, \bibinfo {author} {\bibfnamefont {W.}~\bibnamefont {H\"ansel}}, \bibinfo {author} {\bibfnamefont {M.}~\bibnamefont {Hennrich}},\ and\ \bibinfo {author} {\bibfnamefont {R.}~\bibnamefont {Blatt}},\ }\href {https://doi.org/10.1103/PhysRevLett.106.130506} {\bibfield  {journal} {\bibinfo  {journal} {Phys. Rev. Lett.}\ }\textbf {\bibinfo {volume} {106}},\ \bibinfo {pages} {130506} (\bibinfo {year} {2011})}\BibitemShut {NoStop}%
\bibitem [{\citenamefont {Pezz\`e}\ \emph {et~al.}(2018)\citenamefont {Pezz\`e}, \citenamefont {Smerzi}, \citenamefont {Oberthaler}, \citenamefont {Schmied},\ and\ \citenamefont {Treutlein}}]{RevModPhys.90.035005}%
  \BibitemOpen
  \bibfield  {author} {\bibinfo {author} {\bibfnamefont {L.}~\bibnamefont {Pezz\`e}}, \bibinfo {author} {\bibfnamefont {A.}~\bibnamefont {Smerzi}}, \bibinfo {author} {\bibfnamefont {M.~K.}\ \bibnamefont {Oberthaler}}, \bibinfo {author} {\bibfnamefont {R.}~\bibnamefont {Schmied}},\ and\ \bibinfo {author} {\bibfnamefont {P.}~\bibnamefont {Treutlein}},\ }\href {https://doi.org/10.1103/RevModPhys.90.035005} {\bibfield  {journal} {\bibinfo  {journal} {Rev. Mod. Phys.}\ }\textbf {\bibinfo {volume} {90}},\ \bibinfo {pages} {035005} (\bibinfo {year} {2018})}\BibitemShut {NoStop}%
\bibitem [{\citenamefont {Pont}\ \emph {et~al.}(2024)\citenamefont {Pont}, \citenamefont {Corrielli}, \citenamefont {Fyrillas}, \citenamefont {Agresti}, \citenamefont {Carvacho}, \citenamefont {Maring}, \citenamefont {Emeriau}, \citenamefont {Ceccarelli}, \citenamefont {Albiero}, \citenamefont {Dias~Ferreira}, \citenamefont {Somaschi}, \citenamefont {Senellart}, \citenamefont {Sagnes}, \citenamefont {Morassi}, \citenamefont {Lema{\^\i}tre}, \citenamefont {Senellart}, \citenamefont {Sciarrino}, \citenamefont {Liscidini}, \citenamefont {Belabas},\ and\ \citenamefont {Osellame}}]{Pont2024-se}%
  \BibitemOpen
  \bibfield  {author} {\bibinfo {author} {\bibfnamefont {M.}~\bibnamefont {Pont}}, \bibinfo {author} {\bibfnamefont {G.}~\bibnamefont {Corrielli}}, \bibinfo {author} {\bibfnamefont {A.}~\bibnamefont {Fyrillas}}, \bibinfo {author} {\bibfnamefont {I.}~\bibnamefont {Agresti}}, \bibinfo {author} {\bibfnamefont {G.}~\bibnamefont {Carvacho}}, \bibinfo {author} {\bibfnamefont {N.}~\bibnamefont {Maring}}, \bibinfo {author} {\bibfnamefont {P.-E.}\ \bibnamefont {Emeriau}}, \bibinfo {author} {\bibfnamefont {F.}~\bibnamefont {Ceccarelli}}, \bibinfo {author} {\bibfnamefont {R.}~\bibnamefont {Albiero}}, \bibinfo {author} {\bibfnamefont {P.~H.}\ \bibnamefont {Dias~Ferreira}}, \bibinfo {author} {\bibfnamefont {N.}~\bibnamefont {Somaschi}}, \bibinfo {author} {\bibfnamefont {J.}~\bibnamefont {Senellart}}, \bibinfo {author} {\bibfnamefont {I.}~\bibnamefont {Sagnes}}, \bibinfo {author} {\bibfnamefont {M.}~\bibnamefont {Morassi}}, \bibinfo {author} {\bibfnamefont {A.}~\bibnamefont {Lema{\^\i}tre}}, \bibinfo {author}
  {\bibfnamefont {P.}~\bibnamefont {Senellart}}, \bibinfo {author} {\bibfnamefont {F.}~\bibnamefont {Sciarrino}}, \bibinfo {author} {\bibfnamefont {M.}~\bibnamefont {Liscidini}}, \bibinfo {author} {\bibfnamefont {N.}~\bibnamefont {Belabas}},\ and\ \bibinfo {author} {\bibfnamefont {R.}~\bibnamefont {Osellame}},\ }\href {https://doi.org/10.1038/s41534-024-00830-z} {\bibfield  {journal} {\bibinfo  {journal} {npj Quantum Information}\ }\textbf {\bibinfo {volume} {10}},\ \bibinfo {pages} {50} (\bibinfo {year} {2024})}\BibitemShut {NoStop}%
\bibitem [{\citenamefont {Chu}\ \emph {et~al.}(2026)\citenamefont {Chu}, \citenamefont {Mamaev}, \citenamefont {Koppenh{\"o}fer}, \citenamefont {Yuan},\ and\ \citenamefont {Clerk}}]{Chu2026R}%
  \BibitemOpen
  \bibfield  {author} {\bibinfo {author} {\bibfnamefont {A.}~\bibnamefont {Chu}}, \bibinfo {author} {\bibfnamefont {M.}~\bibnamefont {Mamaev}}, \bibinfo {author} {\bibfnamefont {M.}~\bibnamefont {Koppenh{\"o}fer}}, \bibinfo {author} {\bibfnamefont {M.}~\bibnamefont {Yuan}},\ and\ \bibinfo {author} {\bibfnamefont {A.~A.}\ \bibnamefont {Clerk}},\ }\href {https://doi.org/10.1103/qdh9-2pc7} {\bibfield  {journal} {\bibinfo  {journal} {Phys. Rev. X}\ }\textbf {\bibinfo {volume} {16}},\ \bibinfo {pages} {021047} (\bibinfo {year} {2026})}\BibitemShut {NoStop}%
\bibitem [{\citenamefont {Halati}\ \emph {et~al.}(2025)\citenamefont {Halati}, \citenamefont {Sheikhan}, \citenamefont {Morigi},\ and\ \citenamefont {Kollath}}]{Halati2025C}%
  \BibitemOpen
  \bibfield  {author} {\bibinfo {author} {\bibfnamefont {C.-M.}\ \bibnamefont {Halati}}, \bibinfo {author} {\bibfnamefont {A.}~\bibnamefont {Sheikhan}}, \bibinfo {author} {\bibfnamefont {G.}~\bibnamefont {Morigi}},\ and\ \bibinfo {author} {\bibfnamefont {C.}~\bibnamefont {Kollath}},\ }\href {https://doi.org/10.1103/PhysRevLett.134.073604} {\bibfield  {journal} {\bibinfo  {journal} {Phys. Rev. Lett.}\ }\textbf {\bibinfo {volume} {134}},\ \bibinfo {pages} {073604} (\bibinfo {year} {2025})}\BibitemShut {NoStop}%
\bibitem [{\citenamefont {Kirton}\ \emph {et~al.}(2019)\citenamefont {Kirton}, \citenamefont {Roses}, \citenamefont {Keeling},\ and\ \citenamefont {Dalla~Torre}}]{Kirton2019I}%
  \BibitemOpen
  \bibfield  {author} {\bibinfo {author} {\bibfnamefont {P.}~\bibnamefont {Kirton}}, \bibinfo {author} {\bibfnamefont {M.~M.}\ \bibnamefont {Roses}}, \bibinfo {author} {\bibfnamefont {J.}~\bibnamefont {Keeling}},\ and\ \bibinfo {author} {\bibfnamefont {E.~G.}\ \bibnamefont {Dalla~Torre}},\ }\href {https://doi.org/10.1002/qute.201800043} {\bibfield  {journal} {\bibinfo  {journal} {Advanced Quantum Technologies}\ }\textbf {\bibinfo {volume} {2}},\ \bibinfo {pages} {1800043} (\bibinfo {year} {2019})}\BibitemShut {NoStop}%
\bibitem [{\citenamefont {Kim}\ \emph {et~al.}(2025)\citenamefont {Kim}, \citenamefont {Lee}, \citenamefont {Jeon} \emph {et~al.}}]{Kim2025C}%
  \BibitemOpen
  \bibfield  {author} {\bibinfo {author} {\bibfnamefont {K.-Y.}\ \bibnamefont {Kim}}, \bibinfo {author} {\bibfnamefont {J.~H.}\ \bibnamefont {Lee}}, \bibinfo {author} {\bibfnamefont {W.~B.}\ \bibnamefont {Jeon}}, \emph {et~al.},\ }\href {https://doi.org/10.1038/s41467-025-61629-w} {\bibfield  {journal} {\bibinfo  {journal} {Nature Communications}\ }\textbf {\bibinfo {volume} {16}},\ \bibinfo {pages} {6346} (\bibinfo {year} {2025})}\BibitemShut {NoStop}%
\bibitem [{\citenamefont {Fallas~Padilla}\ \emph {et~al.}(2026)\citenamefont {Fallas~Padilla}, \citenamefont {Kaubruegger}, \citenamefont {Gillman}, \citenamefont {Becker},\ and\ \citenamefont {Rey}}]{Fallas2026G}%
  \BibitemOpen
  \bibfield  {author} {\bibinfo {author} {\bibfnamefont {D.}~\bibnamefont {Fallas~Padilla}}, \bibinfo {author} {\bibfnamefont {R.}~\bibnamefont {Kaubruegger}}, \bibinfo {author} {\bibfnamefont {A.}~\bibnamefont {Gillman}}, \bibinfo {author} {\bibfnamefont {S.}~\bibnamefont {Becker}},\ and\ \bibinfo {author} {\bibfnamefont {A.~M.}\ \bibnamefont {Rey}},\ }\href {https://doi.org/10.1103/s8kc-crf3} {\bibfield  {journal} {\bibinfo  {journal} {Phys. Rev. Res.}\ }\textbf {\bibinfo {volume} {8}},\ \bibinfo {pages} {013244} (\bibinfo {year} {2026})}\BibitemShut {NoStop}%
\bibitem [{\citenamefont {Braun}(2002)}]{Braun2002C}%
  \BibitemOpen
  \bibfield  {author} {\bibinfo {author} {\bibfnamefont {D.}~\bibnamefont {Braun}},\ }\href {https://doi.org/10.1103/PhysRevLett.89.277901} {\bibfield  {journal} {\bibinfo  {journal} {Phys. Rev. Lett.}\ }\textbf {\bibinfo {volume} {89}},\ \bibinfo {pages} {277901} (\bibinfo {year} {2002})}\BibitemShut {NoStop}%
\bibitem [{\citenamefont {Benatti}\ \emph {et~al.}(2003)\citenamefont {Benatti}, \citenamefont {Floreanini},\ and\ \citenamefont {Piani}}]{Benatti2003E}%
  \BibitemOpen
  \bibfield  {author} {\bibinfo {author} {\bibfnamefont {F.}~\bibnamefont {Benatti}}, \bibinfo {author} {\bibfnamefont {R.}~\bibnamefont {Floreanini}},\ and\ \bibinfo {author} {\bibfnamefont {M.}~\bibnamefont {Piani}},\ }\href {https://doi.org/10.1103/PhysRevLett.91.070402} {\bibfield  {journal} {\bibinfo  {journal} {Phys. Rev. Lett.}\ }\textbf {\bibinfo {volume} {91}},\ \bibinfo {pages} {070402} (\bibinfo {year} {2003})}\BibitemShut {NoStop}%
\bibitem [{\citenamefont {Cho}\ \emph {et~al.}(2011)\citenamefont {Cho}, \citenamefont {Bose},\ and\ \citenamefont {Kim}}]{PhysRevLett.106.020504}%
  \BibitemOpen
  \bibfield  {author} {\bibinfo {author} {\bibfnamefont {J.}~\bibnamefont {Cho}}, \bibinfo {author} {\bibfnamefont {S.}~\bibnamefont {Bose}},\ and\ \bibinfo {author} {\bibfnamefont {M.~S.}\ \bibnamefont {Kim}},\ }\href {https://doi.org/10.1103/PhysRevLett.106.020504} {\bibfield  {journal} {\bibinfo  {journal} {Phys. Rev. Lett.}\ }\textbf {\bibinfo {volume} {106}},\ \bibinfo {pages} {020504} (\bibinfo {year} {2011})}\BibitemShut {NoStop}%
\bibitem [{\citenamefont {Barreiro}\ \emph {et~al.}(2011)\citenamefont {Barreiro}, \citenamefont {M{\"u}ller}, \citenamefont {Schindler}, \citenamefont {Nigg}, \citenamefont {Monz}, \citenamefont {Chwalla}, \citenamefont {Hennrich}, \citenamefont {Roos}, \citenamefont {Zoller},\ and\ \citenamefont {Blatt}}]{Barreiro2011-jp}%
  \BibitemOpen
  \bibfield  {author} {\bibinfo {author} {\bibfnamefont {J.~T.}\ \bibnamefont {Barreiro}}, \bibinfo {author} {\bibfnamefont {M.}~\bibnamefont {M{\"u}ller}}, \bibinfo {author} {\bibfnamefont {P.}~\bibnamefont {Schindler}}, \bibinfo {author} {\bibfnamefont {D.}~\bibnamefont {Nigg}}, \bibinfo {author} {\bibfnamefont {T.}~\bibnamefont {Monz}}, \bibinfo {author} {\bibfnamefont {M.}~\bibnamefont {Chwalla}}, \bibinfo {author} {\bibfnamefont {M.}~\bibnamefont {Hennrich}}, \bibinfo {author} {\bibfnamefont {C.~F.}\ \bibnamefont {Roos}}, \bibinfo {author} {\bibfnamefont {P.}~\bibnamefont {Zoller}},\ and\ \bibinfo {author} {\bibfnamefont {R.}~\bibnamefont {Blatt}},\ }\href@noop {} {\bibfield  {journal} {\bibinfo  {journal} {Nature}\ }\textbf {\bibinfo {volume} {470}},\ \bibinfo {pages} {486} (\bibinfo {year} {2011})}\BibitemShut {NoStop}%
\bibitem [{\citenamefont {Ticozzi}\ and\ \citenamefont {Viola}(2014)}]{Ticozzi2014}%
  \BibitemOpen
  \bibfield  {author} {\bibinfo {author} {\bibfnamefont {F.}~\bibnamefont {Ticozzi}}\ and\ \bibinfo {author} {\bibfnamefont {L.}~\bibnamefont {Viola}},\ }\href {https://doi.org/10.26421/QIC14.3-4-5} {\bibfield  {journal} {\bibinfo  {journal} {Quantum Information and Computation}\ }\textbf {\bibinfo {volume} {14}},\ \bibinfo {pages} {265} (\bibinfo {year} {2014})}\BibitemShut {NoStop}%
\bibitem [{\citenamefont {Camalet}(2013)}]{Camalet2013S}%
  \BibitemOpen
  \bibfield  {author} {\bibinfo {author} {\bibfnamefont {S.}~\bibnamefont {Camalet}},\ }\href {https://doi.org/10.1140/epjb/e2013-31010-0} {\bibfield  {journal} {\bibinfo  {journal} {Eur. Phys. J. B}\ }\textbf {\bibinfo {volume} {86}},\ \bibinfo {pages} {176} (\bibinfo {year} {2013})}\BibitemShut {NoStop}%
\bibitem [{\citenamefont {Shao}\ \emph {et~al.}(2017)\citenamefont {Shao}, \citenamefont {Wu}, \citenamefont {Yi},\ and\ \citenamefont {Long}}]{PhysRevA.96.062315}%
  \BibitemOpen
  \bibfield  {author} {\bibinfo {author} {\bibfnamefont {X.~Q.}\ \bibnamefont {Shao}}, \bibinfo {author} {\bibfnamefont {J.~H.}\ \bibnamefont {Wu}}, \bibinfo {author} {\bibfnamefont {X.~X.}\ \bibnamefont {Yi}},\ and\ \bibinfo {author} {\bibfnamefont {G.-L.}\ \bibnamefont {Long}},\ }\href {https://doi.org/10.1103/PhysRevA.96.062315} {\bibfield  {journal} {\bibinfo  {journal} {Phys. Rev. A}\ }\textbf {\bibinfo {volume} {96}},\ \bibinfo {pages} {062315} (\bibinfo {year} {2017})}\BibitemShut {NoStop}%
\bibitem [{\citenamefont {Moses}\ \emph {et~al.}(2023)\citenamefont {Moses}, \citenamefont {Baldwin}, \citenamefont {Allman}, \citenamefont {Ancona}, \citenamefont {Ascarrunz}, \citenamefont {Barnes}, \citenamefont {Bartolotta}, \citenamefont {Bjork}, \citenamefont {Blanchard}, \citenamefont {Bohn}, \citenamefont {Bohnet}, \citenamefont {Brown}, \citenamefont {Burdick}, \citenamefont {Burton}, \citenamefont {Campbell}, \citenamefont {Campora}, \citenamefont {Carron}, \citenamefont {Chambers}, \citenamefont {Chan}, \citenamefont {Chen}, \citenamefont {Chernoguzov}, \citenamefont {Chertkov}, \citenamefont {Colina}, \citenamefont {Curtis}, \citenamefont {Daniel}, \citenamefont {DeCross}, \citenamefont {Deen}, \citenamefont {Delaney}, \citenamefont {Dreiling}, \citenamefont {Ertsgaard}, \citenamefont {Esposito}, \citenamefont {Estey}, \citenamefont {Fabrikant}, \citenamefont {Figgatt}, \citenamefont {Foltz}, \citenamefont {Foss-Feig}, \citenamefont {Francois}, \citenamefont {Gaebler}, \citenamefont {Gatterman},
  \citenamefont {Gilbreth}, \citenamefont {Giles}, \citenamefont {Glynn}, \citenamefont {Hall}, \citenamefont {Hankin}, \citenamefont {Hansen}, \citenamefont {Hayes}, \citenamefont {Higashi}, \citenamefont {Hoffman}, \citenamefont {Horning}, \citenamefont {Hout}, \citenamefont {Jacobs}, \citenamefont {Johansen}, \citenamefont {Jones}, \citenamefont {Karcz}, \citenamefont {Klein}, \citenamefont {Lauria}, \citenamefont {Lee}, \citenamefont {Liefer}, \citenamefont {Lu}, \citenamefont {Lucchetti}, \citenamefont {Lytle}, \citenamefont {Malm}, \citenamefont {Matheny}, \citenamefont {Mathewson}, \citenamefont {Mayer}, \citenamefont {Miller}, \citenamefont {Mills}, \citenamefont {Neyenhuis}, \citenamefont {Nugent}, \citenamefont {Olson}, \citenamefont {Parks}, \citenamefont {Price}, \citenamefont {Price}, \citenamefont {Pugh}, \citenamefont {Ransford}, \citenamefont {Reed}, \citenamefont {Roman}, \citenamefont {Rowe}, \citenamefont {Ryan-Anderson}, \citenamefont {Sanders}, \citenamefont {Sedlacek}, \citenamefont
  {Shevchuk}, \citenamefont {Siegfried}, \citenamefont {Skripka}, \citenamefont {Spaun}, \citenamefont {Sprenkle}, \citenamefont {Stutz}, \citenamefont {Swallows}, \citenamefont {Tobey}, \citenamefont {Tran}, \citenamefont {Tran}, \citenamefont {Vogt}, \citenamefont {Volin}, \citenamefont {Walker}, \citenamefont {Zolot},\ and\ \citenamefont {Pino}}]{Moses2023A}%
  \BibitemOpen
  \bibfield  {author} {\bibinfo {author} {\bibfnamefont {S.~A.}\ \bibnamefont {Moses}}, \bibinfo {author} {\bibfnamefont {C.~H.}\ \bibnamefont {Baldwin}}, \bibinfo {author} {\bibfnamefont {M.~S.}\ \bibnamefont {Allman}}, \bibinfo {author} {\bibfnamefont {R.}~\bibnamefont {Ancona}}, \bibinfo {author} {\bibfnamefont {L.}~\bibnamefont {Ascarrunz}}, \bibinfo {author} {\bibfnamefont {C.}~\bibnamefont {Barnes}}, \bibinfo {author} {\bibfnamefont {J.}~\bibnamefont {Bartolotta}}, \bibinfo {author} {\bibfnamefont {B.}~\bibnamefont {Bjork}}, \bibinfo {author} {\bibfnamefont {P.}~\bibnamefont {Blanchard}}, \bibinfo {author} {\bibfnamefont {M.}~\bibnamefont {Bohn}}, \bibinfo {author} {\bibfnamefont {J.~G.}\ \bibnamefont {Bohnet}}, \bibinfo {author} {\bibfnamefont {N.~C.}\ \bibnamefont {Brown}}, \bibinfo {author} {\bibfnamefont {N.~Q.}\ \bibnamefont {Burdick}}, \bibinfo {author} {\bibfnamefont {W.~C.}\ \bibnamefont {Burton}}, \bibinfo {author} {\bibfnamefont {S.~L.}\ \bibnamefont {Campbell}}, \bibinfo {author}
  {\bibfnamefont {J.~P.}\ \bibnamefont {Campora}}, \bibinfo {author} {\bibfnamefont {C.}~\bibnamefont {Carron}}, \bibinfo {author} {\bibfnamefont {J.}~\bibnamefont {Chambers}}, \bibinfo {author} {\bibfnamefont {J.~W.}\ \bibnamefont {Chan}}, \bibinfo {author} {\bibfnamefont {Y.~H.}\ \bibnamefont {Chen}}, \bibinfo {author} {\bibfnamefont {A.}~\bibnamefont {Chernoguzov}}, \bibinfo {author} {\bibfnamefont {E.}~\bibnamefont {Chertkov}}, \bibinfo {author} {\bibfnamefont {J.}~\bibnamefont {Colina}}, \bibinfo {author} {\bibfnamefont {J.~P.}\ \bibnamefont {Curtis}}, \bibinfo {author} {\bibfnamefont {R.}~\bibnamefont {Daniel}}, \bibinfo {author} {\bibfnamefont {M.}~\bibnamefont {DeCross}}, \bibinfo {author} {\bibfnamefont {D.}~\bibnamefont {Deen}}, \bibinfo {author} {\bibfnamefont {C.}~\bibnamefont {Delaney}}, \bibinfo {author} {\bibfnamefont {J.~M.}\ \bibnamefont {Dreiling}}, \bibinfo {author} {\bibfnamefont {C.~T.}\ \bibnamefont {Ertsgaard}}, \bibinfo {author} {\bibfnamefont {J.}~\bibnamefont {Esposito}}, \bibinfo
  {author} {\bibfnamefont {B.}~\bibnamefont {Estey}}, \bibinfo {author} {\bibfnamefont {M.}~\bibnamefont {Fabrikant}}, \bibinfo {author} {\bibfnamefont {C.}~\bibnamefont {Figgatt}}, \bibinfo {author} {\bibfnamefont {C.}~\bibnamefont {Foltz}}, \bibinfo {author} {\bibfnamefont {M.}~\bibnamefont {Foss-Feig}}, \bibinfo {author} {\bibfnamefont {D.}~\bibnamefont {Francois}}, \bibinfo {author} {\bibfnamefont {J.~P.}\ \bibnamefont {Gaebler}}, \bibinfo {author} {\bibfnamefont {T.~M.}\ \bibnamefont {Gatterman}}, \bibinfo {author} {\bibfnamefont {C.~N.}\ \bibnamefont {Gilbreth}}, \bibinfo {author} {\bibfnamefont {J.}~\bibnamefont {Giles}}, \bibinfo {author} {\bibfnamefont {E.}~\bibnamefont {Glynn}}, \bibinfo {author} {\bibfnamefont {A.}~\bibnamefont {Hall}}, \bibinfo {author} {\bibfnamefont {A.~M.}\ \bibnamefont {Hankin}}, \bibinfo {author} {\bibfnamefont {A.}~\bibnamefont {Hansen}}, \bibinfo {author} {\bibfnamefont {D.}~\bibnamefont {Hayes}}, \bibinfo {author} {\bibfnamefont {B.}~\bibnamefont {Higashi}}, \bibinfo
  {author} {\bibfnamefont {I.~M.}\ \bibnamefont {Hoffman}}, \bibinfo {author} {\bibfnamefont {B.}~\bibnamefont {Horning}}, \bibinfo {author} {\bibfnamefont {J.~J.}\ \bibnamefont {Hout}}, \bibinfo {author} {\bibfnamefont {R.}~\bibnamefont {Jacobs}}, \bibinfo {author} {\bibfnamefont {J.}~\bibnamefont {Johansen}}, \bibinfo {author} {\bibfnamefont {L.}~\bibnamefont {Jones}}, \bibinfo {author} {\bibfnamefont {J.}~\bibnamefont {Karcz}}, \bibinfo {author} {\bibfnamefont {T.}~\bibnamefont {Klein}}, \bibinfo {author} {\bibfnamefont {P.}~\bibnamefont {Lauria}}, \bibinfo {author} {\bibfnamefont {P.}~\bibnamefont {Lee}}, \bibinfo {author} {\bibfnamefont {D.}~\bibnamefont {Liefer}}, \bibinfo {author} {\bibfnamefont {S.~T.}\ \bibnamefont {Lu}}, \bibinfo {author} {\bibfnamefont {D.}~\bibnamefont {Lucchetti}}, \bibinfo {author} {\bibfnamefont {C.}~\bibnamefont {Lytle}}, \bibinfo {author} {\bibfnamefont {A.}~\bibnamefont {Malm}}, \bibinfo {author} {\bibfnamefont {M.}~\bibnamefont {Matheny}}, \bibinfo {author} {\bibfnamefont
  {B.}~\bibnamefont {Mathewson}}, \bibinfo {author} {\bibfnamefont {K.}~\bibnamefont {Mayer}}, \bibinfo {author} {\bibfnamefont {D.~B.}\ \bibnamefont {Miller}}, \bibinfo {author} {\bibfnamefont {M.}~\bibnamefont {Mills}}, \bibinfo {author} {\bibfnamefont {B.}~\bibnamefont {Neyenhuis}}, \bibinfo {author} {\bibfnamefont {L.}~\bibnamefont {Nugent}}, \bibinfo {author} {\bibfnamefont {S.}~\bibnamefont {Olson}}, \bibinfo {author} {\bibfnamefont {J.}~\bibnamefont {Parks}}, \bibinfo {author} {\bibfnamefont {G.~N.}\ \bibnamefont {Price}}, \bibinfo {author} {\bibfnamefont {Z.}~\bibnamefont {Price}}, \bibinfo {author} {\bibfnamefont {M.}~\bibnamefont {Pugh}}, \bibinfo {author} {\bibfnamefont {A.}~\bibnamefont {Ransford}}, \bibinfo {author} {\bibfnamefont {A.~P.}\ \bibnamefont {Reed}}, \bibinfo {author} {\bibfnamefont {C.}~\bibnamefont {Roman}}, \bibinfo {author} {\bibfnamefont {M.}~\bibnamefont {Rowe}}, \bibinfo {author} {\bibfnamefont {C.}~\bibnamefont {Ryan-Anderson}}, \bibinfo {author} {\bibfnamefont
  {S.}~\bibnamefont {Sanders}}, \bibinfo {author} {\bibfnamefont {J.}~\bibnamefont {Sedlacek}}, \bibinfo {author} {\bibfnamefont {P.}~\bibnamefont {Shevchuk}}, \bibinfo {author} {\bibfnamefont {P.}~\bibnamefont {Siegfried}}, \bibinfo {author} {\bibfnamefont {T.}~\bibnamefont {Skripka}}, \bibinfo {author} {\bibfnamefont {B.}~\bibnamefont {Spaun}}, \bibinfo {author} {\bibfnamefont {R.~T.}\ \bibnamefont {Sprenkle}}, \bibinfo {author} {\bibfnamefont {R.~P.}\ \bibnamefont {Stutz}}, \bibinfo {author} {\bibfnamefont {M.}~\bibnamefont {Swallows}}, \bibinfo {author} {\bibfnamefont {R.~I.}\ \bibnamefont {Tobey}}, \bibinfo {author} {\bibfnamefont {A.}~\bibnamefont {Tran}}, \bibinfo {author} {\bibfnamefont {T.}~\bibnamefont {Tran}}, \bibinfo {author} {\bibfnamefont {E.}~\bibnamefont {Vogt}}, \bibinfo {author} {\bibfnamefont {C.}~\bibnamefont {Volin}}, \bibinfo {author} {\bibfnamefont {J.}~\bibnamefont {Walker}}, \bibinfo {author} {\bibfnamefont {A.~M.}\ \bibnamefont {Zolot}},\ and\ \bibinfo {author} {\bibfnamefont
  {J.~M.}\ \bibnamefont {Pino}},\ }\href {https://doi.org/10.1103/PhysRevX.13.041052} {\bibfield  {journal} {\bibinfo  {journal} {Phys. Rev. X}\ }\textbf {\bibinfo {volume} {13}},\ \bibinfo {pages} {041052} (\bibinfo {year} {2023})}\BibitemShut {NoStop}%
\bibitem [{\citenamefont {Guo}\ \emph {et~al.}(2026)\citenamefont {Guo}, \citenamefont {Du},\ and\ \citenamefont {Ma}}]{Guo2026Entangling}%
  \BibitemOpen
  \bibfield  {author} {\bibinfo {author} {\bibfnamefont {Q.}~\bibnamefont {Guo}}, \bibinfo {author} {\bibfnamefont {B.}~\bibnamefont {Du}},\ and\ \bibinfo {author} {\bibfnamefont {R.}~\bibnamefont {Ma}},\ }\href {https://doi.org/10.1103/mr1g-zqtq} {\bibfield  {journal} {\bibinfo  {journal} {Phys. Rev. Res.}\ }\textbf {\bibinfo {volume} {8}},\ \bibinfo {pages} {033175} (\bibinfo {year} {2026})}\BibitemShut {NoStop}%
\bibitem [{\citenamefont {Breuer}\ and\ \citenamefont {Petruccione}(2002)}]{BreuerPetruccione2002}%
  \BibitemOpen
  \bibfield  {author} {\bibinfo {author} {\bibfnamefont {H.-P.}\ \bibnamefont {Breuer}}\ and\ \bibinfo {author} {\bibfnamefont {F.}~\bibnamefont {Petruccione}},\ }\href@noop {} {\emph {\bibinfo {title} {The Theory of Open Quantum Systems}}}\ (\bibinfo  {publisher} {Oxford University Press},\ \bibinfo {year} {2002})\ p.\ \bibinfo {pages} {625}\BibitemShut {NoStop}%
\bibitem [{\citenamefont {Moharramipour}\ \emph {et~al.}(2024)\citenamefont {Moharramipour}, \citenamefont {Lessa}, \citenamefont {Wang}, \citenamefont {Hsieh},\ and\ \citenamefont {Sahu}}]{Moharramipour2024S}%
  \BibitemOpen
  \bibfield  {author} {\bibinfo {author} {\bibfnamefont {A.}~\bibnamefont {Moharramipour}}, \bibinfo {author} {\bibfnamefont {L.~A.}\ \bibnamefont {Lessa}}, \bibinfo {author} {\bibfnamefont {C.}~\bibnamefont {Wang}}, \bibinfo {author} {\bibfnamefont {T.~H.}\ \bibnamefont {Hsieh}},\ and\ \bibinfo {author} {\bibfnamefont {S.}~\bibnamefont {Sahu}},\ }\href {https://doi.org/10.1103/PRXQuantum.5.040336} {\bibfield  {journal} {\bibinfo  {journal} {PRX Quantum}\ }\textbf {\bibinfo {volume} {5}},\ \bibinfo {pages} {040336} (\bibinfo {year} {2024})}\BibitemShut {NoStop}%
\bibitem [{\citenamefont {Li}\ \emph {et~al.}(2025)\citenamefont {Li}, \citenamefont {Pollmann}, \citenamefont {Read},\ and\ \citenamefont {Sala}}]{Li2025H}%
  \BibitemOpen
  \bibfield  {author} {\bibinfo {author} {\bibfnamefont {Y.}~\bibnamefont {Li}}, \bibinfo {author} {\bibfnamefont {F.}~\bibnamefont {Pollmann}}, \bibinfo {author} {\bibfnamefont {N.}~\bibnamefont {Read}},\ and\ \bibinfo {author} {\bibfnamefont {P.}~\bibnamefont {Sala}},\ }\href {https://doi.org/10.1103/PhysRevX.15.011068} {\bibfield  {journal} {\bibinfo  {journal} {Phys. Rev. X}\ }\textbf {\bibinfo {volume} {15}},\ \bibinfo {pages} {011068} (\bibinfo {year} {2025})}\BibitemShut {NoStop}%
\bibitem [{\citenamefont {Westhoff}\ \emph {et~al.}(2026)\citenamefont {Westhoff}, \citenamefont {Moroder}, \citenamefont {Schollw{\"o}ck},\ and\ \citenamefont {Paeckel}}]{Westhoff2025A}%
  \BibitemOpen
  \bibfield  {author} {\bibinfo {author} {\bibfnamefont {P.}~\bibnamefont {Westhoff}}, \bibinfo {author} {\bibfnamefont {M.}~\bibnamefont {Moroder}}, \bibinfo {author} {\bibfnamefont {U.}~\bibnamefont {Schollw{\"o}ck}},\ and\ \bibinfo {author} {\bibfnamefont {S.}~\bibnamefont {Paeckel}},\ }\href@noop {} {\bibfield  {journal} {\bibinfo  {journal} {Physical Review X}\ }\textbf {\bibinfo {volume} {16}},\ \bibinfo {pages} {041020} (\bibinfo {year} {2026})}\BibitemShut {NoStop}%
\bibitem [{Note2()}]{Note2}%
  \BibitemOpen
  \bibinfo {note} {The imaginary part of the half Fourier transform in~\protect \eqref {eq::def_S} produces the Lamb-shift contribution, which we will not need explicitly (see Section~\ref {sec::frombathmem}).}\BibitemShut {Stop}%
\bibitem [{\citenamefont {Seif}\ \emph {et~al.}(2022)\citenamefont {Seif}, \citenamefont {Wang},\ and\ \citenamefont {Clerk}}]{Seif2022D}%
  \BibitemOpen
  \bibfield  {author} {\bibinfo {author} {\bibfnamefont {A.}~\bibnamefont {Seif}}, \bibinfo {author} {\bibfnamefont {Y.-X.}\ \bibnamefont {Wang}},\ and\ \bibinfo {author} {\bibfnamefont {A.~A.}\ \bibnamefont {Clerk}},\ }\href {https://doi.org/10.1103/PhysRevLett.128.070402} {\bibfield  {journal} {\bibinfo  {journal} {Phys. Rev. Lett.}\ }\textbf {\bibinfo {volume} {128}},\ \bibinfo {pages} {070402} (\bibinfo {year} {2022})}\BibitemShut {NoStop}%
\bibitem [{\citenamefont {Lindblad}(1976)}]{Lindblad1976b}%
  \BibitemOpen
  \bibfield  {author} {\bibinfo {author} {\bibfnamefont {G.}~\bibnamefont {Lindblad}},\ }\href {https://doi.org/10.1007/BF01608499} {\bibfield  {journal} {\bibinfo  {journal} {Commun. Math. Phys.}\ }\textbf {\bibinfo {volume} {48}},\ \bibinfo {pages} {119} (\bibinfo {year} {1976})}\BibitemShut {NoStop}%
\bibitem [{\citenamefont {Imamoglu}(1994)}]{Imamoglu1994Stochastic}%
  \BibitemOpen
  \bibfield  {author} {\bibinfo {author} {\bibfnamefont {A.}~\bibnamefont {Imamoglu}},\ }\href {https://doi.org/10.1103/PhysRevA.50.3650} {\bibfield  {journal} {\bibinfo  {journal} {Phys. Rev. A}\ }\textbf {\bibinfo {volume} {50}},\ \bibinfo {pages} {3650} (\bibinfo {year} {1994})}\BibitemShut {NoStop}%
\bibitem [{\citenamefont {Dalton}\ \emph {et~al.}(2001)\citenamefont {Dalton}, \citenamefont {Barnett},\ and\ \citenamefont {Garraway}}]{Dalton2001Theory}%
  \BibitemOpen
  \bibfield  {author} {\bibinfo {author} {\bibfnamefont {B.~J.}\ \bibnamefont {Dalton}}, \bibinfo {author} {\bibfnamefont {S.~M.}\ \bibnamefont {Barnett}},\ and\ \bibinfo {author} {\bibfnamefont {B.~M.}\ \bibnamefont {Garraway}},\ }\href {https://doi.org/10.1103/PhysRevA.64.053813} {\bibfield  {journal} {\bibinfo  {journal} {Phys. Rev. A}\ }\textbf {\bibinfo {volume} {64}},\ \bibinfo {pages} {053813} (\bibinfo {year} {2001})}\BibitemShut {NoStop}%
\bibitem [{\citenamefont {Garraway}(1997)}]{Garraway1997Nonperturbative}%
  \BibitemOpen
  \bibfield  {author} {\bibinfo {author} {\bibfnamefont {B.~M.}\ \bibnamefont {Garraway}},\ }\href {https://doi.org/10.1103/PhysRevA.55.2290} {\bibfield  {journal} {\bibinfo  {journal} {Phys. Rev. A}\ }\textbf {\bibinfo {volume} {55}},\ \bibinfo {pages} {2290} (\bibinfo {year} {1997})}\BibitemShut {NoStop}%
\bibitem [{\citenamefont {Pleasance}\ \emph {et~al.}(2020)\citenamefont {Pleasance}, \citenamefont {Garraway},\ and\ \citenamefont {Petruccione}}]{Pleasance2020Generalized}%
  \BibitemOpen
  \bibfield  {author} {\bibinfo {author} {\bibfnamefont {G.}~\bibnamefont {Pleasance}}, \bibinfo {author} {\bibfnamefont {B.~M.}\ \bibnamefont {Garraway}},\ and\ \bibinfo {author} {\bibfnamefont {F.}~\bibnamefont {Petruccione}},\ }\href {https://doi.org/10.1103/PhysRevResearch.2.043058} {\bibfield  {journal} {\bibinfo  {journal} {Phys. Rev. Research}\ }\textbf {\bibinfo {volume} {2}},\ \bibinfo {pages} {043058} (\bibinfo {year} {2020})}\BibitemShut {NoStop}%
\bibitem [{\citenamefont {Debecker}\ \emph {et~al.}(2024{\natexlab{a}})\citenamefont {Debecker}, \citenamefont {Martin},\ and\ \citenamefont {Damanet}}]{Debecker2024C}%
  \BibitemOpen
  \bibfield  {author} {\bibinfo {author} {\bibfnamefont {B.}~\bibnamefont {Debecker}}, \bibinfo {author} {\bibfnamefont {J.}~\bibnamefont {Martin}},\ and\ \bibinfo {author} {\bibfnamefont {F.}~\bibnamefont {Damanet}},\ }\href {https://doi.org/10.1103/PhysRevLett.133.140403} {\bibfield  {journal} {\bibinfo  {journal} {Phys. Rev. Lett.}\ }\textbf {\bibinfo {volume} {133}},\ \bibinfo {pages} {140403} (\bibinfo {year} {2024}{\natexlab{a}})}\BibitemShut {NoStop}%
\bibitem [{\citenamefont {Blais}\ \emph {et~al.}(2021)\citenamefont {Blais}, \citenamefont {Grimsmo}, \citenamefont {Girvin},\ and\ \citenamefont {Wallraff}}]{Blais2021Circuit}%
  \BibitemOpen
  \bibfield  {author} {\bibinfo {author} {\bibfnamefont {A.}~\bibnamefont {Blais}}, \bibinfo {author} {\bibfnamefont {A.~L.}\ \bibnamefont {Grimsmo}}, \bibinfo {author} {\bibfnamefont {S.~M.}\ \bibnamefont {Girvin}},\ and\ \bibinfo {author} {\bibfnamefont {A.}~\bibnamefont {Wallraff}},\ }\href {https://doi.org/10.1103/RevModPhys.93.025005} {\bibfield  {journal} {\bibinfo  {journal} {Rev. Mod. Phys.}\ }\textbf {\bibinfo {volume} {93}},\ \bibinfo {pages} {025005} (\bibinfo {year} {2021})}\BibitemShut {NoStop}%
\bibitem [{\citenamefont {Debecker}\ \emph {et~al.}(2026)\citenamefont {Debecker}, \citenamefont {Serrano-Ens{\'a}stiga}, \citenamefont {Bastin}, \citenamefont {Damanet},\ and\ \citenamefont {Martin}}]{Debecker2026S}%
  \BibitemOpen
  \bibfield  {author} {\bibinfo {author} {\bibfnamefont {B.}~\bibnamefont {Debecker}}, \bibinfo {author} {\bibfnamefont {E.}~\bibnamefont {Serrano-Ens{\'a}stiga}}, \bibinfo {author} {\bibfnamefont {T.}~\bibnamefont {Bastin}}, \bibinfo {author} {\bibfnamefont {F.}~\bibnamefont {Damanet}},\ and\ \bibinfo {author} {\bibfnamefont {J.}~\bibnamefont {Martin}},\ }\href {https://arxiv.org/abs/2603.20091} {\bibinfo {title} {Symmetric resourceful steady states via non-markovian dissipation}} (\bibinfo {year} {2026}),\ \Eprint {https://arxiv.org/abs/2603.20091} {arXiv:2603.20091 [quant-ph]} \BibitemShut {NoStop}%
\bibitem [{\citenamefont {Debecker}\ \emph {et~al.}(2024{\natexlab{b}})\citenamefont {Debecker}, \citenamefont {Martin},\ and\ \citenamefont {Damanet}}]{Debecker2024S}%
  \BibitemOpen
  \bibfield  {author} {\bibinfo {author} {\bibfnamefont {B.}~\bibnamefont {Debecker}}, \bibinfo {author} {\bibfnamefont {J.}~\bibnamefont {Martin}},\ and\ \bibinfo {author} {\bibfnamefont {F.}~\bibnamefont {Damanet}},\ }\href {https://doi.org/10.1103/PhysRevA.110.042201} {\bibfield  {journal} {\bibinfo  {journal} {Phys. Rev. A}\ }\textbf {\bibinfo {volume} {110}},\ \bibinfo {pages} {042201} (\bibinfo {year} {2024}{\natexlab{b}})}\BibitemShut {NoStop}%
\bibitem [{\citenamefont {Prosen}(2008)}]{Prosen2008T}%
  \BibitemOpen
  \bibfield  {author} {\bibinfo {author} {\bibfnamefont {T.}~\bibnamefont {Prosen}},\ }\href {https://doi.org/10.1088/1367-2630/10/4/043026} {\bibfield  {journal} {\bibinfo  {journal} {New Journal of Physics}\ }\textbf {\bibinfo {volume} {10}},\ \bibinfo {pages} {043026} (\bibinfo {year} {2008})}\BibitemShut {NoStop}%
\bibitem [{\citenamefont {Prosen}\ and\ \citenamefont {Seligman}(2010)}]{Prosen2010Q}%
  \BibitemOpen
  \bibfield  {author} {\bibinfo {author} {\bibfnamefont {T.}~\bibnamefont {Prosen}}\ and\ \bibinfo {author} {\bibfnamefont {T.~H.}\ \bibnamefont {Seligman}},\ }\href {https://doi.org/10.1088/1751-8113/43/39/392004} {\bibfield  {journal} {\bibinfo  {journal} {Journal of Physics A: Mathematical and Theoretical}\ }\textbf {\bibinfo {volume} {43}},\ \bibinfo {pages} {392004} (\bibinfo {year} {2010})}\BibitemShut {NoStop}%
\bibitem [{\citenamefont {Barthel}\ and\ \citenamefont {Zhang}(2022)}]{Barthel_2022}%
  \BibitemOpen
  \bibfield  {author} {\bibinfo {author} {\bibfnamefont {T.}~\bibnamefont {Barthel}}\ and\ \bibinfo {author} {\bibfnamefont {Y.}~\bibnamefont {Zhang}},\ }\href {https://doi.org/10.1088/1742-5468/ac8e5c} {\bibfield  {journal} {\bibinfo  {journal} {Journal of Statistical Mechanics: Theory and Experiment}\ }\textbf {\bibinfo {volume} {2022}},\ \bibinfo {pages} {113101} (\bibinfo {year} {2022})}\BibitemShut {NoStop}%
\bibitem [{\citenamefont {Bu{\v{c}}a}\ and\ \citenamefont {Prosen}(2012)}]{Buca2012}%
  \BibitemOpen
  \bibfield  {author} {\bibinfo {author} {\bibfnamefont {B.}~\bibnamefont {Bu{\v{c}}a}}\ and\ \bibinfo {author} {\bibfnamefont {T.}~\bibnamefont {Prosen}},\ }\href {https://doi.org/10.1088/1367-2630/14/7/073007} {\bibfield  {journal} {\bibinfo  {journal} {New Journal of Physics}\ }\textbf {\bibinfo {volume} {14}},\ \bibinfo {pages} {073007} (\bibinfo {year} {2012})}\BibitemShut {NoStop}%
\bibitem [{\citenamefont {Gühne}\ and\ \citenamefont {Tóth}(2009)}]{Guhne2009E}%
  \BibitemOpen
  \bibfield  {author} {\bibinfo {author} {\bibfnamefont {O.}~\bibnamefont {Gühne}}\ and\ \bibinfo {author} {\bibfnamefont {G.}~\bibnamefont {Tóth}},\ }\href {https://doi.org/https://doi.org/10.1016/j.physrep.2009.02.004} {\bibfield  {journal} {\bibinfo  {journal} {Physics Reports}\ }\textbf {\bibinfo {volume} {474}},\ \bibinfo {pages} {1} (\bibinfo {year} {2009})}\BibitemShut {NoStop}%
\bibitem [{\citenamefont {Lu}\ \emph {et~al.}(2007)\citenamefont {Lu}, \citenamefont {Zhou}, \citenamefont {G{\"u}hne}, \citenamefont {Gao}, \citenamefont {Zhang}, \citenamefont {Yuan}, \citenamefont {Goebel}, \citenamefont {Yang},\ and\ \citenamefont {Pan}}]{Lu2007E}%
  \BibitemOpen
  \bibfield  {author} {\bibinfo {author} {\bibfnamefont {C.-Y.}\ \bibnamefont {Lu}}, \bibinfo {author} {\bibfnamefont {X.-Q.}\ \bibnamefont {Zhou}}, \bibinfo {author} {\bibfnamefont {O.}~\bibnamefont {G{\"u}hne}}, \bibinfo {author} {\bibfnamefont {W.-B.}\ \bibnamefont {Gao}}, \bibinfo {author} {\bibfnamefont {J.}~\bibnamefont {Zhang}}, \bibinfo {author} {\bibfnamefont {Z.-S.}\ \bibnamefont {Yuan}}, \bibinfo {author} {\bibfnamefont {A.}~\bibnamefont {Goebel}}, \bibinfo {author} {\bibfnamefont {T.}~\bibnamefont {Yang}},\ and\ \bibinfo {author} {\bibfnamefont {J.-W.}\ \bibnamefont {Pan}},\ }\href {https://doi.org/10.1038/nphys507} {\bibfield  {journal} {\bibinfo  {journal} {Nature Physics}\ }\textbf {\bibinfo {volume} {3}},\ \bibinfo {pages} {91} (\bibinfo {year} {2007})}\BibitemShut {NoStop}%
\bibitem [{\citenamefont {Wu}\ \emph {et~al.}(2026)\citenamefont {Wu}, \citenamefont {Shi}, \citenamefont {Sun}, \citenamefont {Wang}, \citenamefont {Gu}, \citenamefont {Chiribella}, \citenamefont {Zhao},\ and\ \citenamefont {Wu}}]{Wu2026E}%
  \BibitemOpen
  \bibfield  {author} {\bibinfo {author} {\bibfnamefont {D.}~\bibnamefont {Wu}}, \bibinfo {author} {\bibfnamefont {F.}~\bibnamefont {Shi}}, \bibinfo {author} {\bibfnamefont {J.-C.}\ \bibnamefont {Sun}}, \bibinfo {author} {\bibfnamefont {B.-W.}\ \bibnamefont {Wang}}, \bibinfo {author} {\bibfnamefont {X.-M.}\ \bibnamefont {Gu}}, \bibinfo {author} {\bibfnamefont {G.}~\bibnamefont {Chiribella}}, \bibinfo {author} {\bibfnamefont {Q.}~\bibnamefont {Zhao}},\ and\ \bibinfo {author} {\bibfnamefont {J.}~\bibnamefont {Wu}},\ }\href {https://doi.org/10.1103/w6fl-nx59} {\bibfield  {journal} {\bibinfo  {journal} {Phys. Rev. Lett.}\ }\textbf {\bibinfo {volume} {137}},\ \bibinfo {pages} {100201} (\bibinfo {year} {2026})}\BibitemShut {NoStop}%
\bibitem [{\citenamefont {Gühne}\ and\ \citenamefont {Seevinck}(2010)}]{Guhne2010S}%
  \BibitemOpen
  \bibfield  {author} {\bibinfo {author} {\bibfnamefont {O.}~\bibnamefont {Gühne}}\ and\ \bibinfo {author} {\bibfnamefont {M.}~\bibnamefont {Seevinck}},\ }\href {https://doi.org/10.1088/1367-2630/12/5/053002} {\bibfield  {journal} {\bibinfo  {journal} {New Journal of Physics}\ }\textbf {\bibinfo {volume} {12}},\ \bibinfo {pages} {053002} (\bibinfo {year} {2010})}\BibitemShut {NoStop}%
\bibitem [{\citenamefont {Gittsovich}\ \emph {et~al.}(2010)\citenamefont {Gittsovich}, \citenamefont {Hyllus},\ and\ \citenamefont {G{\"u}hne}}]{Gittsovich2010M}%
  \BibitemOpen
  \bibfield  {author} {\bibinfo {author} {\bibfnamefont {O.}~\bibnamefont {Gittsovich}}, \bibinfo {author} {\bibfnamefont {P.}~\bibnamefont {Hyllus}},\ and\ \bibinfo {author} {\bibfnamefont {O.}~\bibnamefont {G{\"u}hne}},\ }\href {https://doi.org/10.1103/PhysRevA.82.032306} {\bibfield  {journal} {\bibinfo  {journal} {Phys. Rev. A}\ }\textbf {\bibinfo {volume} {82}},\ \bibinfo {pages} {032306} (\bibinfo {year} {2010})}\BibitemShut {NoStop}%
\bibitem [{\citenamefont {T{\'o}th}\ and\ \citenamefont {G{\"u}hne}(2005)}]{Toth2005D}%
  \BibitemOpen
  \bibfield  {author} {\bibinfo {author} {\bibfnamefont {G.}~\bibnamefont {T{\'o}th}}\ and\ \bibinfo {author} {\bibfnamefont {O.}~\bibnamefont {G{\"u}hne}},\ }\href {https://doi.org/10.1103/PhysRevLett.94.060501} {\bibfield  {journal} {\bibinfo  {journal} {Phys. Rev. Lett.}\ }\textbf {\bibinfo {volume} {94}},\ \bibinfo {pages} {060501} (\bibinfo {year} {2005})}\BibitemShut {NoStop}%
\bibitem [{\citenamefont {Knips}\ \emph {et~al.}(2016)\citenamefont {Knips}, \citenamefont {Schwemmer}, \citenamefont {Klein}, \citenamefont {Wie{\ifmmode \acute{s}\else \'{s}\fi{}}niak},\ and\ \citenamefont {Weinfurter}}]{Knips2016M}%
  \BibitemOpen
  \bibfield  {author} {\bibinfo {author} {\bibfnamefont {L.}~\bibnamefont {Knips}}, \bibinfo {author} {\bibfnamefont {C.}~\bibnamefont {Schwemmer}}, \bibinfo {author} {\bibfnamefont {N.}~\bibnamefont {Klein}}, \bibinfo {author} {\bibfnamefont {M.}~\bibnamefont {Wie{\ifmmode \acute{s}\else \'{s}\fi{}}niak}},\ and\ \bibinfo {author} {\bibfnamefont {H.}~\bibnamefont {Weinfurter}},\ }\href {https://doi.org/10.1103/PhysRevLett.117.210504} {\bibfield  {journal} {\bibinfo  {journal} {Phys. Rev. Lett.}\ }\textbf {\bibinfo {volume} {117}},\ \bibinfo {pages} {210504} (\bibinfo {year} {2016})}\BibitemShut {NoStop}%
\bibitem [{\citenamefont {Li}\ \emph {et~al.}(2026)\citenamefont {Li}, \citenamefont {Dai}, \citenamefont {Mu{\~{n}}oz-Arias}, \citenamefont {Reuer}, \citenamefont {Huber},\ and\ \citenamefont {Friis}}]{Li2026}%
  \BibitemOpen
  \bibfield  {author} {\bibinfo {author} {\bibfnamefont {N.~K.~H.}\ \bibnamefont {Li}}, \bibinfo {author} {\bibfnamefont {X.}~\bibnamefont {Dai}}, \bibinfo {author} {\bibfnamefont {M.~H.}\ \bibnamefont {Mu{\~{n}}oz-Arias}}, \bibinfo {author} {\bibfnamefont {K.}~\bibnamefont {Reuer}}, \bibinfo {author} {\bibfnamefont {M.}~\bibnamefont {Huber}},\ and\ \bibinfo {author} {\bibfnamefont {N.}~\bibnamefont {Friis}},\ }\href {https://doi.org/10.1038/s41467-026-69320-4} {\bibfield  {journal} {\bibinfo  {journal} {Nature Communications}\ }\textbf {\bibinfo {volume} {17}},\ \bibinfo {pages} {1707} (\bibinfo {year} {2026})}\BibitemShut {NoStop}%
\bibitem [{\citenamefont {Pagano}\ \emph {et~al.}(2025)\citenamefont {Pagano}, \citenamefont {Adamczyk},\ and\ \citenamefont {So}}]{pagano2025varenna}%
  \BibitemOpen
  \bibfield  {author} {\bibinfo {author} {\bibfnamefont {G.}~\bibnamefont {Pagano}}, \bibinfo {author} {\bibfnamefont {W.}~\bibnamefont {Adamczyk}},\ and\ \bibinfo {author} {\bibfnamefont {V.}~\bibnamefont {So}},\ }\href@noop {} {\bibinfo {title} {Fundamentals of trapped ions and quantum simulation of chemical dynamics}} (\bibinfo {year} {2025}),\ \Eprint {https://arxiv.org/abs/2505.20412} {arXiv:2505.20412 [quant-ph]} \BibitemShut {NoStop}%
\bibitem [{\citenamefont {Monroe}\ \emph {et~al.}(2021)\citenamefont {Monroe}, \citenamefont {Campbell}, \citenamefont {Duan}, \citenamefont {Gong}, \citenamefont {Gorshkov}, \citenamefont {Hess}, \citenamefont {Islam}, \citenamefont {Kim}, \citenamefont {Linke}, \citenamefont {Pagano}, \citenamefont {Richerme}, \citenamefont {Senko},\ and\ \citenamefont {Yao}}]{monroe2021programmable}%
  \BibitemOpen
  \bibfield  {author} {\bibinfo {author} {\bibfnamefont {C.}~\bibnamefont {Monroe}}, \bibinfo {author} {\bibfnamefont {W.~C.}\ \bibnamefont {Campbell}}, \bibinfo {author} {\bibfnamefont {L.-M.}\ \bibnamefont {Duan}}, \bibinfo {author} {\bibfnamefont {Z.-X.}\ \bibnamefont {Gong}}, \bibinfo {author} {\bibfnamefont {A.~V.}\ \bibnamefont {Gorshkov}}, \bibinfo {author} {\bibfnamefont {P.~W.}\ \bibnamefont {Hess}}, \bibinfo {author} {\bibfnamefont {R.}~\bibnamefont {Islam}}, \bibinfo {author} {\bibfnamefont {K.}~\bibnamefont {Kim}}, \bibinfo {author} {\bibfnamefont {N.~M.}\ \bibnamefont {Linke}}, \bibinfo {author} {\bibfnamefont {G.}~\bibnamefont {Pagano}}, \bibinfo {author} {\bibfnamefont {P.}~\bibnamefont {Richerme}}, \bibinfo {author} {\bibfnamefont {C.}~\bibnamefont {Senko}},\ and\ \bibinfo {author} {\bibfnamefont {N.~Y.}\ \bibnamefont {Yao}},\ }\href {https://doi.org/10.1103/RevModPhys.93.025001} {\bibfield  {journal} {\bibinfo  {journal} {Rev. Mod. Phys.}\ }\textbf {\bibinfo {volume} {93}},\ \bibinfo {pages}
  {025001} (\bibinfo {year} {2021})}\BibitemShut {NoStop}%
\bibitem [{\citenamefont {Lemmer}\ \emph {et~al.}(2018)\citenamefont {Lemmer}, \citenamefont {Cormick}, \citenamefont {Tamascelli}, \citenamefont {Schaetz}, \citenamefont {Huelga},\ and\ \citenamefont {Plenio}}]{lemmer2018engineering}%
  \BibitemOpen
  \bibfield  {author} {\bibinfo {author} {\bibfnamefont {A.}~\bibnamefont {Lemmer}}, \bibinfo {author} {\bibfnamefont {C.}~\bibnamefont {Cormick}}, \bibinfo {author} {\bibfnamefont {D.}~\bibnamefont {Tamascelli}}, \bibinfo {author} {\bibfnamefont {T.}~\bibnamefont {Schaetz}}, \bibinfo {author} {\bibfnamefont {S.~F.}\ \bibnamefont {Huelga}},\ and\ \bibinfo {author} {\bibfnamefont {M.~B.}\ \bibnamefont {Plenio}},\ }\href@noop {} {\bibfield  {journal} {\bibinfo  {journal} {New Journal of Physics}\ }\textbf {\bibinfo {volume} {20}},\ \bibinfo {pages} {073002} (\bibinfo {year} {2018})}\BibitemShut {NoStop}%
\bibitem [{\citenamefont {Marijanovi{\'c}}\ \emph {et~al.}(2026)\citenamefont {Marijanovi{\'c}}, \citenamefont {Bhattacharya}, \citenamefont {Menon}, \citenamefont {So}, \citenamefont {Pagano},\ and\ \citenamefont {Demler}}]{Marijanovic2026}%
  \BibitemOpen
  \bibfield  {author} {\bibinfo {author} {\bibfnamefont {F.}~\bibnamefont {Marijanovi{\'c}}}, \bibinfo {author} {\bibfnamefont {U.}~\bibnamefont {Bhattacharya}}, \bibinfo {author} {\bibfnamefont {A.}~\bibnamefont {Menon}}, \bibinfo {author} {\bibfnamefont {V.}~\bibnamefont {So}}, \bibinfo {author} {\bibfnamefont {G.}~\bibnamefont {Pagano}},\ and\ \bibinfo {author} {\bibfnamefont {E.}~\bibnamefont {Demler}},\ }\href {https://doi.org/10.48550/arXiv.2610.18970} {\bibinfo {title} {{Programmable Polaron Models and Tomography in Trapped-Ion Quantum Simulators}}} (\bibinfo {year} {2026}),\ \Eprint {https://arxiv.org/abs/2610.18970} {arXiv:2610.18970 [quant-ph]} \BibitemShut {NoStop}%
\bibitem [{\citenamefont {Kyprianidis}\ \emph {et~al.}(2024)\citenamefont {Kyprianidis}, \citenamefont {Rasmusson},\ and\ \citenamefont {Richerme}}]{Kyprianidis2024interaction}%
  \BibitemOpen
  \bibfield  {author} {\bibinfo {author} {\bibfnamefont {A.}~\bibnamefont {Kyprianidis}}, \bibinfo {author} {\bibfnamefont {A.~J.}\ \bibnamefont {Rasmusson}},\ and\ \bibinfo {author} {\bibfnamefont {P.}~\bibnamefont {Richerme}},\ }\href {https://doi.org/10.1088/1367-2630/ad264d} {\bibfield  {journal} {\bibinfo  {journal} {New Journal of Physics}\ }\textbf {\bibinfo {volume} {26}},\ \bibinfo {pages} {023033} (\bibinfo {year} {2024})}\BibitemShut {NoStop}%
\bibitem [{\citenamefont {De}\ \emph {et~al.}(2025)\citenamefont {De}, \citenamefont {Cook}, \citenamefont {Ali}, \citenamefont {Collins}, \citenamefont {Morong}, \citenamefont {Paz}, \citenamefont {Titum}, \citenamefont {Pagano}, \citenamefont {Gorshkov}, \citenamefont {Maghrebi},\ and\ \citenamefont {Monroe}}]{De2025NonEquilibrium}%
  \BibitemOpen
  \bibfield  {author} {\bibinfo {author} {\bibfnamefont {A.}~\bibnamefont {De}}, \bibinfo {author} {\bibfnamefont {P.}~\bibnamefont {Cook}}, \bibinfo {author} {\bibfnamefont {M.}~\bibnamefont {Ali}}, \bibinfo {author} {\bibfnamefont {K.}~\bibnamefont {Collins}}, \bibinfo {author} {\bibfnamefont {W.}~\bibnamefont {Morong}}, \bibinfo {author} {\bibfnamefont {D.}~\bibnamefont {Paz}}, \bibinfo {author} {\bibfnamefont {P.}~\bibnamefont {Titum}}, \bibinfo {author} {\bibfnamefont {G.}~\bibnamefont {Pagano}}, \bibinfo {author} {\bibfnamefont {A.~V.}\ \bibnamefont {Gorshkov}}, \bibinfo {author} {\bibfnamefont {M.}~\bibnamefont {Maghrebi}},\ and\ \bibinfo {author} {\bibfnamefont {C.}~\bibnamefont {Monroe}},\ }\href {https://doi.org/10.1038/s41467-025-63398-y} {\bibfield  {journal} {\bibinfo  {journal} {Nature Communications}\ }\textbf {\bibinfo {volume} {16}},\ \bibinfo {pages} {7939} (\bibinfo {year} {2025})}\BibitemShut {NoStop}%
\bibitem [{\citenamefont {De}\ \emph {et~al.}(2026)\citenamefont {De}, \citenamefont {Lerose}, \citenamefont {Luo}, \citenamefont {Surace}, \citenamefont {Schuckert}, \citenamefont {Bennewitz}, \citenamefont {Ware}, \citenamefont {Morong}, \citenamefont {Collins}, \citenamefont {Davoudi}, \citenamefont {Gorshkov}, \citenamefont {Katz},\ and\ \citenamefont {Monroe}}]{De2026StringBreaking}%
  \BibitemOpen
  \bibfield  {author} {\bibinfo {author} {\bibfnamefont {A.}~\bibnamefont {De}}, \bibinfo {author} {\bibfnamefont {A.}~\bibnamefont {Lerose}}, \bibinfo {author} {\bibfnamefont {D.}~\bibnamefont {Luo}}, \bibinfo {author} {\bibfnamefont {F.~M.}\ \bibnamefont {Surace}}, \bibinfo {author} {\bibfnamefont {A.}~\bibnamefont {Schuckert}}, \bibinfo {author} {\bibfnamefont {E.~R.}\ \bibnamefont {Bennewitz}}, \bibinfo {author} {\bibfnamefont {B.}~\bibnamefont {Ware}}, \bibinfo {author} {\bibfnamefont {W.}~\bibnamefont {Morong}}, \bibinfo {author} {\bibfnamefont {K.~S.}\ \bibnamefont {Collins}}, \bibinfo {author} {\bibfnamefont {Z.}~\bibnamefont {Davoudi}}, \bibinfo {author} {\bibfnamefont {A.~V.}\ \bibnamefont {Gorshkov}}, \bibinfo {author} {\bibfnamefont {O.}~\bibnamefont {Katz}},\ and\ \bibinfo {author} {\bibfnamefont {C.}~\bibnamefont {Monroe}},\ }\bibfield  {journal} {\bibinfo  {journal} {Nature Physics}\ }\href {https://doi.org/10.1038/s41567-026-03422-0} {10.1038/s41567-026-03422-0} (\bibinfo {year}
  {2026})\BibitemShut {NoStop}%
\bibitem [{\citenamefont {So}\ \emph {et~al.}(2026{\natexlab{b}})\citenamefont {So}, \citenamefont {Zhu}, \citenamefont {Duraisamy~Suganthi}, \citenamefont {Menon}, \citenamefont {Tomaras}, \citenamefont {Zhuravel}, \citenamefont {Pu},\ and\ \citenamefont {Pagano}}]{so2026finitetemp}%
  \BibitemOpen
  \bibfield  {author} {\bibinfo {author} {\bibfnamefont {V.}~\bibnamefont {So}}, \bibinfo {author} {\bibfnamefont {M.}~\bibnamefont {Zhu}}, \bibinfo {author} {\bibfnamefont {M.}~\bibnamefont {Duraisamy~Suganthi}}, \bibinfo {author} {\bibfnamefont {A.}~\bibnamefont {Menon}}, \bibinfo {author} {\bibfnamefont {G.}~\bibnamefont {Tomaras}}, \bibinfo {author} {\bibfnamefont {R.}~\bibnamefont {Zhuravel}}, \bibinfo {author} {\bibfnamefont {H.}~\bibnamefont {Pu}},\ and\ \bibinfo {author} {\bibfnamefont {G.}~\bibnamefont {Pagano}},\ }\href {https://doi.org/10.1103/mf8d-91ws} {\bibfield  {journal} {\bibinfo  {journal} {Phys. Rev. Lett.}\ }\textbf {\bibinfo {volume} {137}},\ \bibinfo {pages} {050604} (\bibinfo {year} {2026}{\natexlab{b}})}\BibitemShut {NoStop}%
\bibitem [{\citenamefont {Turner}\ \emph {et~al.}(2018)\citenamefont {Turner}, \citenamefont {Michailidis}, \citenamefont {Abanin}, \citenamefont {Serbyn},\ and\ \citenamefont {Papi{\'{c}}}}]{Turner2018W}%
  \BibitemOpen
  \bibfield  {author} {\bibinfo {author} {\bibfnamefont {C.~J.}\ \bibnamefont {Turner}}, \bibinfo {author} {\bibfnamefont {A.~A.}\ \bibnamefont {Michailidis}}, \bibinfo {author} {\bibfnamefont {D.~A.}\ \bibnamefont {Abanin}}, \bibinfo {author} {\bibfnamefont {M.}~\bibnamefont {Serbyn}},\ and\ \bibinfo {author} {\bibfnamefont {Z.}~\bibnamefont {Papi{\'{c}}}},\ }\href {https://doi.org/10.1038/s41567-018-0137-5} {\bibfield  {journal} {\bibinfo  {journal} {Nature Physics}\ }\textbf {\bibinfo {volume} {14}},\ \bibinfo {pages} {745} (\bibinfo {year} {2018})}\BibitemShut {NoStop}%
\bibitem [{\citenamefont {Pocklington}\ and\ \citenamefont {Clerk}(2024)}]{Pocklington2024}%
  \BibitemOpen
  \bibfield  {author} {\bibinfo {author} {\bibfnamefont {A.}~\bibnamefont {Pocklington}}\ and\ \bibinfo {author} {\bibfnamefont {A.~A.}\ \bibnamefont {Clerk}},\ }\href {https://doi.org/10.1103/PRXQuantum.5.040305} {\bibfield  {journal} {\bibinfo  {journal} {PRX Quantum}\ }\textbf {\bibinfo {volume} {5}},\ \bibinfo {pages} {040305} (\bibinfo {year} {2024})}\BibitemShut {NoStop}%
\bibitem [{\citenamefont {Norris}(1997)}]{Norris1997}%
  \BibitemOpen
  \bibfield  {author} {\bibinfo {author} {\bibfnamefont {J.~R.}\ \bibnamefont {Norris}},\ }\href {https://doi.org/10.1017/CBO9780511810633} {\emph {\bibinfo {title} {{Markov Chains}}}},\ \bibinfo {series} {Cambridge Series in Statistical and Probabilistic Mathematics}, Vol.~\bibinfo {volume} {2}\ (\bibinfo  {publisher} {Cambridge University Press},\ \bibinfo {address} {Cambridge},\ \bibinfo {year} {1997})\BibitemShut {NoStop}%
\bibitem [{\citenamefont {Deschamps}\ \emph {et~al.}(2016)\citenamefont {Deschamps}, \citenamefont {Fagnola}, \citenamefont {Sasso},\ and\ \citenamefont {Umanit{\`a}}}]{Deschamps2016S}%
  \BibitemOpen
  \bibfield  {author} {\bibinfo {author} {\bibfnamefont {J.}~\bibnamefont {Deschamps}}, \bibinfo {author} {\bibfnamefont {F.}~\bibnamefont {Fagnola}}, \bibinfo {author} {\bibfnamefont {E.}~\bibnamefont {Sasso}},\ and\ \bibinfo {author} {\bibfnamefont {V.}~\bibnamefont {Umanit{\`a}}},\ }\href {https://doi.org/10.1142/S0129055X16500033} {\bibfield  {journal} {\bibinfo  {journal} {Reviews in Mathematical Physics}\ }\textbf {\bibinfo {volume} {28}},\ \bibinfo {pages} {1650003} (\bibinfo {year} {2016})},\ \Eprint {https://arxiv.org/abs/https://doi.org/10.1142/S0129055X16500033} {https://doi.org/10.1142/S0129055X16500033} \BibitemShut {NoStop}%
\bibitem [{\citenamefont {Casagrande}\ \emph {et~al.}(2021)\citenamefont {Casagrande}, \citenamefont {Poletti},\ and\ \citenamefont {Landi}}]{Casagrande2021}%
  \BibitemOpen
  \bibfield  {author} {\bibinfo {author} {\bibfnamefont {H.~P.}\ \bibnamefont {Casagrande}}, \bibinfo {author} {\bibfnamefont {D.}~\bibnamefont {Poletti}},\ and\ \bibinfo {author} {\bibfnamefont {G.~T.}\ \bibnamefont {Landi}},\ }\href {https://doi.org/10.1016/j.cpc.2021.108060} {\bibfield  {journal} {\bibinfo  {journal} {Computer Physics Communications}\ }\textbf {\bibinfo {volume} {267}},\ \bibinfo {pages} {108060} (\bibinfo {year} {2021})}\BibitemShut {NoStop}%
\bibitem [{\citenamefont {K{\"o}hler}\ \emph {et~al.}(2021)\citenamefont {K{\"o}hler}, \citenamefont {Stolpp},\ and\ \citenamefont {Paeckel}}]{Koehler_2021}%
  \BibitemOpen
  \bibfield  {author} {\bibinfo {author} {\bibfnamefont {T.}~\bibnamefont {K{\"o}hler}}, \bibinfo {author} {\bibfnamefont {J.}~\bibnamefont {Stolpp}},\ and\ \bibinfo {author} {\bibfnamefont {S.}~\bibnamefont {Paeckel}},\ }\href {https://doi.org/10.21468/scipostphys.10.3.058} {\bibfield  {journal} {\bibinfo  {journal} {SciPost Physics}\ }\textbf {\bibinfo {volume} {10}},\ \bibinfo {pages} {058} (\bibinfo {year} {2021})}\BibitemShut {NoStop}%
\bibitem [{\citenamefont {Minganti}\ and\ \citenamefont {Huybrechts}(2022)}]{Minganti2022}%
  \BibitemOpen
  \bibfield  {author} {\bibinfo {author} {\bibfnamefont {F.}~\bibnamefont {Minganti}}\ and\ \bibinfo {author} {\bibfnamefont {D.}~\bibnamefont {Huybrechts}},\ }\href {https://doi.org/10.22331/q-2022-02-10-649} {\bibfield  {journal} {\bibinfo  {journal} {Quantum}\ }\textbf {\bibinfo {volume} {6}},\ \bibinfo {pages} {649} (\bibinfo {year} {2022})}\BibitemShut {NoStop}%
\bibitem [{\citenamefont {Grundner}\ \emph {et~al.}(2024)\citenamefont {Grundner}, \citenamefont {Westhoff}, \citenamefont {Kugler}, \citenamefont {Parcollet},\ and\ \citenamefont {Schollw\"ock}}]{Grundner2024}%
  \BibitemOpen
  \bibfield  {author} {\bibinfo {author} {\bibfnamefont {M.}~\bibnamefont {Grundner}}, \bibinfo {author} {\bibfnamefont {P.}~\bibnamefont {Westhoff}}, \bibinfo {author} {\bibfnamefont {F.~B.}\ \bibnamefont {Kugler}}, \bibinfo {author} {\bibfnamefont {O.}~\bibnamefont {Parcollet}},\ and\ \bibinfo {author} {\bibfnamefont {U.}~\bibnamefont {Schollw\"ock}},\ }\href {https://doi.org/10.1103/PhysRevB.109.155124} {\bibfield  {journal} {\bibinfo  {journal} {Phys. Rev. B}\ }\textbf {\bibinfo {volume} {109}},\ \bibinfo {pages} {155124} (\bibinfo {year} {2024})}\BibitemShut {NoStop}%
\bibitem [{\citenamefont {Paeckel}(2026)}]{Paeckel2026NS}%
  \BibitemOpen
  \bibfield  {author} {\bibinfo {author} {\bibfnamefont {S.}~\bibnamefont {Paeckel}},\ }\href {https://doi.org/10.1103/bx76-hps4} {\bibfield  {journal} {\bibinfo  {journal} {Phys. Rev. Lett.}\ }\textbf {\bibinfo {volume} {136}},\ \bibinfo {pages} {160401} (\bibinfo {year} {2026})}\BibitemShut {NoStop}%
\bibitem [{\citenamefont {Schollw{\"o}ck}(2011)}]{Schollwoeck2011}%
  \BibitemOpen
  \bibfield  {author} {\bibinfo {author} {\bibfnamefont {U.}~\bibnamefont {Schollw{\"o}ck}},\ }\href {https://www.sciencedirect.com/science/article/pii/S0003491610001752} {\bibfield  {journal} {\bibinfo  {journal} {Ann. Phys.}\ }\textbf {\bibinfo {volume} {326}},\ \bibinfo {pages} {96} (\bibinfo {year} {2011})}\BibitemShut {NoStop}%
\bibitem [{\citenamefont {Am-Shallem}\ \emph {et~al.}(2015)\citenamefont {Am-Shallem}, \citenamefont {Levy}, \citenamefont {Schaefer},\ and\ \citenamefont {Kosloff}}]{Amshallem2015}%
  \BibitemOpen
  \bibfield  {author} {\bibinfo {author} {\bibfnamefont {M.}~\bibnamefont {Am-Shallem}}, \bibinfo {author} {\bibfnamefont {A.}~\bibnamefont {Levy}}, \bibinfo {author} {\bibfnamefont {I.}~\bibnamefont {Schaefer}},\ and\ \bibinfo {author} {\bibfnamefont {R.}~\bibnamefont {Kosloff}},\ }\href {https://arxiv.org/abs/1510.08634} {\bibinfo {title} {Three approaches for representing lindblad dynamics by a matrix-vector notation}} (\bibinfo {year} {2015}),\ \Eprint {https://arxiv.org/abs/1510.08634} {arXiv:1510.08634 [quant-ph]} \BibitemShut {NoStop}%
\bibitem [{\citenamefont {Yang}\ and\ \citenamefont {White}(2020)}]{Yang2020}%
  \BibitemOpen
  \bibfield  {author} {\bibinfo {author} {\bibfnamefont {M.}~\bibnamefont {Yang}}\ and\ \bibinfo {author} {\bibfnamefont {S.~R.}\ \bibnamefont {White}},\ }\href {https://doi.org/10.1103%2Fphysrevb.102.094315} {\bibfield  {journal} {\bibinfo  {journal} {Phys. Rev. B}\ }\textbf {\bibinfo {volume} {102}},\ \bibinfo {pages} {094315} (\bibinfo {year} {2020})}\BibitemShut {NoStop}%
\bibitem [{\citenamefont {So}\ \emph {et~al.}(2026{\natexlab{c}})\citenamefont {So}, \citenamefont {Zhu}, \citenamefont {Pu},\ and\ \citenamefont {Pagano}}]{So2026C}%
  \BibitemOpen
  \bibfield  {author} {\bibinfo {author} {\bibfnamefont {V.}~\bibnamefont {So}}, \bibinfo {author} {\bibfnamefont {M.}~\bibnamefont {Zhu}}, \bibinfo {author} {\bibfnamefont {H.}~\bibnamefont {Pu}},\ and\ \bibinfo {author} {\bibfnamefont {G.}~\bibnamefont {Pagano}},\ }\href {https://doi.org/10.48550/arXiv.2610.12579} {\bibinfo {title} {{Correlated Pair Creation with Positive- and Negative-Frequency Oscillators in a Trapped-Ion Quantum Simulator}}} (\bibinfo {year} {2026}{\natexlab{c}}),\ \Eprint {https://arxiv.org/abs/2610.12579} {arXiv:2610.12579 [quant-ph]} \BibitemShut {NoStop}%
\end{thebibliography}%

\end{document}